\documentclass[usenatbib,a4paper]{mn2e}  

\usepackage{graphicx}
\usepackage{txfonts}
\usepackage{natbib}
\usepackage{supertabular}
\usepackage{longtable}
\usepackage{url}
\usepackage{dcolumn}
\usepackage{placeins}
\usepackage{multirow}

\begin{document}

\newcommand{\zabs}{\ensuremath{z_{\rm abs}}}
\newcommand{\zem}{\ensuremath{z_{\rm em}}}
\newcommand{\zqso}{\ensuremath{z_{\rm QSO}}}
\newcommand{\zgal}{\ensuremath{z_{\rm gal}}}
\newcommand{\HH}{\mbox{H$_2$}}
\newcommand{\HD}{\mbox{HD}}
\newcommand{\DD}{\mbox{D$_2$}}
\newcommand{\CO}{\mbox{CO}}
\newcommand{\dla}{damped Lyman-$\alpha$}
\newcommand{\Dla}{Damped Lyman-$\alpha$}
\newcommand{\lya}{Ly-$\alpha$}
\newcommand{\lyb}{Ly-$\beta$}
\newcommand{\lyg}{Ly-$\gamma$}
\newcommand{\Ha}{H$\alpha$}
\newcommand{\Hb}{H$\beta$}
\newcommand{\Hg}{H$\gamma$}
\newcommand{\CaII}{\mbox{Ca\,{\sc ii}}}
\newcommand{\FeII}{\mbox{Fe\,{\sc ii}}}
\newcommand{\HI}{\mbox{H\,{\sc i}}}
\newcommand{\MgI}{\mbox{Mg\,{\sc i}}}
\newcommand{\MgII}{\mbox{Mg\,{\sc ii}}}
\newcommand{\MnII}{\mbox{Mn\,{\sc ii}}}
\newcommand{\NaI}{\mbox{Na\,{\sc i}}}
\newcommand{\OII}{\mbox{O\,{\sc ii}}}
\newcommand{\OIII}{\mbox{O\,{\sc iii}}}
\newcommand{\Ho}{\ensuremath{H_0}}
\newcommand{\abs}[1]{\left| #1 \right|} 
\newcommand{\avg}[1]{\left< #1 \right>} 
\newcommand{\kms}{\ensuremath{{\rm km\,s^{-1}}}}
\newcommand{\cmsq}{\ensuremath{{\rm cm}^{-2}}}
\newcommand{\Pks}{\ensuremath{{\rm P}_{\rm KS}}}
\newcommand{\WMgII}{\ensuremath{W_{\lambda 2796}}}
\newcommand{\WMgI}{\ensuremath{W_{\lambda 2852}}}
\newcommand{\WFeII}{\ensuremath{W_{\lambda 2600}}}

\newcommand{\lstaro}{\ensuremath{L_{\rm [OIII]}^\star}}

\title{Quasars probing intermediate redshift star-forming galaxies}

\author[P. Noterdaeme, R. Srianand and V. Mohan]{P. Noterdaeme, R. Srianand and V. Mohan
\thanks{E-mail:[pasquiern, anand, vmohan]@iucaa.ernet.in} \\
Inter-University Centre for Astronomy and Astrophysics, Post Bag 4, Ganeshkhind, 411\,007 Pune, India
}

\date{}

\maketitle

\begin{abstract}

We present a sample of 46 [\OIII]-emitting galaxies at $z<0.8$  
detected in the fibre spectra of quasars from the Sloan Digital Sky Survey, Data Release 7 (SDSS-DR7) 
through an automatic search procedure.
We also detect [\OII] and \Hb\ emission lines from most of 
these galaxies in the SDSS spectra. 
We study both the emission and absorption properties of a sub-sample of 17 galaxies in the redshift 
range $z=0.4$-0.7, where \MgII\ lines are covered by the SDSS spectra.
The measured lower-limits on the star-formation rates 
of these galaxies are in the range 0.2-20~M$_{\odot}$\,yr$^{-1}$. The emission line 
luminosities and (O/H) metallicities from R23 measured in this sample are similar to what is found in normal 
galaxies at these redshifts. 
Thus, this constitutes a unique sample of intermediate redshift star-forming
galaxies where we can study the QSO absorber - galaxy connection. 
Strong \MgII\ ($\WMgII\ga1$~{\AA}) as well as \MgI\ absorption lines are detected in the 
QSO spectra at the redshift 
of most of these galaxies. 
Strong \FeII\ ($\WFeII>1$~{\AA}) absorption lines are also generally detected 
whenever the appropriate wavelength ranges are covered. 
This suggests that most of these systems could be bona-fide Damped Lyman-$\alpha$ systems.
We investigate various possible relations between the \MgII\ rest equivalent widths and 
the emission line properties. 
We find a possible (2\,$\sigma$) correlation between the emission-line metallicity of 
the galaxies and the \MgII\ rest equivalent width of the absorbers ($\log ({\rm O/H})+12=0.1\WMgII+8.27$), 
which could be a consequence of an underlying mass-metallicity relation.
However, [\OIII]-selected \MgII\ systems represent only a minor fraction of the 
strong \MgII\ absorbers. 
We find this cannot be attributed to biases related either to the spectral signal-to-noise ratio 
or to the brightness of the QSOs.
We measure the average observed fluxes (collected into the SDSS fibre) 
of the [\OII] and [\OIII] lines associated to \MgII-selected systems through stacking 
technique. We find that the average lumiosities of emission lines are higher for systems 
with larger \WMgII. The stacked luminosities are found to be below the typical detection 
limit in individual spectra, 
indicating that faint galaxies can contribute appreciably to 
the observed population of strong \MgII\ absorbers at intermediate redshifts. We also present 
long-slit spectroscopic observations of SDSS\,J113108+202151, the most luminous line-emitting 
galaxy in our $z\ge 0.4$ sample. Surprisingly, we find that the line-emitting region 
does not coincide with the nearby extended bright galaxy with consistent photometric redshift 
seen in the SDSS image.

\end{abstract}


\section{Introduction}

The study of intervening absorption lines seen in the spectra of bright distant objects
is one of the most sensitive and powerful probe for understanding the 
early evolution of galaxies.  
Indeed at low and 
intermediate redshifts ($z \la 1$), the connections between QSO absorption systems 
and galaxies are mainly investigated for \MgII-selected systems. These absorbers 
are found to be statistically associated with relatively bright field galaxies
seen within a few tens of kpc to the QSO line of sight \citep{Bergeron91, 
Steidel95}. These studies 
established that \MgII\ absorbers provide an unbiased way to detect 
normal galaxies at different redshifts. 
However, the success rate of detecting \MgII\
absorption in the spectrum of QSOs that have known foreground galaxies
with redshift measurements is much less than one \citep[][]{Bechtold92, 
Bowen95, Tripp05}. These studies
suggest that the gaseous halos around galaxies may be less uniformly
populated than what was thought before \citep[see][]{Kacprzak08}. 
Also it is not necessary that the galaxies responsible for \MgII\ 
absorption are always bright $L_{\star}$ galaxies. 

Integral field spectroscopy seems to be a promising technique for the 
study of galaxies associated to quasar absorption line systems and allowed \citet{Bouche07} 
to detect \Ha\ emission associated to 14 strong $z\sim1$ \MgII\ absorbers
(with impact parameters in the range 1-40~kpc), indicating 
large star formation rates (1-20~M$_{\odot}$\,yr$^{-1}$).
Another possibility is to directly search for galaxy light from \MgII\ absorbers 
in special cases where the quasar flux at short wavelengths is switched-off 
by a higher redshift Lyman-limit system \citep{Christensen09}. Till now, 
such observations have only resulted in stringent upper-limits on the broad-band 
luminosity of the related galaxies.

It has also been proposed that other classes of QSO absorbers, such as 
those characterised by strong \CaII\ absorption lines, 
could select the most metal-rich gas \citep[see, e.g.][]{Wild06,Nestor08}, 
hence probing more central parts of high-redshift galaxies.
\citet{Wild07} have statistically detected [\OII] emission 
associated to strong \MgII- and \CaII-selected absorbers 
by stacking SDSS quasar spectra.
However, only a few direct detections of emission lines from absorbing galaxies have been reported so far.
\citet{Zych07} presented direct imaging and long-slit spectroscopic observations of five quasars 
with strong \CaII\ systems at $z<0.5$. They detected [\OII], [\OIII], \Ha\ and \Hb\ emission lines 
at the redshift of the absorbers. The luminosity of the corresponding galaxies is high, $L \simeq L_{\star}$, 
with star-formation rates in the range 0.3-30~M$_{\odot}$\,yr$^{-1}$.

When galaxies are detected with some projected separation to the QSO 
sight-line it is not obvious whether one is detecting the halo gas 
associated with the galaxy or one is probing the correlation length of 
metals in the IGM with respect to the bright galaxies. 
The contribution of possible faint galaxies closer to 
the line of sight (i.e. within the point spread function of the QSO) 
that remain undetected is also not well explored.
Therefore, even after two decades of intense research 
activity to establish the \MgII\ absorber-galaxy relationship, 
there are still open questions in this field that need to be answered. 

Here, we present direct detections of emission lines from intervening 
star-forming galaxies (with impact parameters $\le 10$~kpc and redshifts 
in the range 0.1$\le z \le $0.8) in quasar spectra from the Sloan Digital Sky Survey 
(SDSS) Data Release 7. For objects at $z\ge 0.4$, the SDSS spectra allow us to
search for \MgII\ absorption originating from these galaxies.
Our approach is very different from all previous ones in the sense that we do 
not make any pre-selection of galaxies based on QSO absorption lines. 
On the contrary, we take advantage of SDSS fibre spectra, by first searching 
for galaxy emission lines on top of background quasars spectra, then looking 
for the associated absorption lines. Throughout the paper, we adopt a 
$\Lambda$CDM cosmology with $\Omega_m=0.3$, $\Omega_{\Lambda}=0.7$ and 
$\Ho=70\,\kms\,$Mpc$^{-1}$ \citep[e.g.][]{Spergel03}.

\section{Search for intervening galaxies}

We are mostly interested in detecting normal star forming galaxies close to
the line of sight of background QSOs. We focus on the redshift range $z=0.4$-0.7 
as most prominent emission lines, as well as the \MgII\ absorption 
lines, fall in the spectral range covered in the SDSS spectrum.
In particular, [\OIII] doublet lines are useful because the [\OIII]$\lambda5007$ is one of 
the strongest lines in the optical range and is accompanied by the close-by [\OIII]$\lambda4959$ 
line with a fixed intensity ratio, allowing for an easy identification 
using an automated procedure. In this section we derive a rough 
estimate of the expected number of [\OIII] emitters that can be detected
from the SDSS QSO spectra, describe our automatic routine to detect 
galaxies and discuss the bias due to QSO luminosity.

\subsection{Detectability and number density of faint galaxies}

Here we try to get a rough estimate of the number of 
galaxies with different [\OIII] luminosities, at $0.4 \le$\zgal$\le 0.8$
that will be within the SDSS fibre centred around a distant QSO. 
The expected peak flux of the [\OIII]$\lambda$5007 emission from a \lstaro\ 
galaxy \citep[$\log \lstaro ($erg\,s$^{-1})$~=~41.95;][]{Hippelein03} at $z=0.5$ will 
be of the order of $F_{\star}(z=0.5)\sim 10^{-16}$~erg\,s$^{-1}$\,cm$^{-2}$\,{\AA}$^{-1}$, 
assuming a line width FWHM$\simeq5{\AA}$.
Note that, throughout the paper, $\lstaro$ refers to the Schechter parameter 
of the [\OIII]-luminosity function. 
As we aim here at detecting low luminosity galaxies, we should reach a detection limit of 
the order of $10^{-17}$~erg\,s$^{-1}$\,cm$^{-2}$\,{\AA}$^{-1}$.

If we assume the dominant noise for the detection of an emission line is the photon noise
from the quasar continuum (i.e. we ignore the background noise) then the signal-to-noise 
ratio in the peak of the emission line, SNR$_l$, can be written as:

\begin{equation}
\label{eqn_detectability}
{\rm SNR}_l= {{N_l} \over \sqrt{(N_{c}+N_{l})}},
\end{equation}

\noindent where $N_l$ and $N_c$ are the counts in the line and in the continuum respectively.
The signal-to-noise ratio of the continuum, SNR$_c$, is equal to $N_c/\sqrt{N_c}$ and 
the count ratio is equal to the flux ratio ($N_l/N_c=F_l/F_c$). Therefore, we can write

\begin{equation}
\label{eqn_detectabilitytwo}
{\rm SNR}_l= {\rm SNR}_c {{F_l/F_c} \over \sqrt{1+F_l/F_c}}.
\end{equation}
\par\noindent
The detectability of the emission line will  not only depend on the [\OIII] line flux ($F_l$) 
but also on the quasar flux ($F_c$). 
For a given emission line flux, the minimum signal-to-noise ratio required to detect
the [\OIII] emission will depend upon the magnitude of the quasar. That is, the 3\,$\sigma$ 
detection of the [\OIII]$\lambda$5007 emission line arising from a $L_{0}=0.2\lstaro$ 
galaxy towards a $i=19$ quasar will require a spectrum with signal-to-noise ratio $\ge 10$. 
In turn, detecting the same galaxy towards a $i=17$ quasar will require a signal-to-noise 
ratio higher than about 45.

Next, we estimate the probability of a given line of sight to pass very close or through 
a galaxy with line luminosity greater than a limiting luminosity $L_{0}$. Integrating the [\OIII] 
luminosity function over the luminosities gives the number density of galaxies:

\begin{equation}
 \label{n}
n_{L>L_0}=\int_{L_0}^{\infty}\Phi(L)\,dL = \Phi_{\star}\Gamma\left({\alpha+1,{{L_0}\over{\lstaro}}}\right),
\end{equation}

\noindent where $\Phi_{\star}$, $\alpha$ and $\lstaro$ are the Schechter parameters of the 
luminosity function, taken from Table 5 of \citet{Ly07} and $\Gamma(a,b)$ is the incomplete 
gamma function. We can estimate the number of galaxies with an impact parameter less than 
the SDSS fibre radius ($1.5\arcsec$, i.e. $r\sim$10~kpc at $z\sim0.5$): 
\begin{equation}
 \label{nn}
N_{L>L_0}=\pi r^2 n_{L>L_0},
\end{equation}
For $L_{0}=0.2\lstaro$, this gives us a number density per unit distance 
$N_{L>0.2L\star}\sim1-2\times10^{-6}$~Mpc$^{-1}$.
That is, for a line of sight to a quasar probing $z\sim 0.4-0.6$, the 
probability of a $L\ge0.2\lstaro$ galaxy being at an impact parameter less than 
the SDSS fibre radius will be about $5-10\times 10^{-4}$. We need to consider
this as a very conservative upper limit as while estimating this number
we have not considered (1) any bias due to the luminosity of the background QSO {
(see Sect.~\ref{qsolum})},
(2) the emission line attenuation due to dust internal to the galaxy, (3) the 
fibre losses (i.e. the fact that only a fraction of [\OIII] flux may go through 
the fibre) and (4) the colour selection of QSOs missing dusty sightlines 
\citep{Noterdaeme09co}, as expected from galaxies at very low impact parameters. 
From these simple-minded calculations, one can already see that a huge number 
of quasar spectra ($> 10^4$) with adequate signal-to-noise ratios will be required 
to detect a handful star-forming galaxies with an impact parameter $\le 10$ kpc 
along the line of sight to distant QSOs. The availability of a large number of 
high quality QSO spectra in the SDSS database makes it a realistic possibility to 
build a sample of such star forming galaxies to bridge the connection between strong 
\MgII\ systems and star-forming galaxies. 

\subsection{Search for galaxy [\OIII] lines in SDSS quasar spectra}

We employ a correlation analysis to select emission line galaxies close to the line 
of sight to 98\,978 quasars from the Sloan Digital Sky Survey II, Data Release 
7, without any prior knowledge of the absorption properties of the galaxies. 
As a first step we iteratively fit the quasar continuum by applying Savitsky-Golay 
filtering and removing  deviant pixels.
We then cross-correlate the continuum-subtracted quasar spectra with a template profile 
of [\OIII]$\lambda\lambda$4959,5007 generated from the SDSS galaxy template spectrum.

We restrict our search out of the Lyman-$\alpha$ forest (in particular for high redshift
QSOs) and below $\lambda=8400$~{\AA} (i.e. $\zgal<0.68$) to avoid the most crowded telluric 
line regions. 
Whenever a high correlation is found ($C>0.9$), we check the [\OIII]$\lambda$5007 
(resp. [\OIII]$\lambda$4959) are detected at more than 2 (resp. 1)~$\sigma$ level. 
Note that because of the galactic template used, wide emission lines arising 
from AGNs are not picked-up by our procedure.
Each candidate was then inspected visually for the presence of other emission lines
in particular \Hb, \Hg\ and [\OII]$\lambda\lambda$3726,3729. 
Spurious detections were 
identified and removed from the sample. 
We list all the 44 confirmed emission line
galaxies in the redshift range 0.1$\le$\zgal $\le$0.68 in Table~\ref{qso}.
Seventeen of these galaxies are in the redshift range ideally suited for detecting
\MgII\ absorption in the SDSS spectrum itself. These objects form the main sample
of this present study.

In addition, we performed an automatic search for the [\OII]$\lambda\lambda3726,3729$ emission 
doublet\footnote{The two lines are always blended at the SDSS resolution} associated to \MgII\ 
systems to detect higher redshifts systems, for which telluric lines make any search 
based on [\OIII] a difficult task. {\MgII\ systems were found by an automatic 
procedure based on correlation analysis, similar to that used by \citet{Noterdaeme09dla} to search 
for DLAs.}
This provided us with two additional galaxies, at 
$z=0.669$ and 0.788 towards SDSS\,J120908+022734 and SDSS\,J161728+061604, respectively.
The detections are summarised in Table~\ref{qso}. We concentrate, in this 
paper, on the sample of 17 [\OIII]-selected galaxies at 
$0.4<\zgal<0.7$ for which the expected positions of the \MgII\,$\lambda\lambda$2796,2803 
absorption lines are also covered. Lower redshift systems are also provided in Table~\ref{qso} 
for the interested reader. 

\begin{table}
\caption{Sample of galaxies with [\OIII] emission around QSO sightline \label{qso}}
\begin{tabular}{c c c c c}
\hline
\hline
QSO            &plate,MJD,fiber & $\zqso$ & $\zgal$$^a$ & selection \\
\hline                 
J023914-070557 &0455,51909,562 &0.715&  0.342    & [\OIII]\\ 
J080216+143506 &2266,53679,154 &1.562&  0.141    & [\OIII]\\ 
J080808+064108 &1756,53080,184 &2.108&{\bf 0.433}& [\OIII]\\  
J081154+202148 &1925,53327,437 &1.247&{\bf 0.445}& [\OIII]\\
J082057+400326 &0760,52264,075 &0.588&  0.301    & [\OIII]\\ 
J085113+071959 &1299,52972,256 &1.654&  0.130    & [\OIII]\\ 
J091417+325955 &1592,52990,275 &4.656&{\bf 0.444}& [\OIII]\\
J092913+302225 &1939,53389,537 &1.815&{\bf 0.439}& [\OIII]\\  
J094041+341535 &1594,52992,048 &1.718&{\bf 0.447}& [\OIII]\\  
J094335-004322 &0266,51630,125 &0.271&  0.099    & [\OIII]\\ 
J094759+120537 &1742,53053,624 &1.287&  0.257    & [\OIII]\\ 
J095228+032616 &0571,52286,144 &1.290&{\bf 0.419}& [\OIII]\\  
J101246+171419 &2587,54138,556 &1.103&  0.112    & [\OIII]\\
J103309+205956 &2376,53770,427 &1.113&  0.361    & [\OIII]\\ 
J104223+092708 &1240,52734,584 &1.108&{\bf 0.592}& [\OIII]\\ 
J104257+074850 &1000,52643,522 &2.666&  0.033    & [\OIII]\\ 
J110224+573512 &0950,52378,241 &1.620&  0.293    & [\OIII]\\ 
J111343+184002 &2490,54179,023 &2.062&  0.169    & [\OIII]\\
J112146+021757 &0511,52636,175 &1.274&  0.263    & [\OIII]\\ 
J113002+602628 &0952,52409,403 &0.374&  0.061    & [\OIII]\\ 
J113108+202151 &2502,54180,371 &1.763&{\bf 0.563}& [\OIII]\\  
J114340+520303 &0881,52368,313 &1.816&  0.132    & [\OIII]\\ 
J120538+604057 &0954,52405,535 &0.537&{\bf 0.434}& [\OIII]\\  
J120908+022734 &0517,52024,541 &1.226&{\bf 0.669}& \MgII+[\OII] \\  
J121510+141802 &1765,53466,068 &1.299&{\bf 0.421}& [\OIII]\\  
J122752+165522 &2598,54232,188 &3.348&{\bf 0.565}& [\OIII]\\  
J125339+175832 &2601,54144,625 &0.505&{\bf 0.401}& [\OIII]\\ 
J131804+522510 &1667,53430,420 &2.990&  0.393    & [\OIII]\\ 
J132542+255525 &2244,53795,141 &1.439&{\bf 0.433}& [\OIII]\\
J133733+445129 &1465,53082,300 &1.168&  0.159    & [\OIII]\\ 
J132918+630424 &0603,52056,089 &0.987&  0.366    & [\OIII]\\ 
J140103-005030 &0301,51942,052 &0.927&  0.357    & [\OIII]\\ 
J140159+414156 &1346,52822,060 &1.701&  0.120    & [\OIII]\\ 
J142421+453523 &1287,52728,101 &1.614&{\bf 0.421}& [\OIII]\\  
J143458+504118 &1046,52460,542 &1.485&  0.199    & [\OIII]\\ 
J144412+022301 &0536,52024,626 &1.215&  0.140    & [\OIII]\\ 
J145240+544345 &1163,52669,505 &1.520&  0.102    & [\OIII]\\ 
J150140+571026 &0610,52056,133 &1.799&  0.103    & [\OIII]\\ 
J154542+505759 &0796,52401,495 &0.942&{\bf 0.525}& [\OIII]\\  
J160521+510740 &0620,52375,535 &1.229&  0.099    & [\OIII]\\ 
J161016+500728 &0623,52051,304 &0.239&  0.127    & [\OIII]\\ 
J161607+212401 &1853,53566,192 &1.115&  0.335    & [\OIII]\\ 
J161728+061604 &1731,53884,113 &1.244&{\bf 0.788}& \MgII+[\OII] \\  
J165508+224150 &1415,52885,478 &0.621&{\bf 0.453}& [\OIII]\\  
J165632+414617 &0631,52079,118 &2.163&{\bf 0.662}& [\OIII]\\  
J235621+002906 &0387,51791,343 &1.049&  0.331    & [\OIII]\\ 
\hline
\end{tabular}
\footnotesize
$a$ The redshifts of galaxies studied in this paper, $\zgal\ge0.4$, are marked in bold face.
\normalsize
\end{table}

\subsection{Effect of QSO luminosity \label{qsolum}}

\begin{figure}
 \includegraphics[bb=20 180 550 690,width=\hsize]{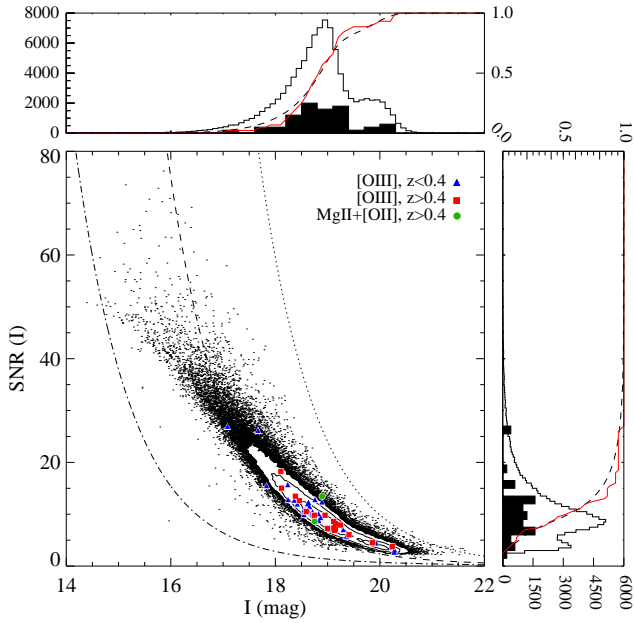} 
\caption{$i$-band signal-to-noise ratio and magnitude of the searched quasars. 
{Contours are drawn in the region with highest density of points for presentation purpose.}
The upper and right panels show, respectively the magnitude and the signal-to-noise 
ratio distributions for the whole quasar sample (unfilled histogram) and for quasars 
selected upon the presence of intervening [\OIII] emission line (filled histogram). The filled 
histograms have been scaled for presentation purpose only. Dotted 
(resp. solid) curves on the top and right panels represent the normalised cumulative 
distributions for the whole quasar sample (resp. quasars with [\OIII]-selected 
intervening galaxies). The three curves in the main panel represent the minimum 
signal-to-noise required for the 1\,$\sigma$ detection of a emission line with peak flux 
$F_{l}=1/5\times 10^{-17}$ (dotted), $10^{-17}$ (dashed) and $5\times10^{-17}$ 
(dashed-dotted)~erg\,s$^{-1}$\,cm$^{-2}$\,{\AA}$^{-1}$.
\label{snrmag}}
\end{figure}

In Fig.~\ref{snrmag} we present the fibre magnitudes and spectral signal-to-noise ratios 
in $i$-band of the whole quasar sample searched for [\OIII] emission lines and compare to that 
of quasars 
with detected intervening emission lines. It is somehow surprising to see that the detections 
do not occupy any preferred region of the SNR-magnitude diagram. This is better seen in the upper 
(resp. right) panel where we compare the differential and cumulative distributions of 
the magnitudes (resp. signal-to-noise ratios) of QSOs with and without a foreground 
[\OIII]-emitting galaxy. Indeed, a double-side Kolmogorov-Smirnov test 
shows that the probability of the magnitudes of the quasar with [\OIII]-selected intervening 
galaxies to arise from the same parent population as the whole quasar sample is high ($\Pks=0.88$). 
The same test on signal-to-noise ratio gives $\Pks=0.60$.

Following Eq.~\ref{eqn_detectabilitytwo}, we overplot the S/N ratio required for the 
1\,$\sigma$ detection of an emission line with peak intensity 
$F_{l}=10^{-17}$~erg\,s$^{-1}$\,cm$^{-2}$\,{\AA}$^{-1}$ (or equivalently the 3\,$\sigma$ 
detection of $F_{l}=3\times 10^{-17}$~erg\,s$^{-1}$\,cm$^{-2}$\,{\AA}$^{-1}$) as a function 
of the $i$-band magnitude of the background quasar. 
The spectral S/N ratios and $i$-band magnitudes of the quasars roughly follow this relation.
{While detections are naturally made easy towards faint quasars, the steeply increasing S/N ratio 
of SDSS spectra with the quasar brightness also allows for detections towards relatively bright quasars.}
However, the signal-to-noise ratios reached by the SDSS for the very bright quasars ($i<17$) are 
still not sufficiently high to detect intervening emission lines with 
$F_{l}\sim10^{-17}$~erg\,s$^{-1}$\,cm$^{-2}$\,{\AA}$^{-1}$. As can be seen in the top panel of 
Fig.~\ref{snrmag}, quasars with $i<17$ are rare anyway and do not contribute much to the 
statistics.

\section{Emission line analysis}

The [\OII]$\lambda\lambda$3726,3729 doublet and  \Hb\ are the 
other strong lines
that are expected in the wavelength range covered by the SDSS spectrum
for the redshift range of our interest.
We detect [\OII]$\lambda\lambda$3726,3729 in all the cases except
for the $z = 0.445$ galaxy along the line of sight towards J081154+202148.
However, it is obvious from Fig.~\ref{sample_fig} that the S/N in the 
wavelength range of [\OII] emission is poor. The \Hb\ line is detected in
all but two cases.
Here we perform a detailed analysis of \Hb, [\OIII] and [\OII] emission lines
through Gaussian fitting.  As a first step we subtract the continuum
emission that includes the continuum light from both the QSO and the galaxy. 
The unabsorbed continuum (including 
the quasar broad emission lines but excluding intervening emission lines) is 
determined accurately by manually adjusting a spline function to the observed 
spectrum.
Then we simultaneously fit the detected emission lines using an IDL 
code based on MPFIT \citep{Markwardt09}, which performs $\chi^2$-minimisation 
by Levenberg-Marquardt technique. We use a single redshift for all emission 
lines. The [\OII]$\lambda\lambda3726,3729$ doublet is fitted with a double 
Gaussian. Note that although the two lines are always blended at the SDSS spectral 
resolution it is still possible to distinguish the two corresponding peaks. 
As the flux ratio of these two lines depends on the kinetic temperature 
of the gas and the electron density \citep[e.g.][]{Kisielius09}, we fit 
the blend with two components allowing for the relative flux ratio to be 
a free parameter but keeping the line widths tied. 
In the case of the forbidden lines [\OIII]$\lambda\lambda$4959,5007, 
we impose the line ratio to follow the theoretical value of 3 
\citep[e.g.][]{Storey00}. The results of our fits are shown in 
Fig.~\ref{sample_fig}.
The integrated line intensities of [\OIII], [\OII] and \Hb\ lines are 
then measured from the fitted Gaussian parameters. In the following, 
we always refer to [\OIII] flux (or luminosity) as the sum of
fluxes (or luminosity) of 
[\OIII]$\lambda$5007 and [\OIII]$\lambda$4959. Similarly
in the case of  [\OII], we use the sum of [\OII]$\lambda$3727 and 
[\OII]$\lambda$3730. The total emission line flux together with the 
associated errors are given in Table~\ref{sfr}. In the case of 
non-detection we give 2$\sigma$ upper limits.

\newcounter{prev_fig}
\setcounter{prev_fig}{\value{figure}}

\begin{figure*}
\centering
\begin{tabular}{ccc}
\multicolumn{1}{l}{\hspace{0.7cm} \FeII, \MnII \hspace{2.7cm} \MgII \hspace{0.25cm} \MgI} & \multicolumn{1}{l}{\hspace{0.5cm} [\OII] \hspace{1.2cm} \Hb\ \hspace{1.8cm} [\OIII]} & \\

\includegraphics[bb=219  0  425 790,angle=90,clip=,width=0.395\hsize]{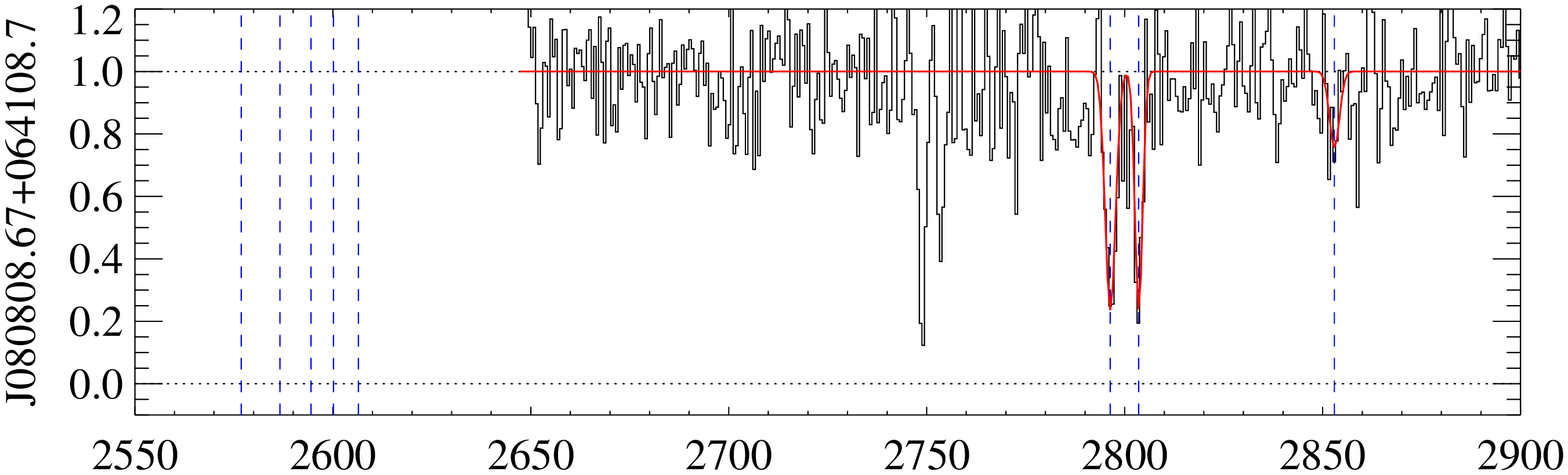} &
\includegraphics[bb=219  0  425 750,angle=90,clip=,width=0.375\hsize]{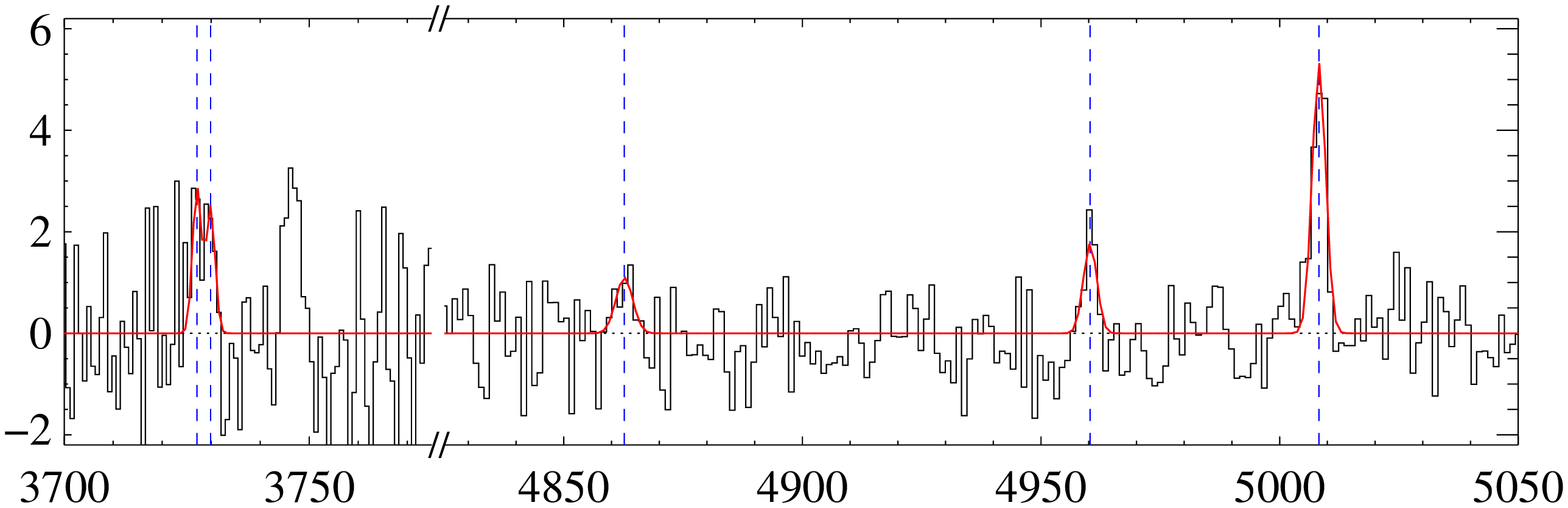} &
\includegraphics[bb=219 519 425 722,angle=90,clip=,width=0.1015\hsize]{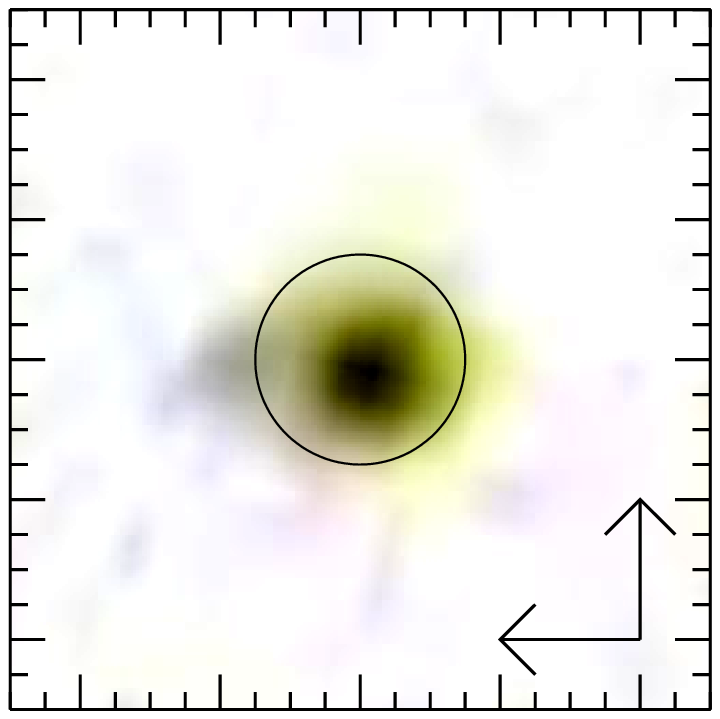} \\

\includegraphics[bb=219  0  425 790,angle=90,clip=,width=0.395\hsize ]{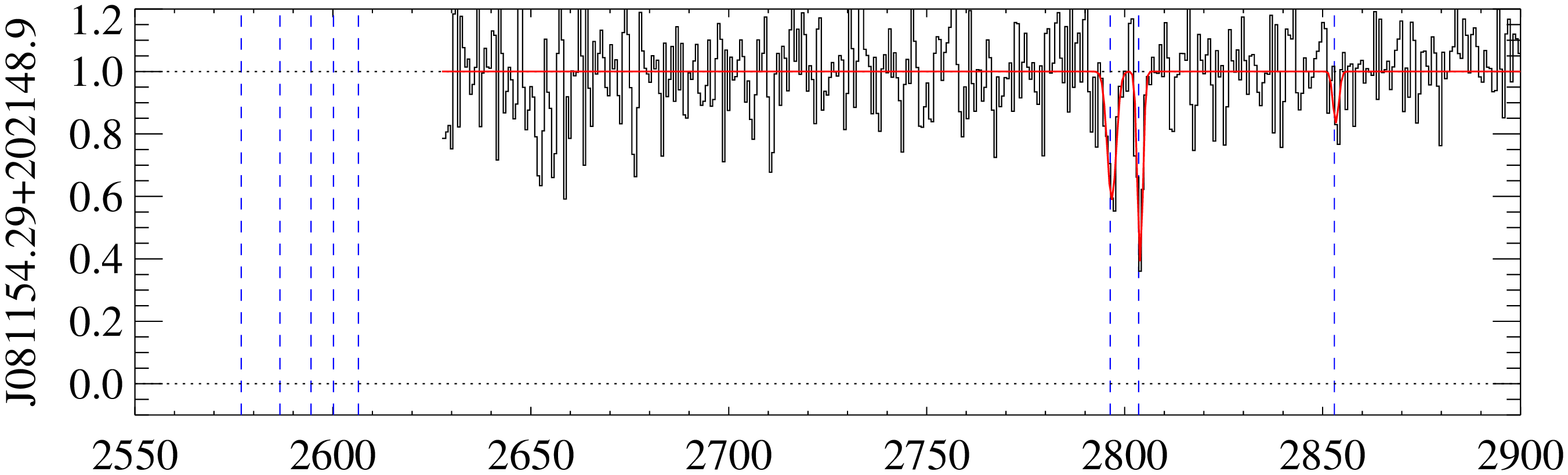} &
\includegraphics[bb=219  0  425 750,angle=90,clip=,width=0.375\hsize ]{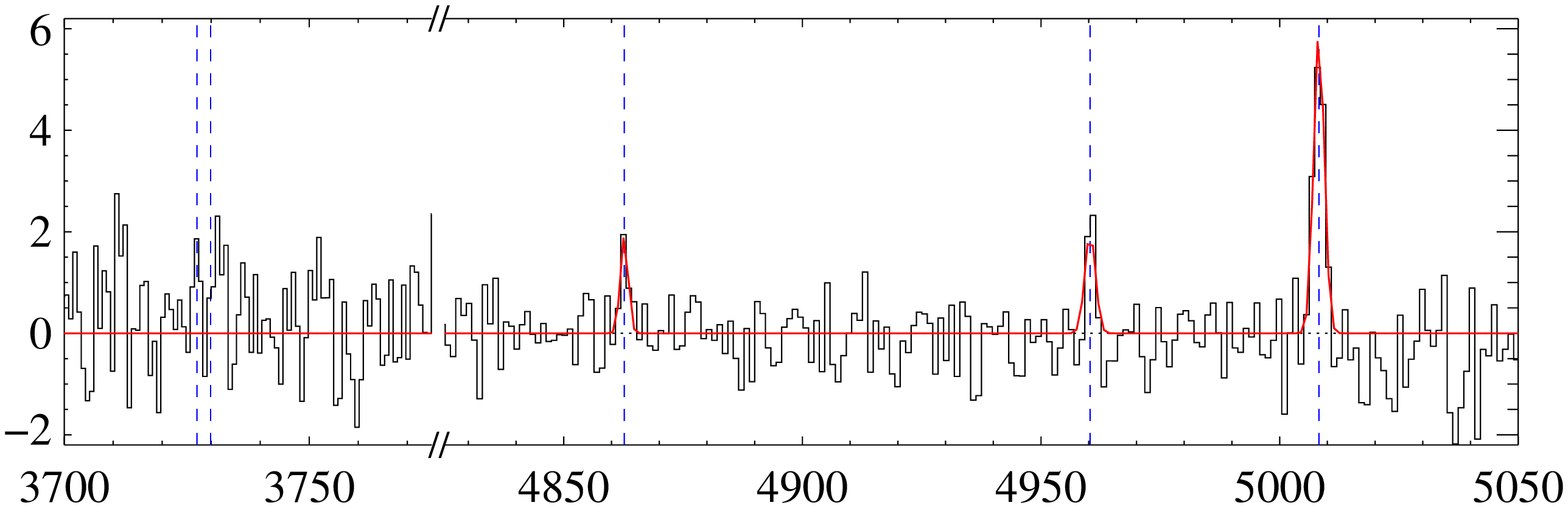} &
\includegraphics[bb=219 519 425 722,angle=90,clip=,width=0.1015\hsize]{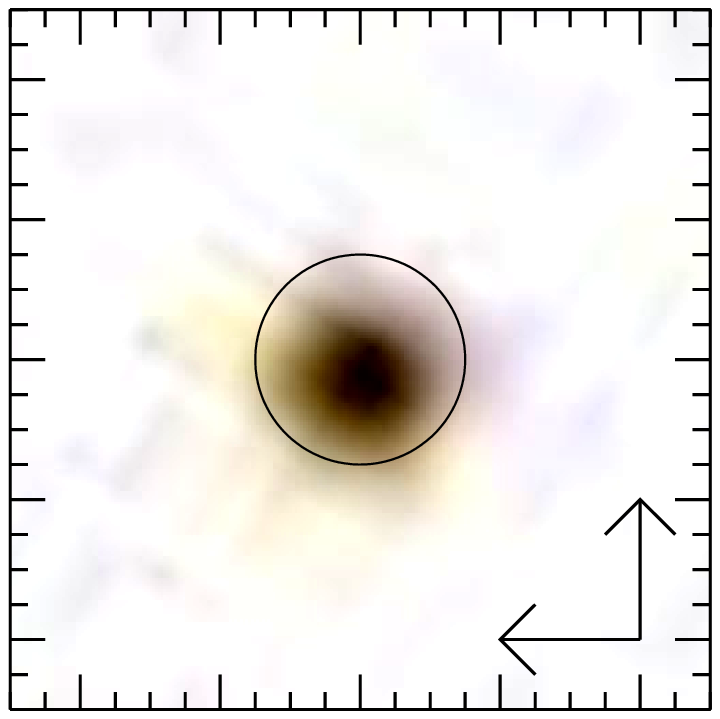} \\

\includegraphics[bb=219 0   425 790,angle=90,clip=,width=0.395\hsize ]{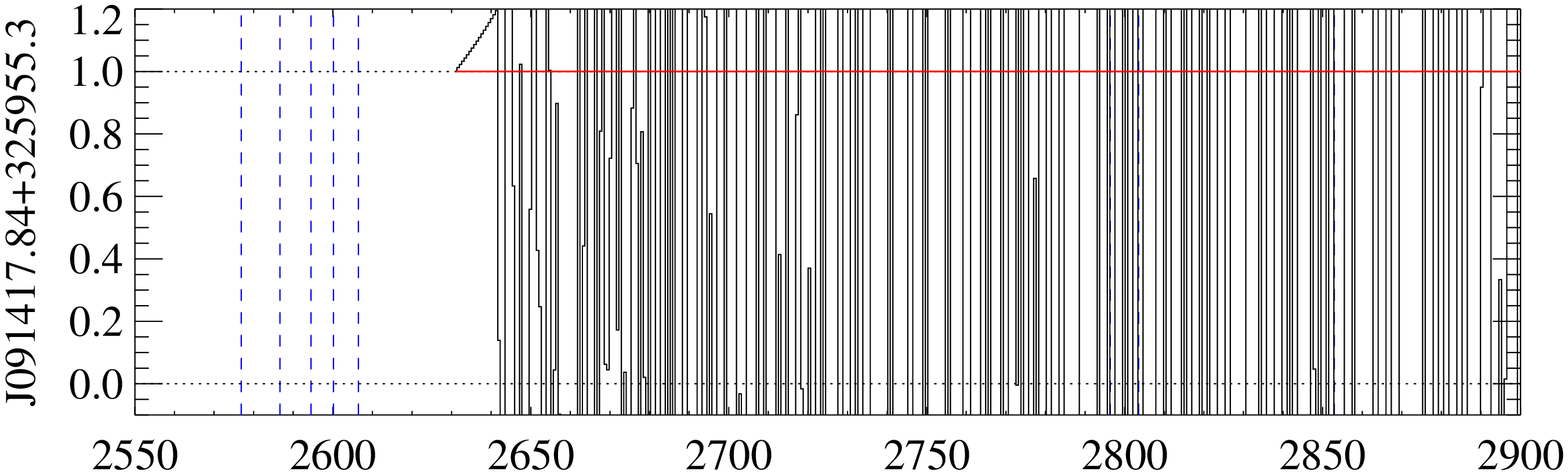} &
\includegraphics[bb=219 0   425 750,angle=90,clip=,width=0.375\hsize ]{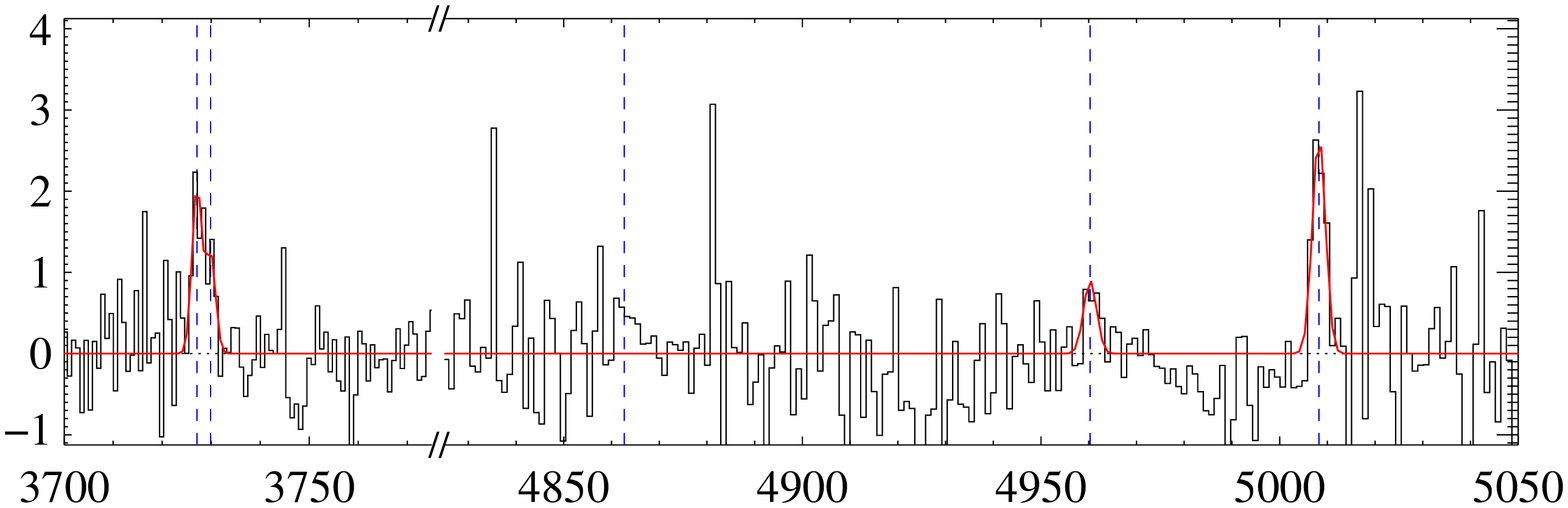} &
\includegraphics[bb=219 519 425 722,angle=90,clip=,width=0.1015\hsize]{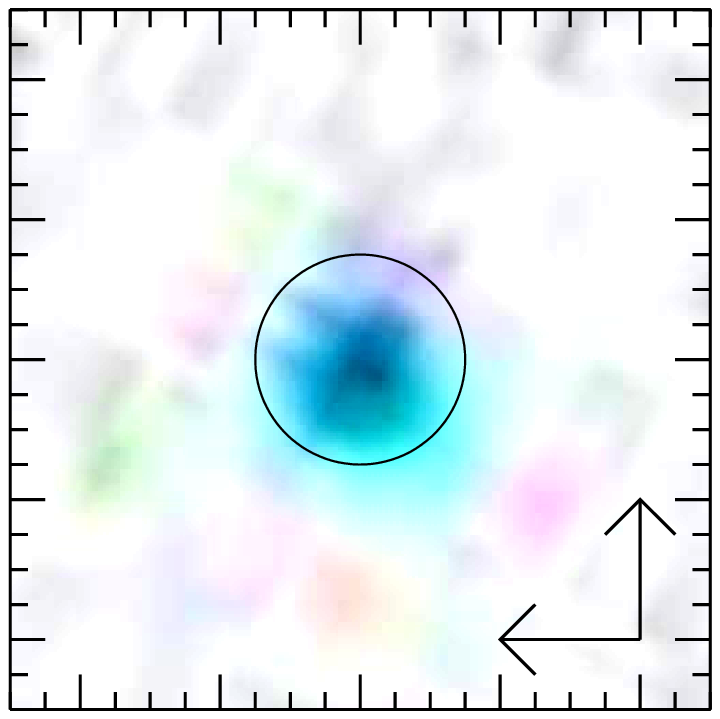} \\

\includegraphics[bb=219 0   425 790,angle=90,clip=,width=0.395\hsize ]{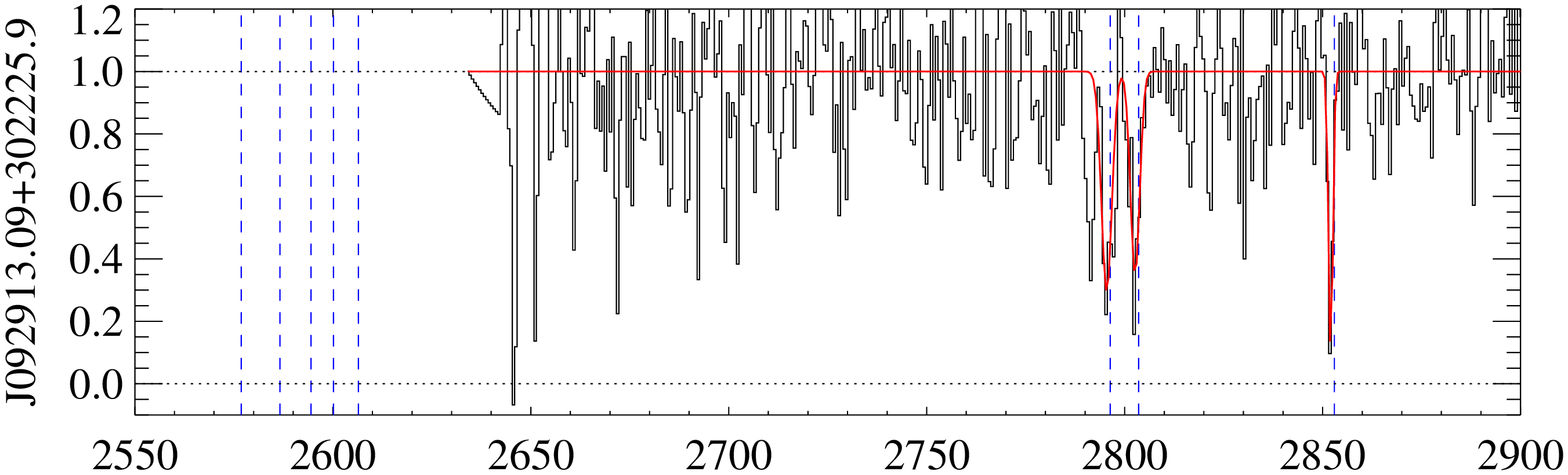} &
\includegraphics[bb=219 0   425 750,angle=90,clip=,width=0.375\hsize ]{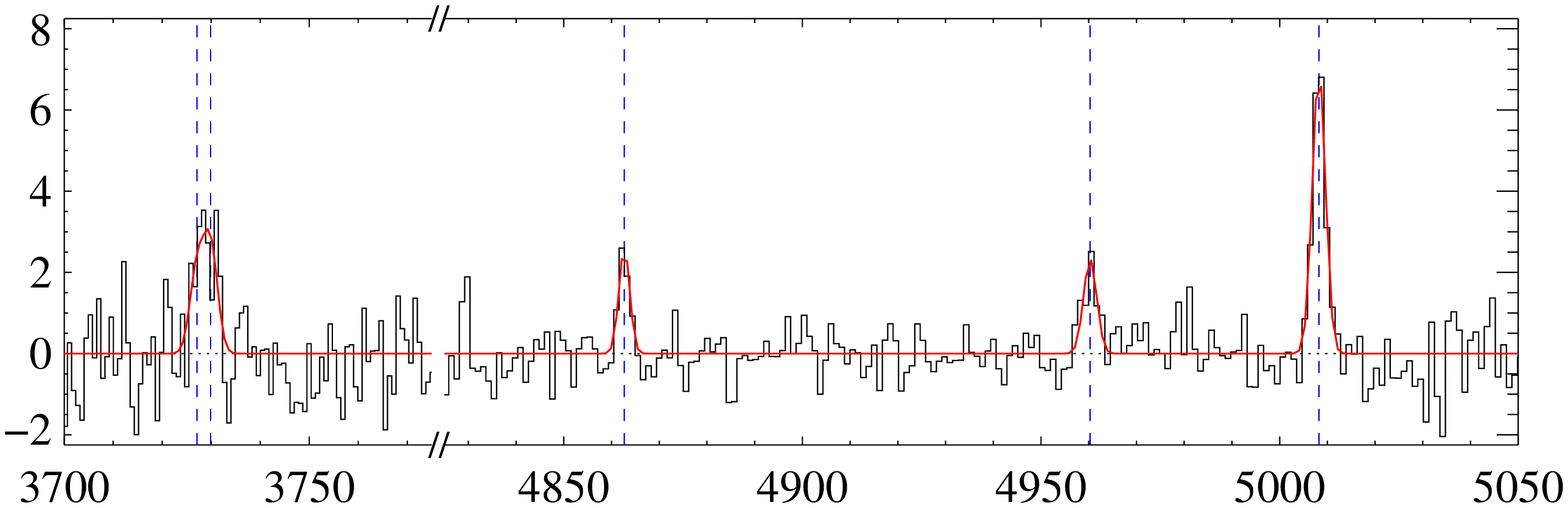} &
\includegraphics[bb=219 519 425 722,angle=90,clip=,width=0.1015\hsize]{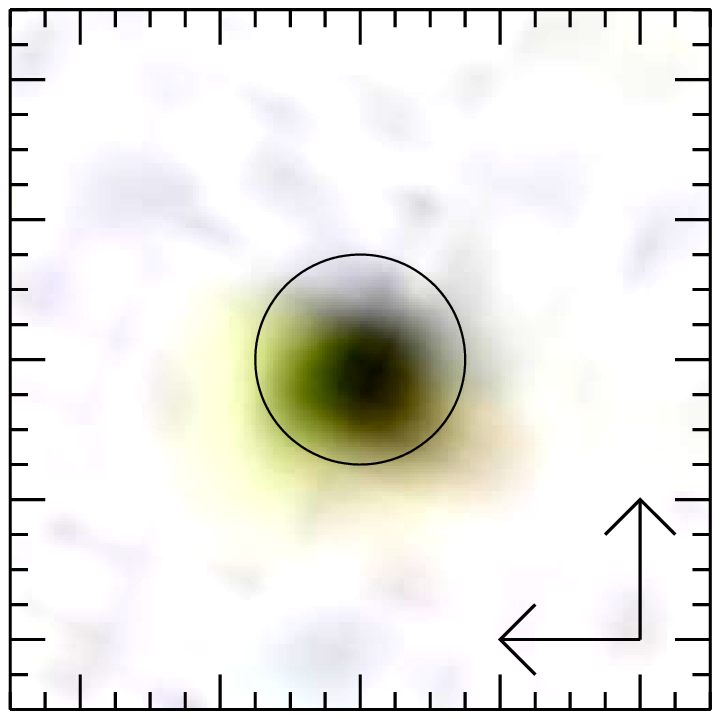} \\

\includegraphics[bb=219 0   425 790,angle=90,clip=,width=0.395\hsize ]{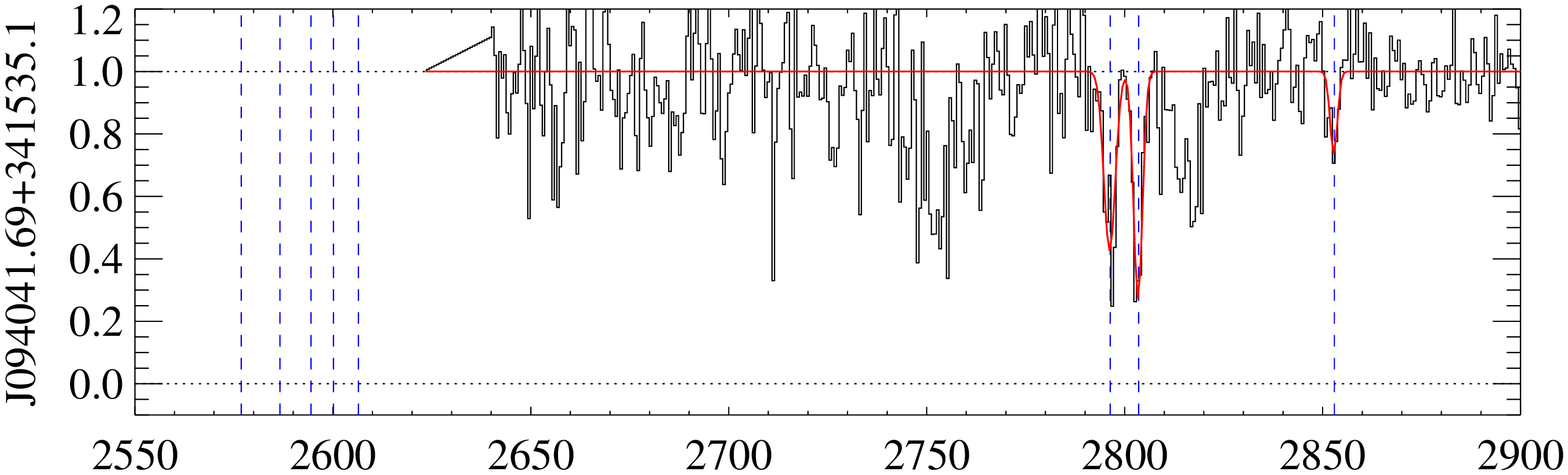} &
\includegraphics[bb=219 0   425 750,angle=90,clip=,width=0.375\hsize ]{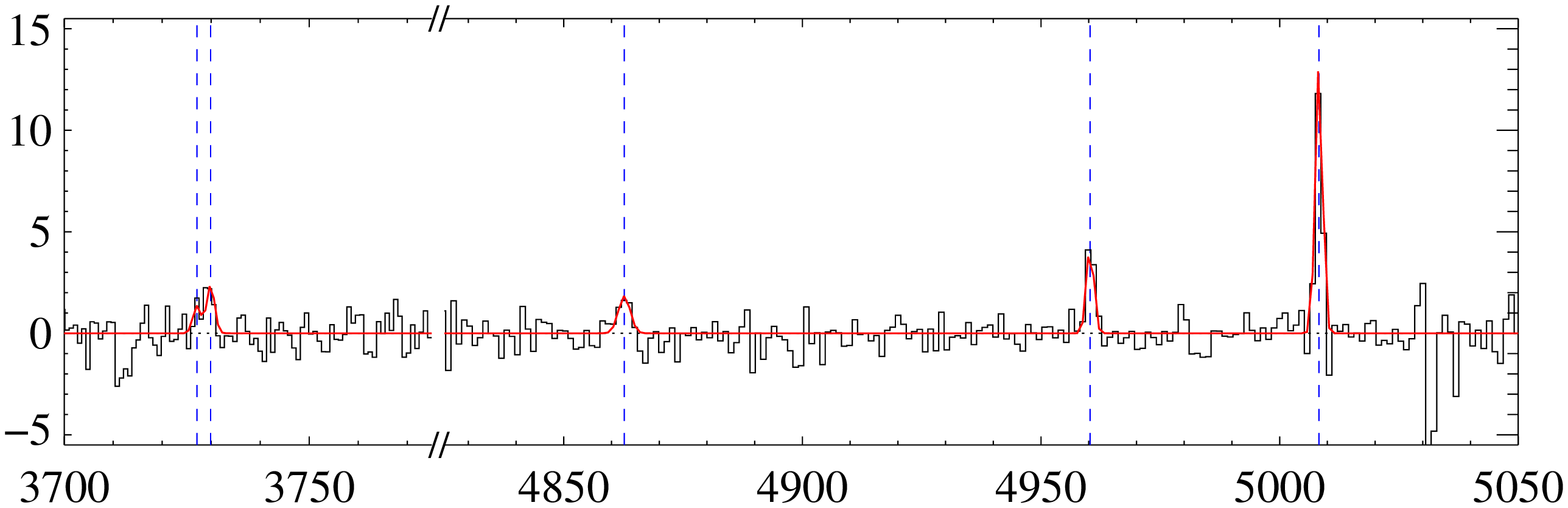} &
\includegraphics[bb=219 519 425 722,angle=90,clip=,width=0.1015\hsize]{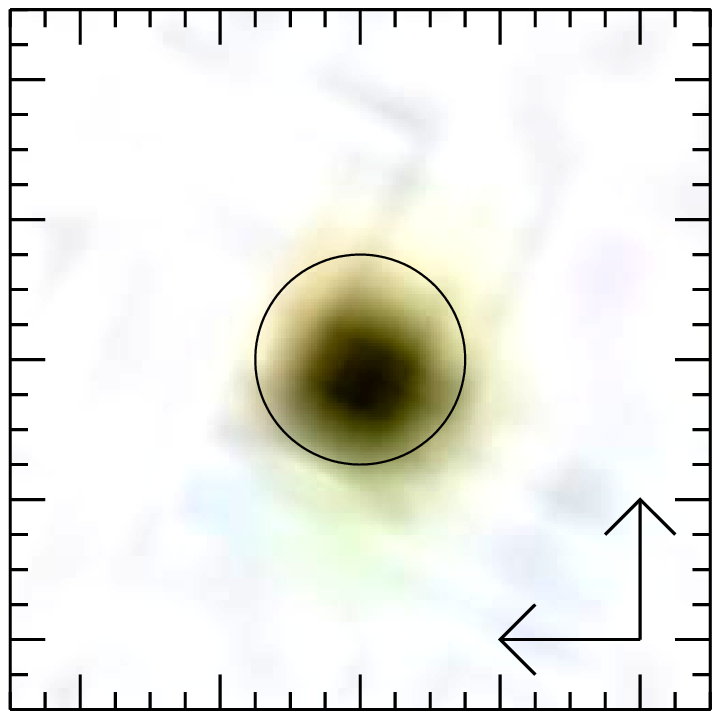} \\

\includegraphics[bb=219 0   425 790,angle=90,clip=,width=0.395\hsize ]{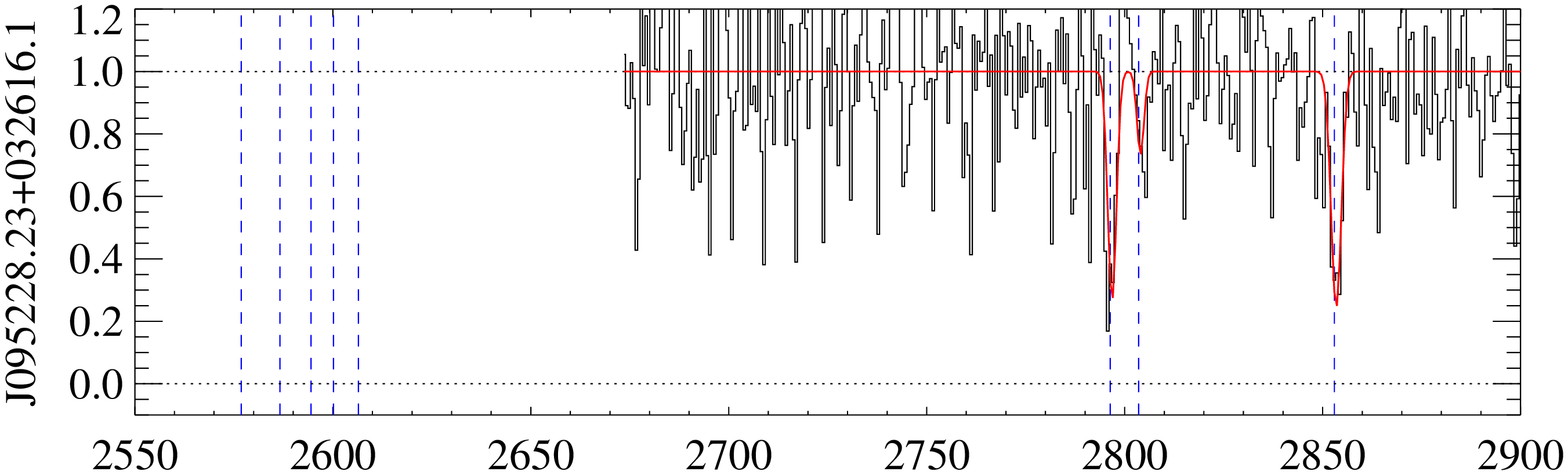} &
\includegraphics[bb=219 0   425 750,angle=90,clip=,width=0.375\hsize ]{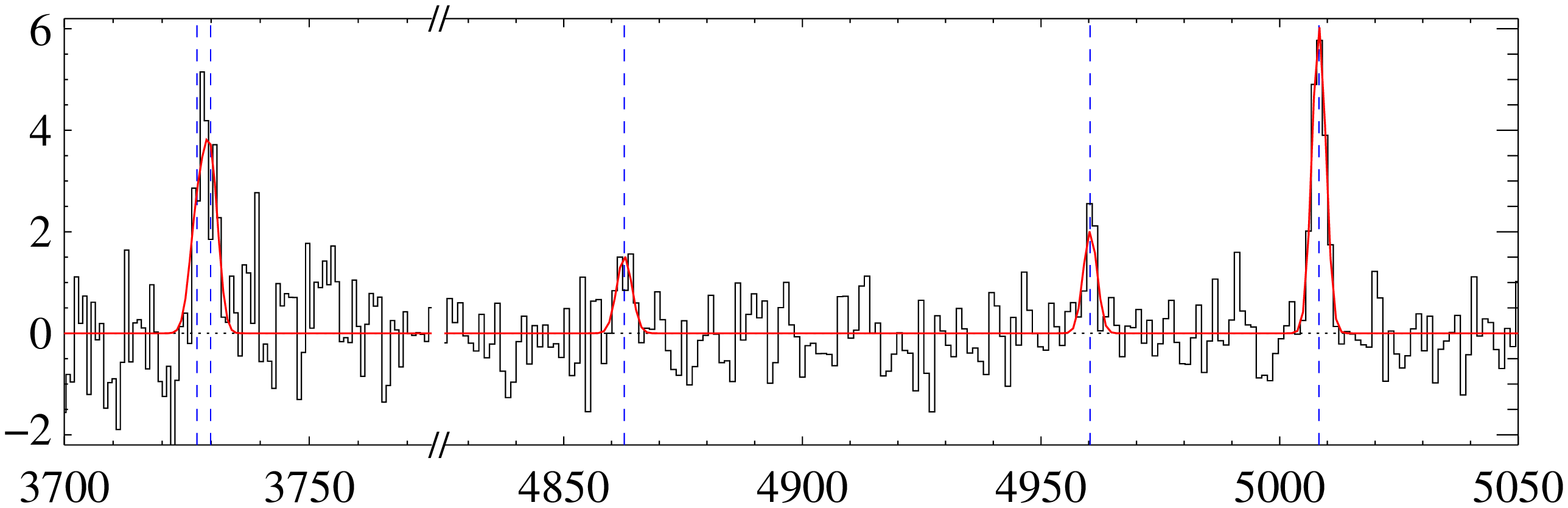} &
\includegraphics[bb=219 519 425 722,angle=90,clip=,width=0.1015\hsize]{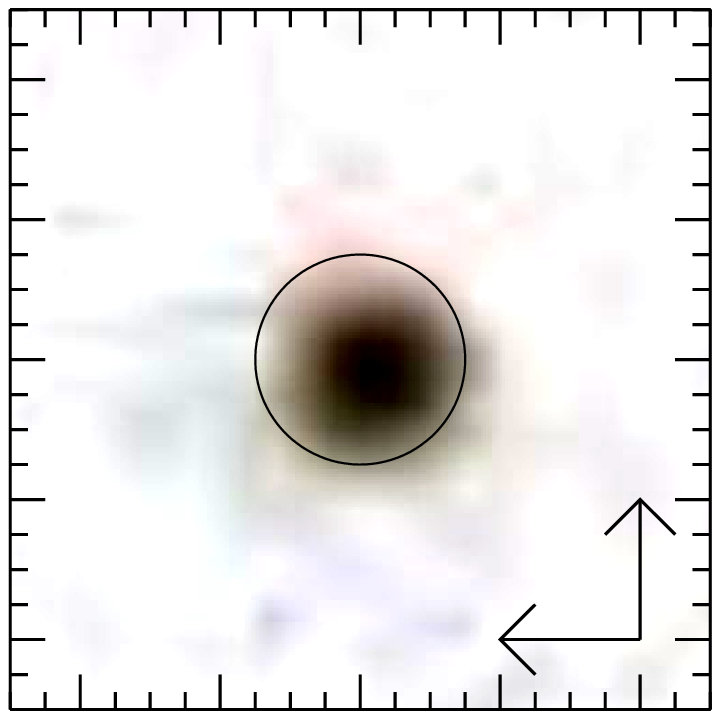} \\

\includegraphics[bb=219 0   425 790,angle=90,clip=,width=0.395\hsize ]{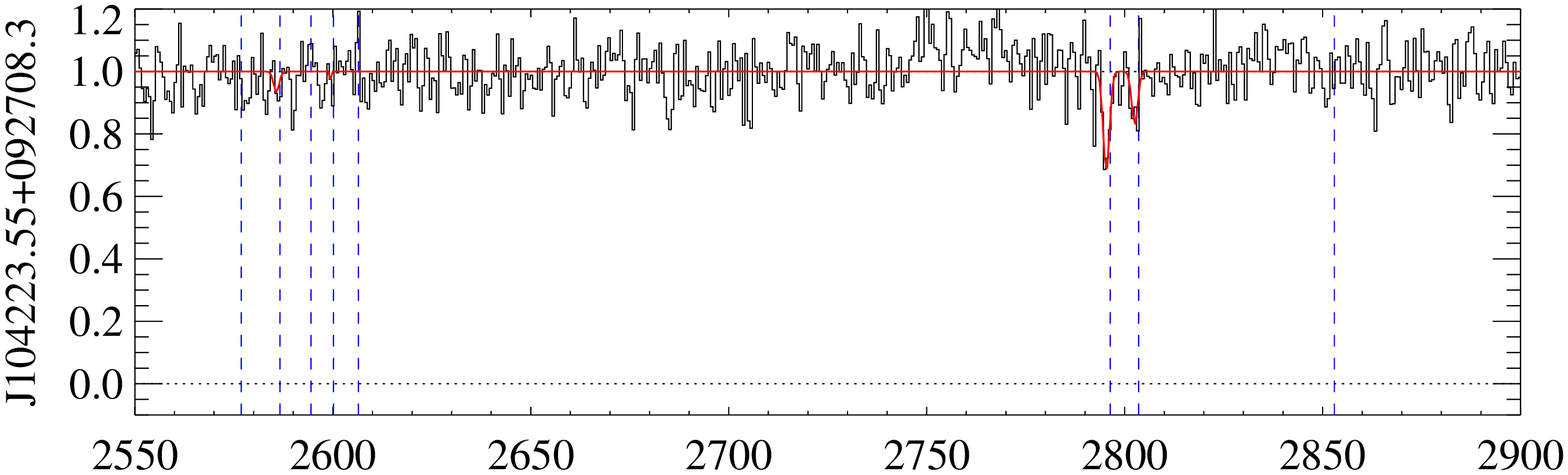} &
\includegraphics[bb=219 0   425 750,angle=90,clip=,width=0.375\hsize ]{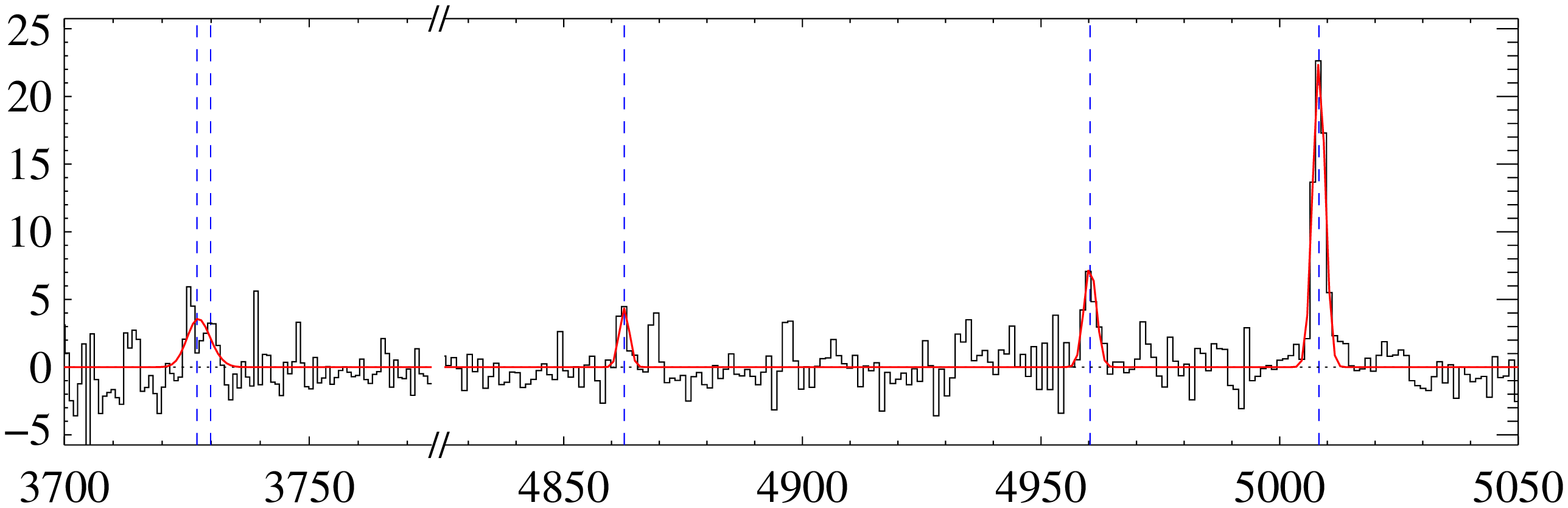} &
\includegraphics[bb=219 519 425 722,angle=90,clip=,width=0.1015\hsize]{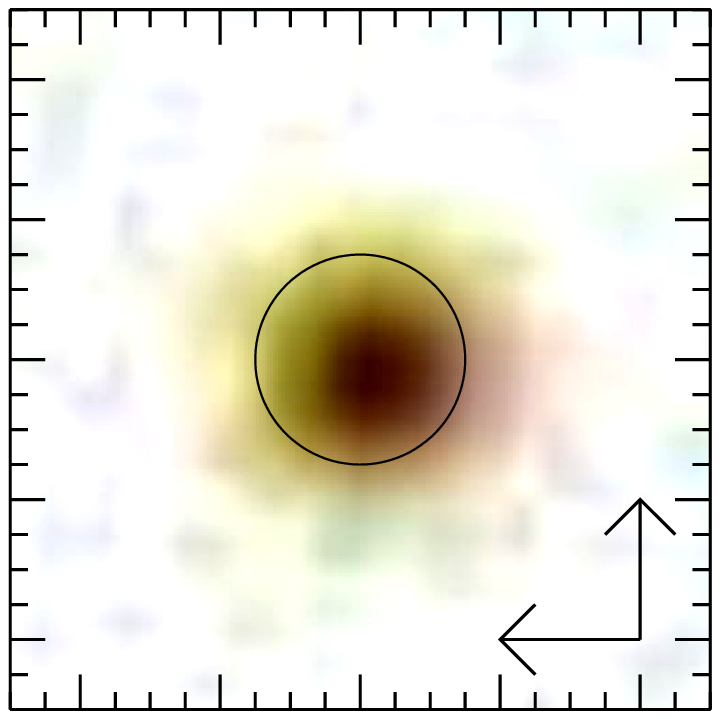} \\

\includegraphics[bb=219 0   425 790,angle=90,clip=,width=0.395\hsize ]{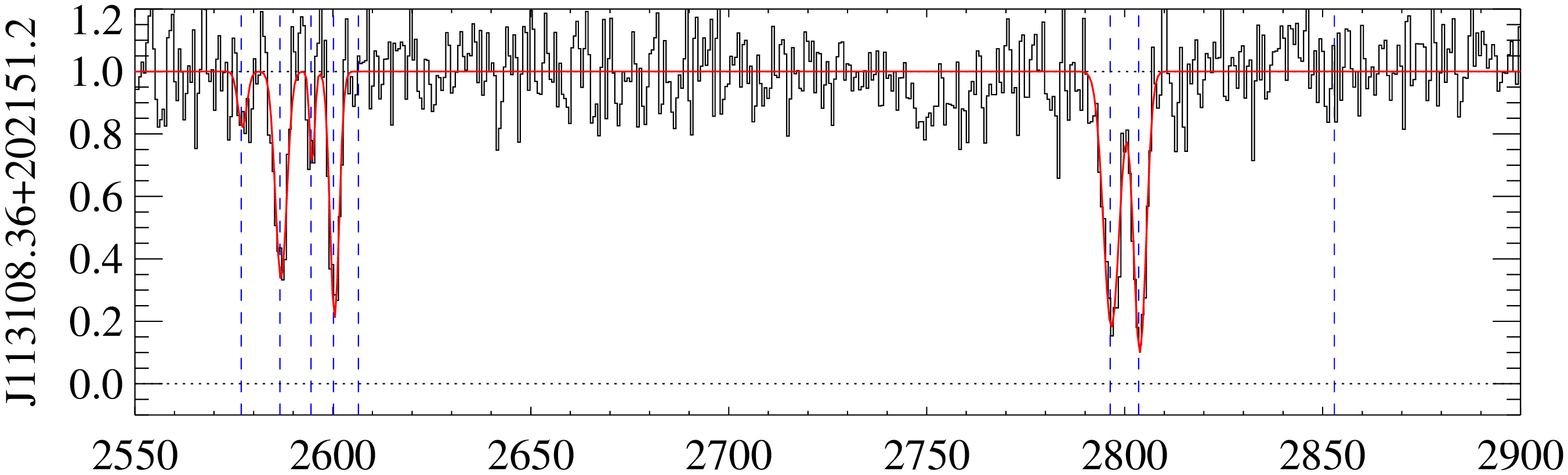} &
\includegraphics[bb=219 0   425 750,angle=90,clip=,width=0.375\hsize ]{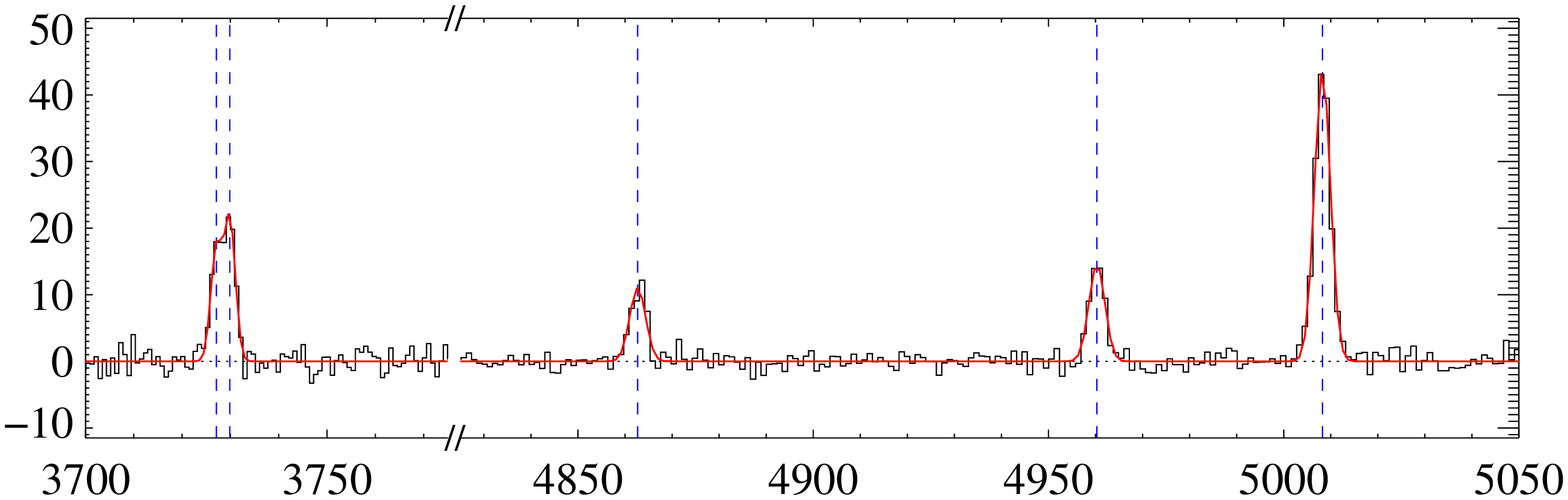} &
\includegraphics[bb=219 519 425 722,angle=90,clip=,width=0.1015\hsize]{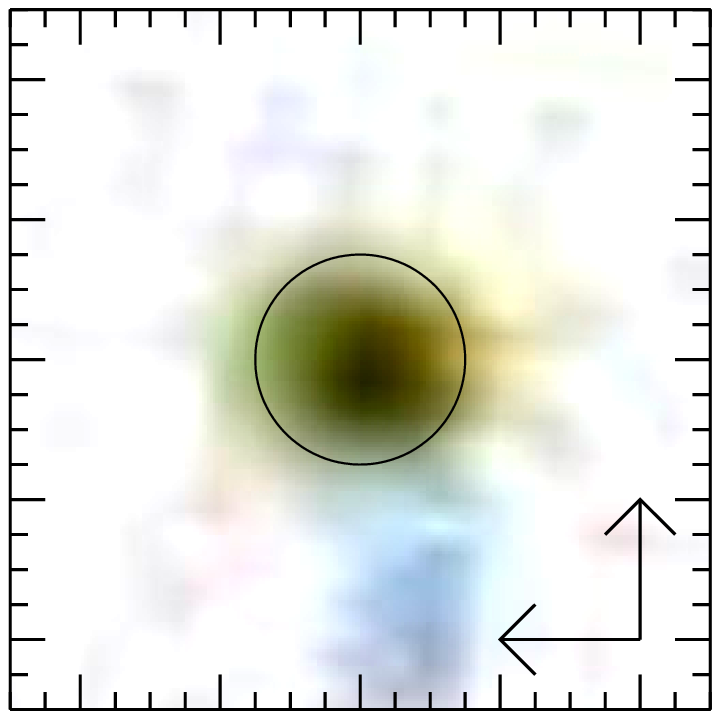} \\

\includegraphics[bb=219 0   425 790,angle=90,clip=,width=0.395\hsize ]{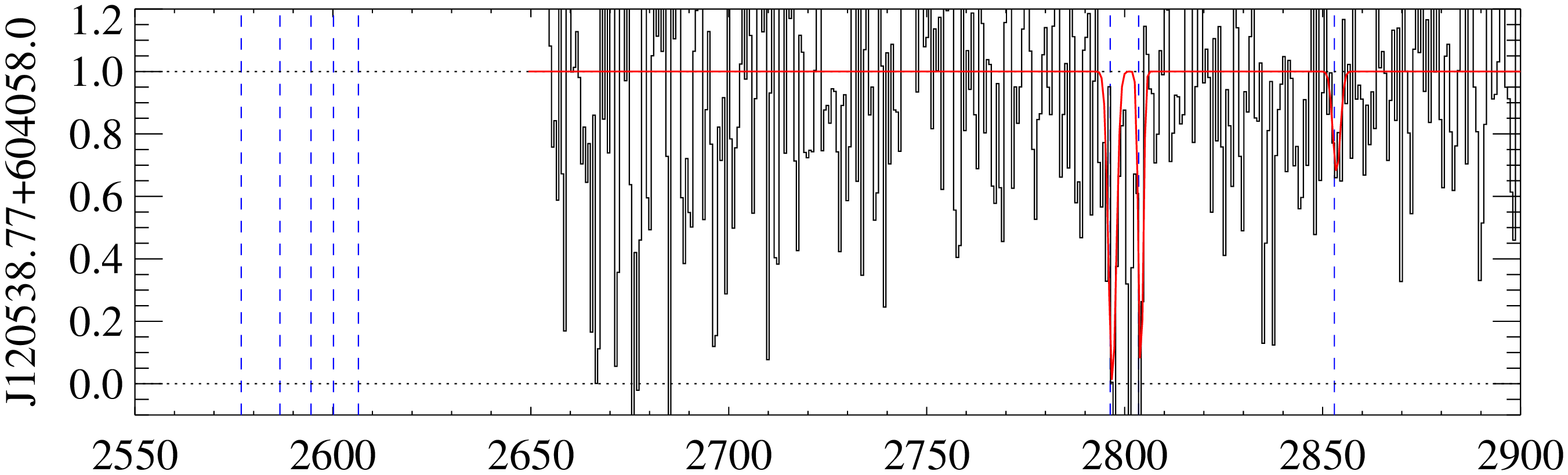} &
\includegraphics[bb=219 0   425 750,angle=90,clip=,width=0.375\hsize ]{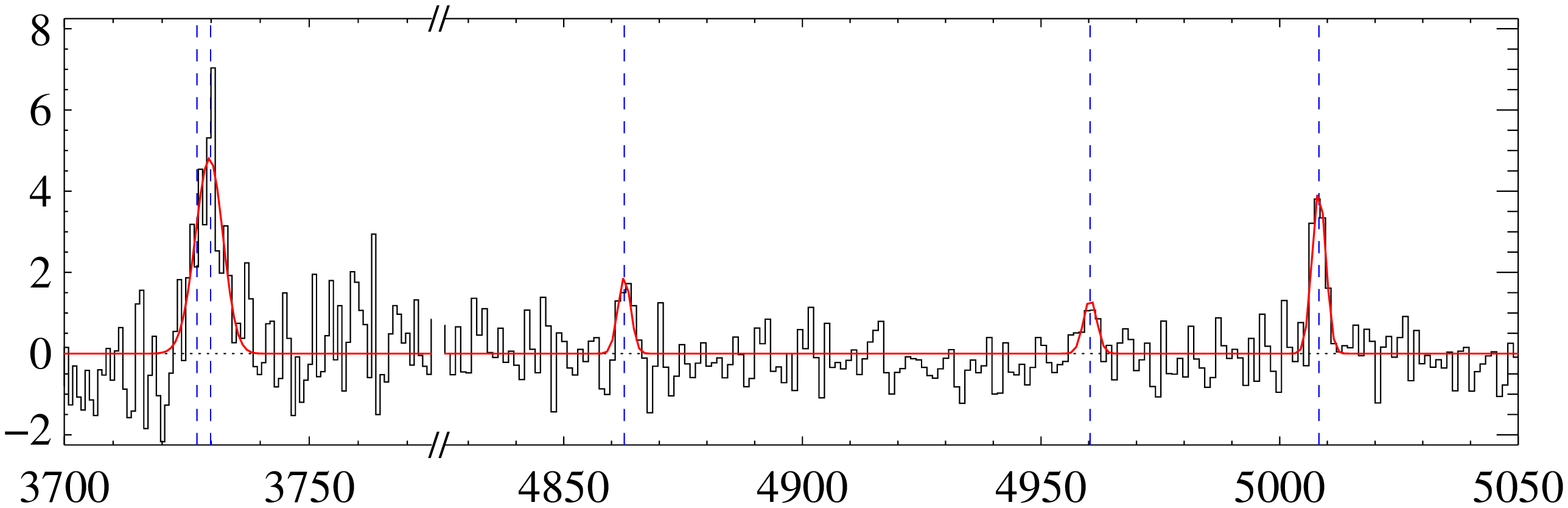} &
\includegraphics[bb=219 519 425 722,angle=90,clip=,width=0.1015\hsize]{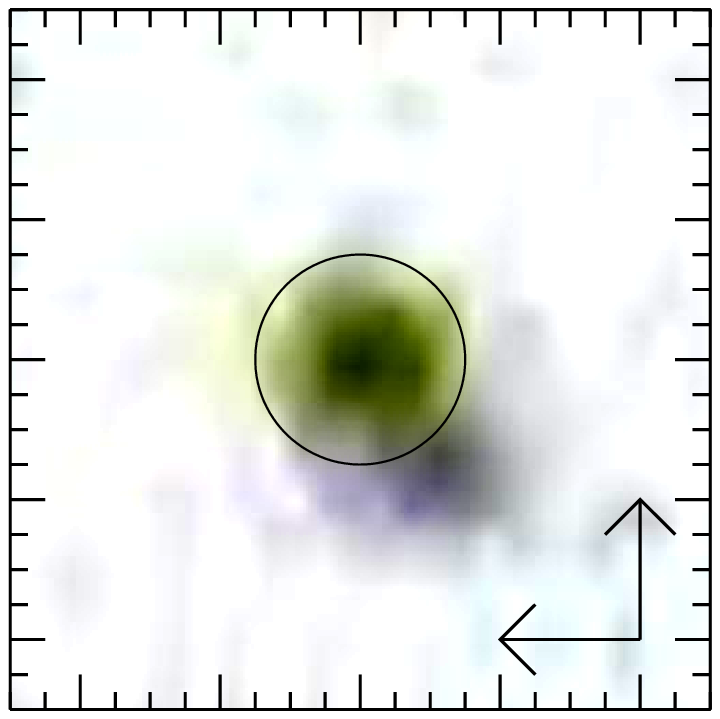} \\

\includegraphics[bb=190 0   425 790,angle=90,clip=,width=0.395\hsize ]{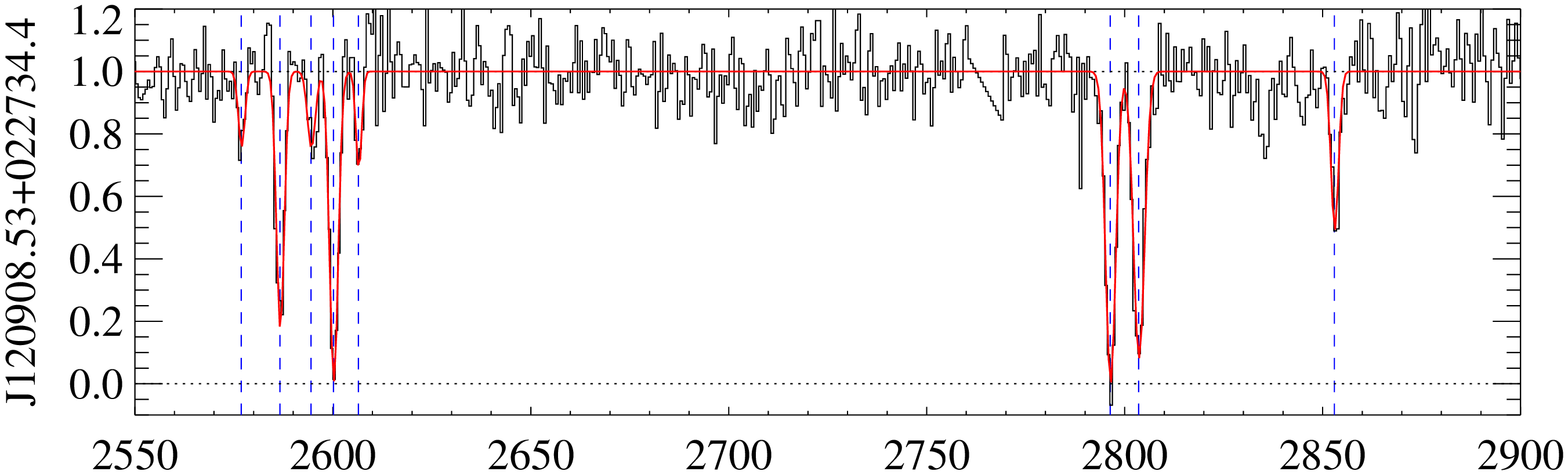} &
\includegraphics[bb=190 0   425 750,angle=90,clip=,width=0.375\hsize ]{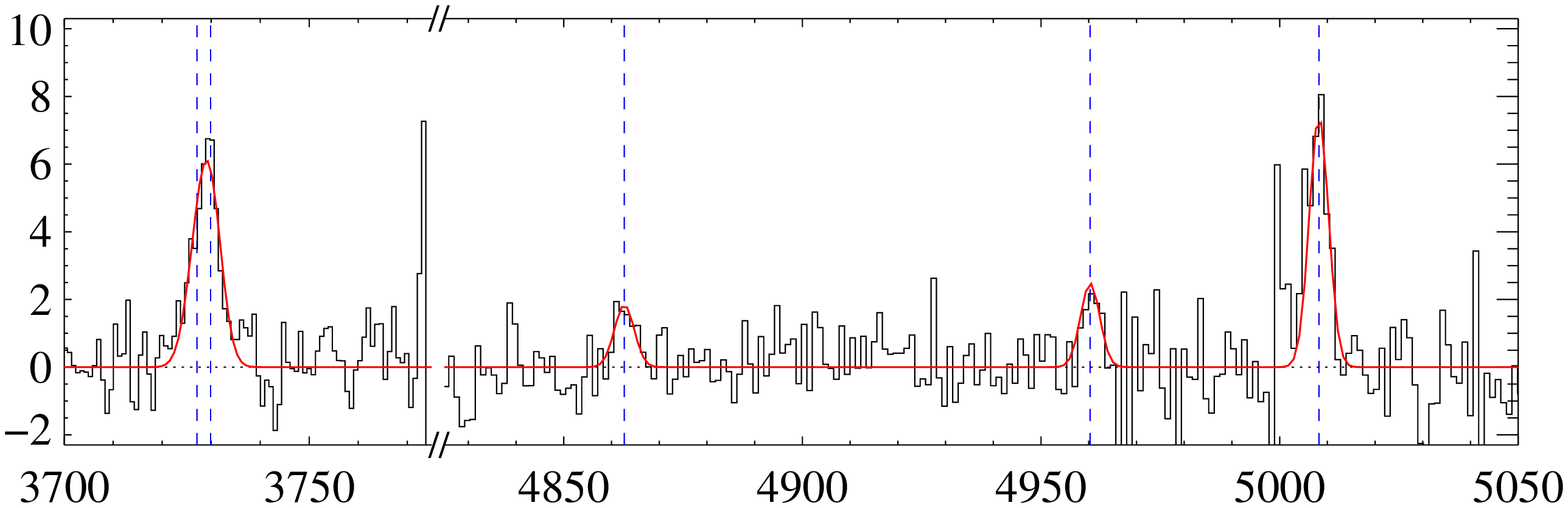} &
\includegraphics[bb=190 519 425 722,angle=90,clip=,width=0.1015\hsize]{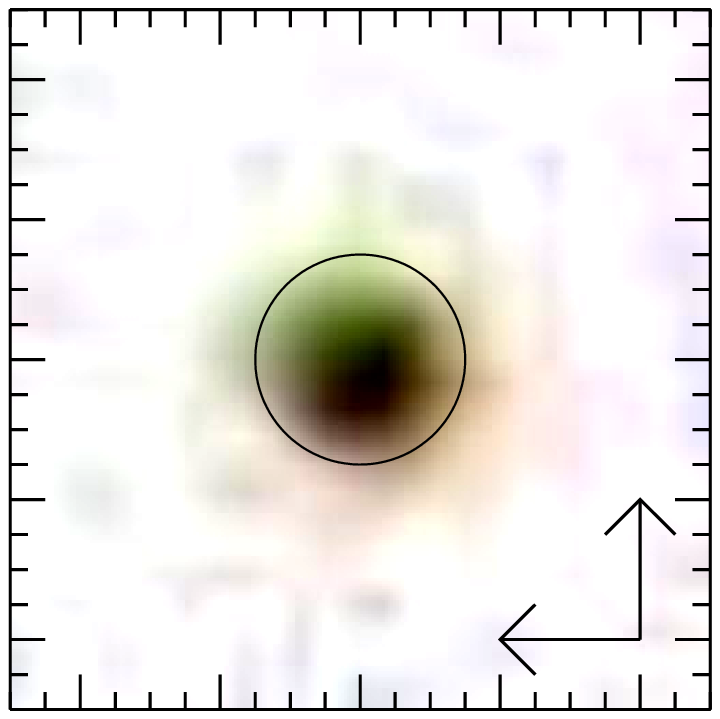} \\

Rest wavelength~({\AA}) & Rest wavelength~({\AA}) & $\leftarrow$ 10 \arcsec $\rightarrow$ \\

\end{tabular}
\caption{Absorption (left) and emission (middle) lines from the intervening galaxies. {Note that in the left 
panels (absorption), the spectrum is normalised by dividing the observed spectrum by the QSO continuum, while in 
the middle panels (emission), the QSO continuum is subtracted from the observed spectrum and the flux-scale is 
in units of 10$^{-17}$~erg\,s$^{-1}$\,{\AA}$^{-1}$. Best fitted absorption and emission lines are overplotted.}
The SDSS 
images of the QSOs are shown in the right panels. The black circle represents the position of the 3\arcsec-diameter 
SDSS fibre. North is top and East is left. \label{sample_fig}
}
\end{figure*}

\setcounter{figure}{\value{prev_fig}}
\begin{figure*}
\centering
\begin{tabular}{ccc}
\multicolumn{1}{l}{\hspace{0.7cm} \FeII, \MnII \hspace{2.7cm} \MgII \hspace{0.25cm} \MgI} & \multicolumn{1}{l}{\hspace{0.5cm} [\OII] \hspace{1.2cm} \Hb\ \hspace{1.8cm} [\OIII]} & \\

\includegraphics[bb=219 0   425 790,angle=90,clip=,width=0.395\hsize ]{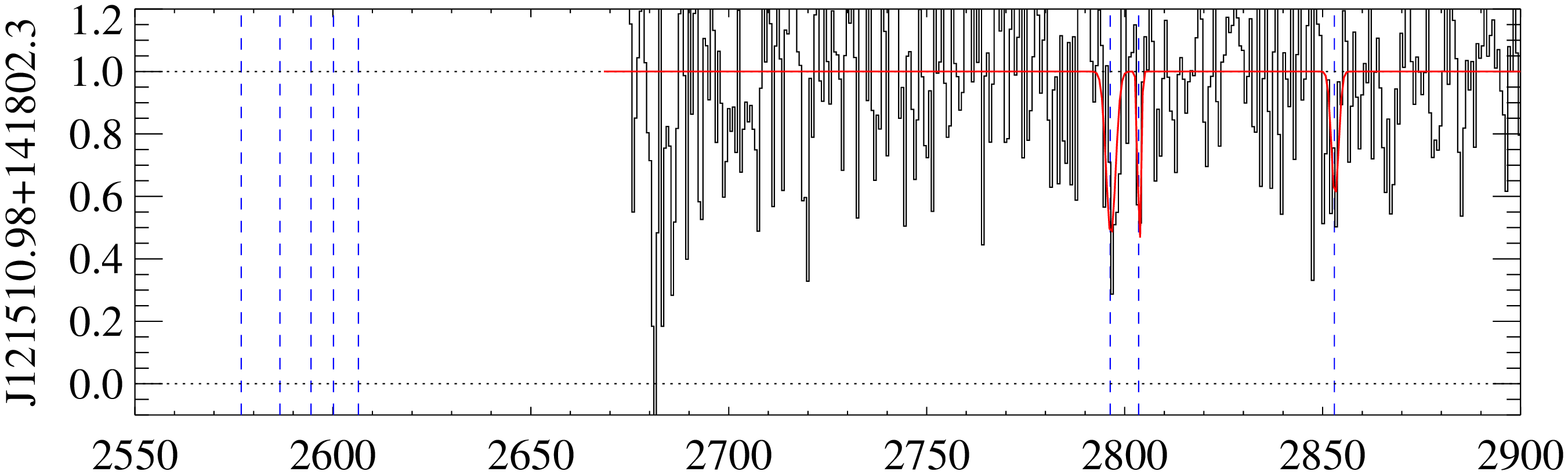} &
\includegraphics[bb=219 0   425 750,angle=90,clip=,width=0.375\hsize ]{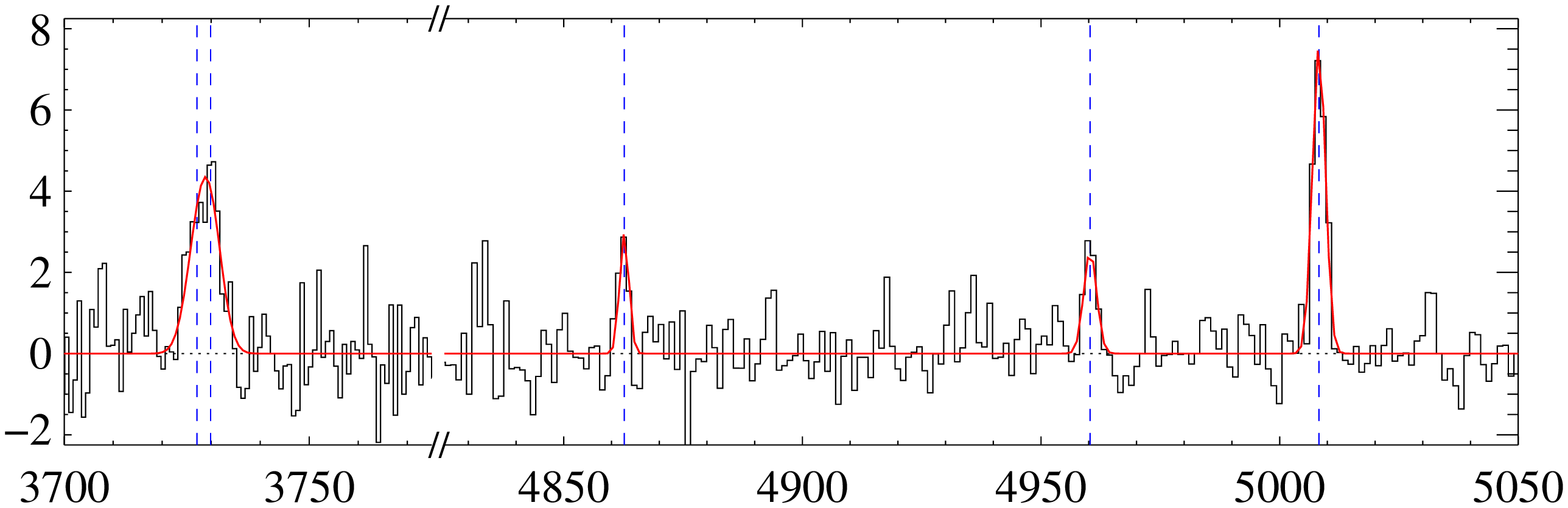} &
\includegraphics[bb=219 519 425 722,angle=90,clip=,width=0.1015\hsize]{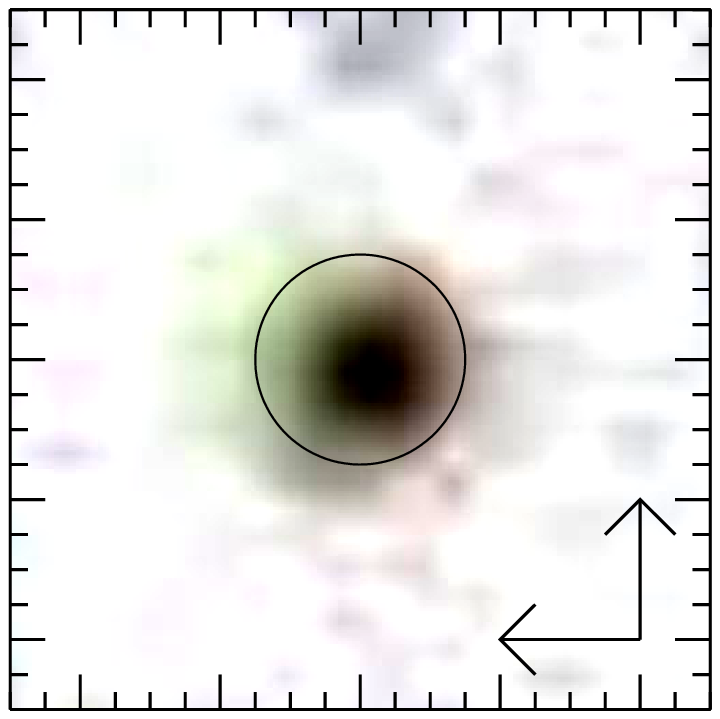} \\

\includegraphics[bb=219 0   425 790,angle=90,clip=,width=0.395\hsize ]{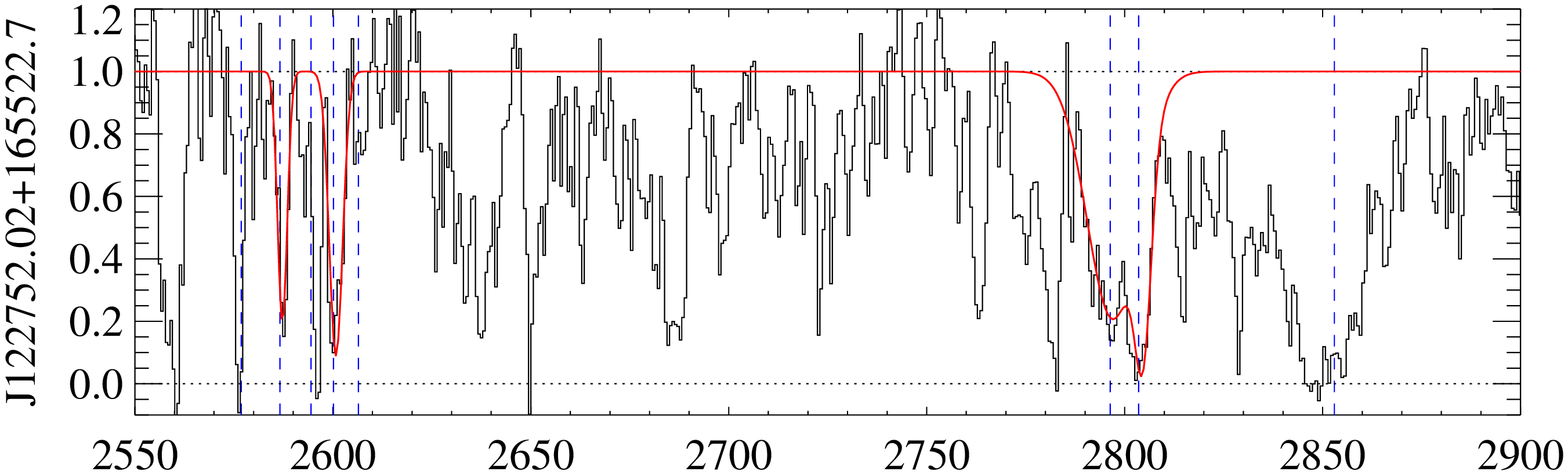} &
\includegraphics[bb=219 0   425 750,angle=90,clip=,width=0.375\hsize ]{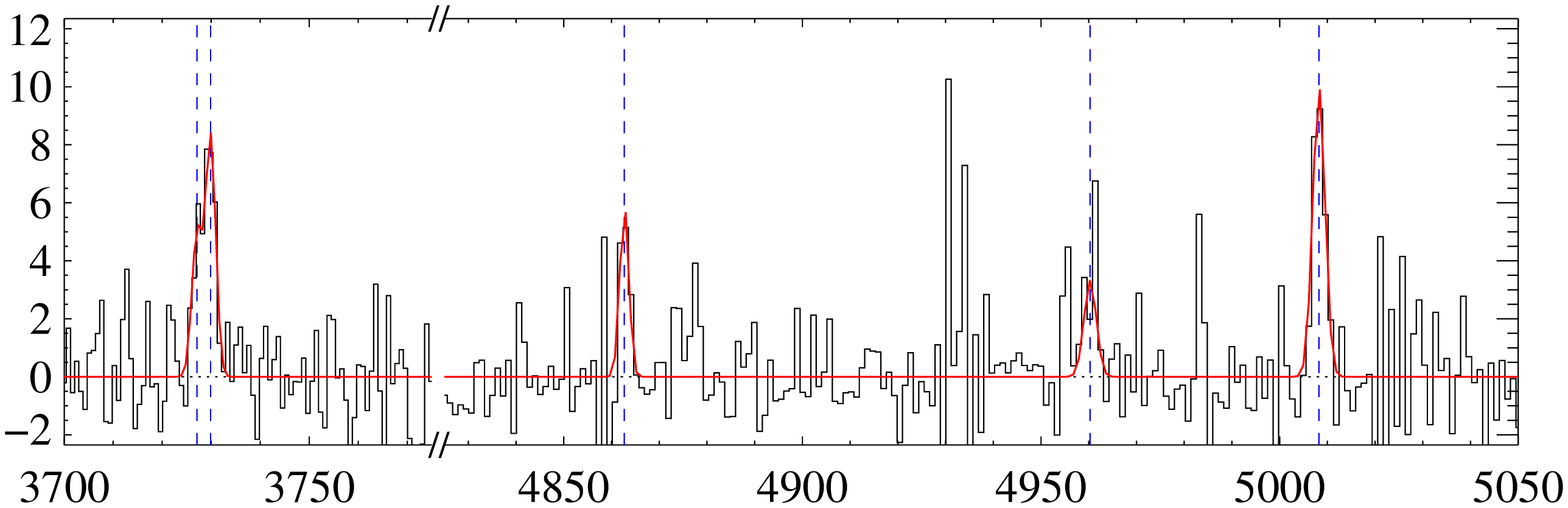} &
\includegraphics[bb=219 519 425 722,angle=90,clip=,width=0.1015\hsize]{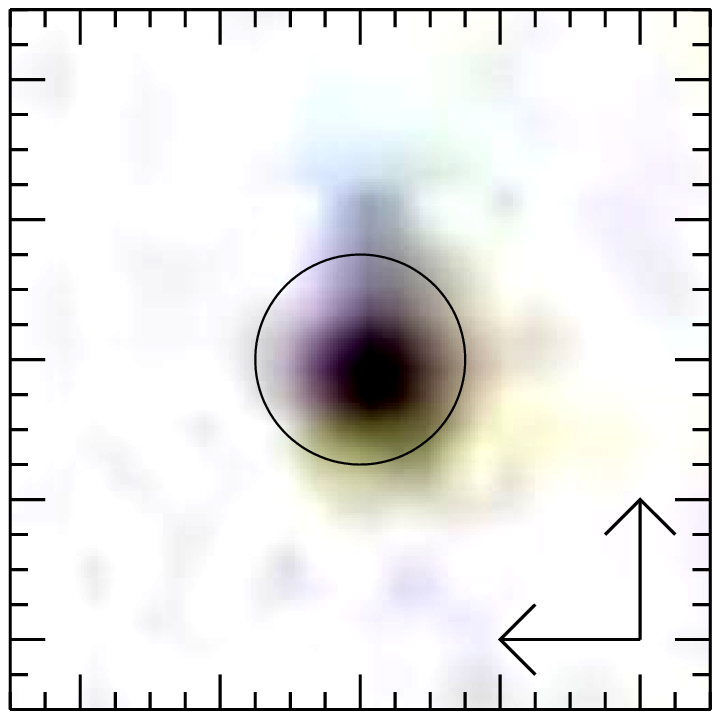} \\

\includegraphics[bb=219 0   425 790,angle=90,clip=,width=0.395\hsize ]{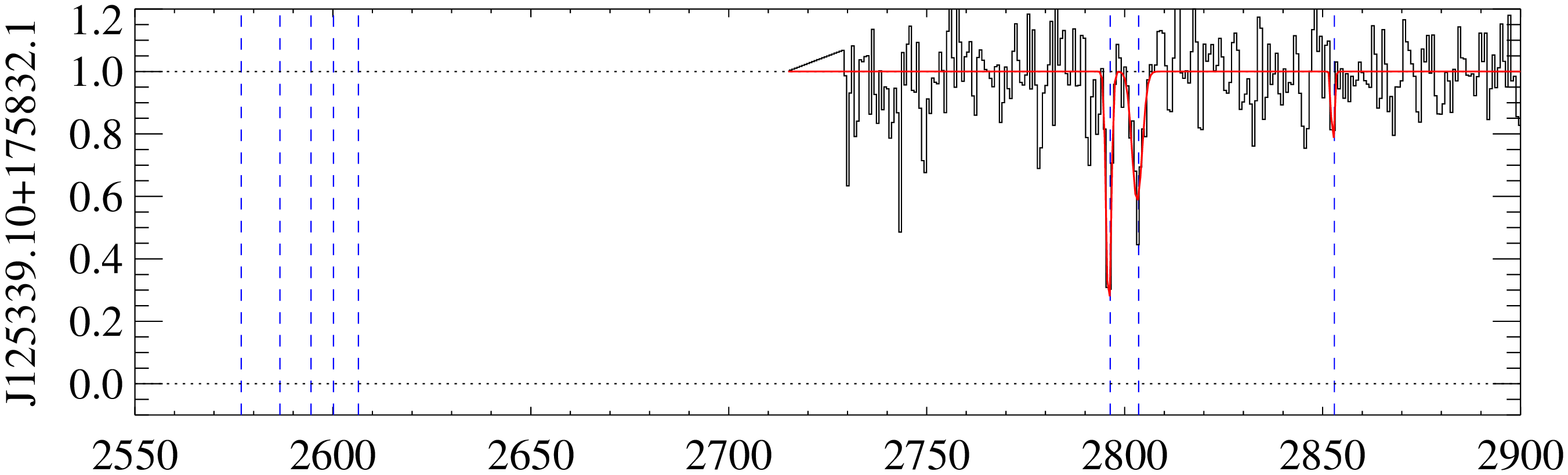} &
\includegraphics[bb=219 0   425 750,angle=90,clip=,width=0.375\hsize ]{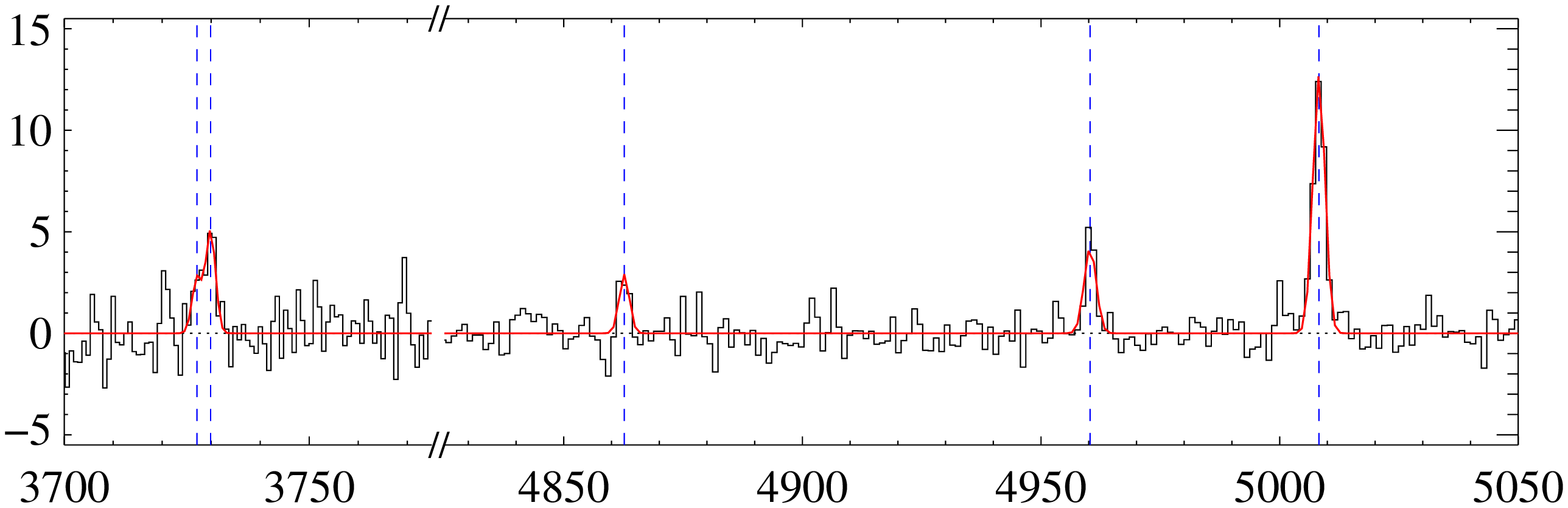} &
\includegraphics[bb=219 519 425 722,angle=90,clip=,width=0.1015\hsize]{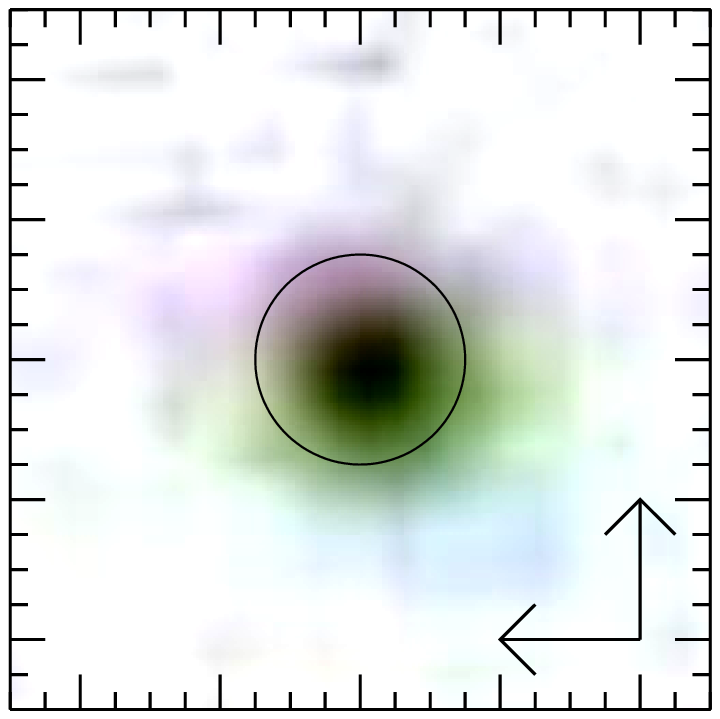} \\

\includegraphics[bb=219 0   425 790,angle=90,clip=,width=0.395\hsize ]{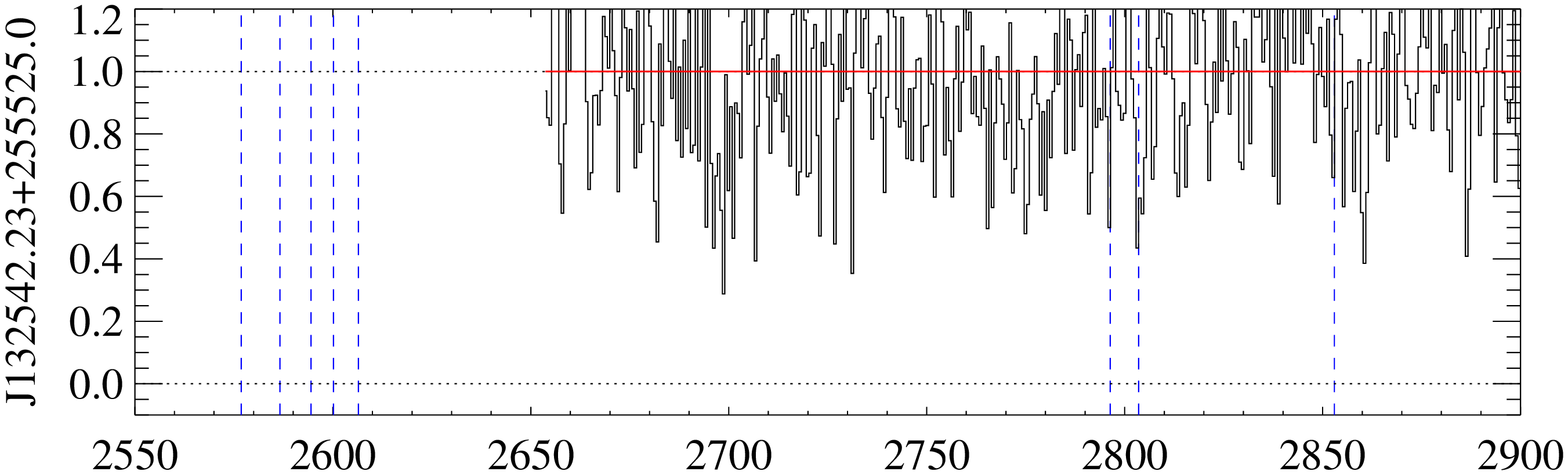} &
\includegraphics[bb=219 0   425 750,angle=90,clip=,width=0.375\hsize ]{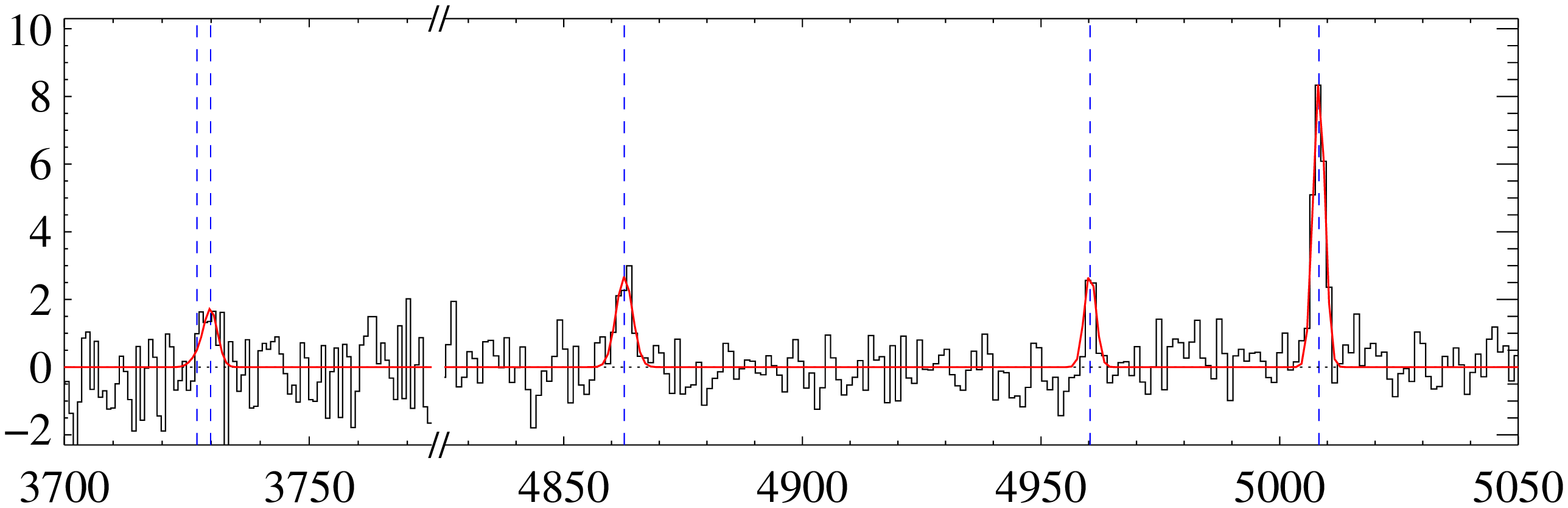} &
\includegraphics[bb=219 519 425 722,angle=90,clip=,width=0.1015\hsize]{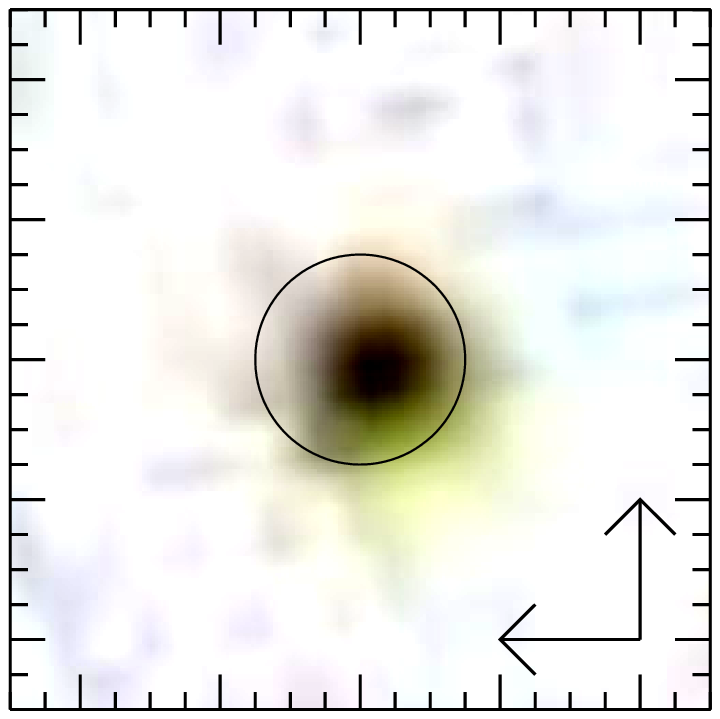} \\

\includegraphics[bb=219 0   425 790,angle=90,clip=,width=0.395\hsize ]{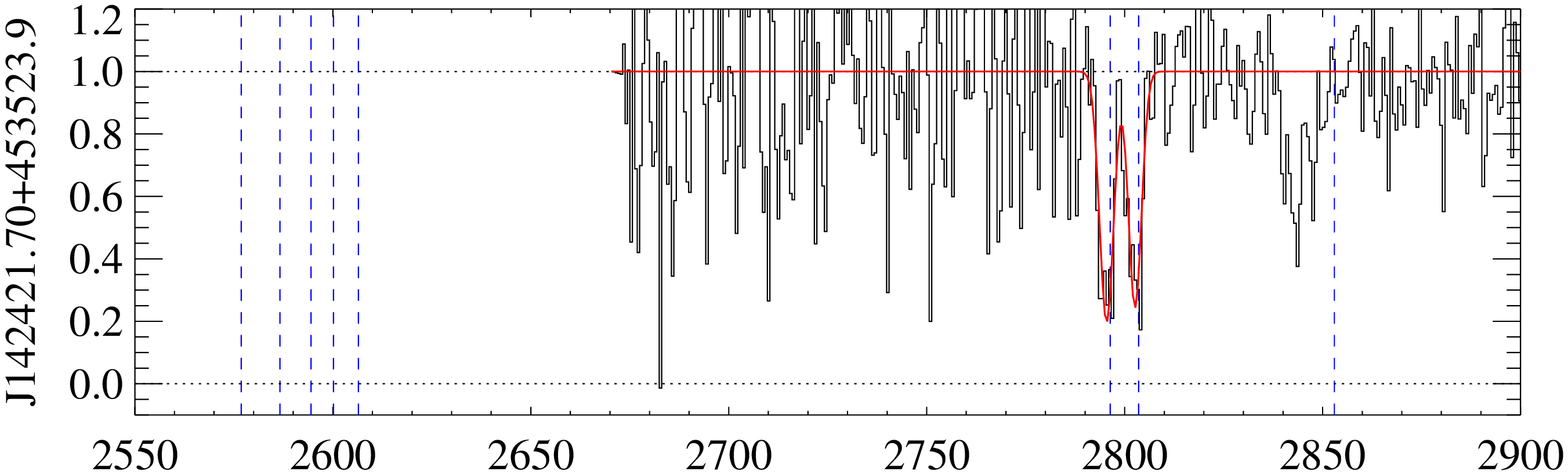} &
\includegraphics[bb=219 0   425 750,angle=90,clip=,width=0.375\hsize ]{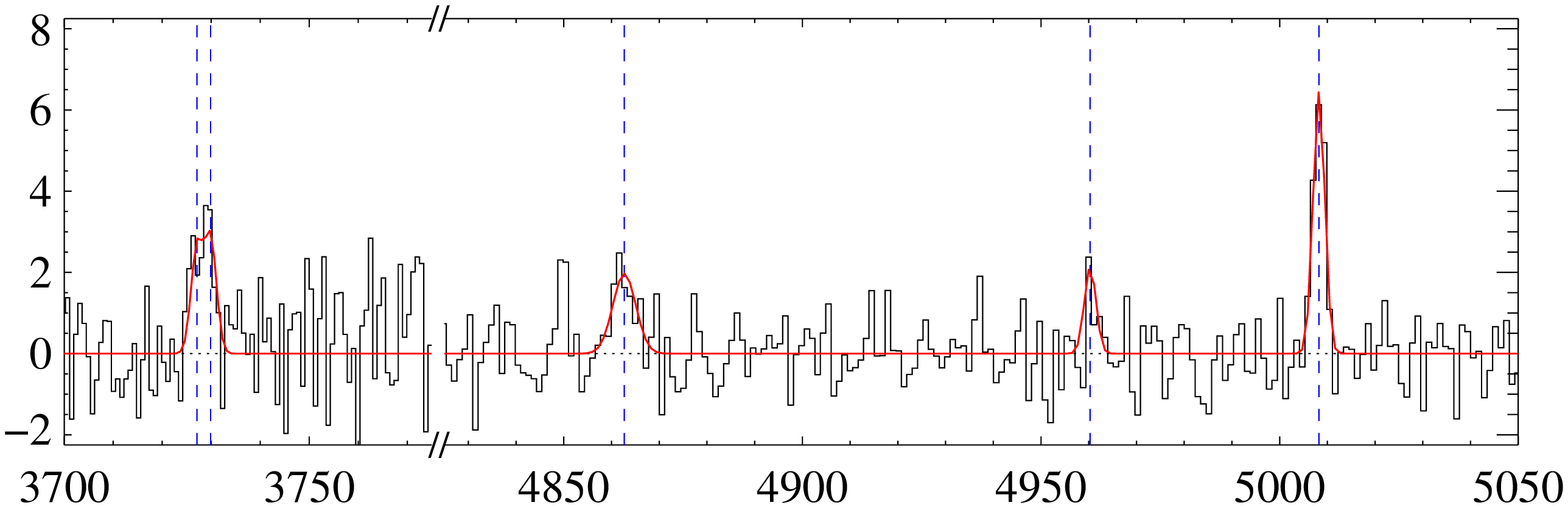} &
\includegraphics[bb=219 519 425 722,angle=90,clip=,width=0.1015\hsize]{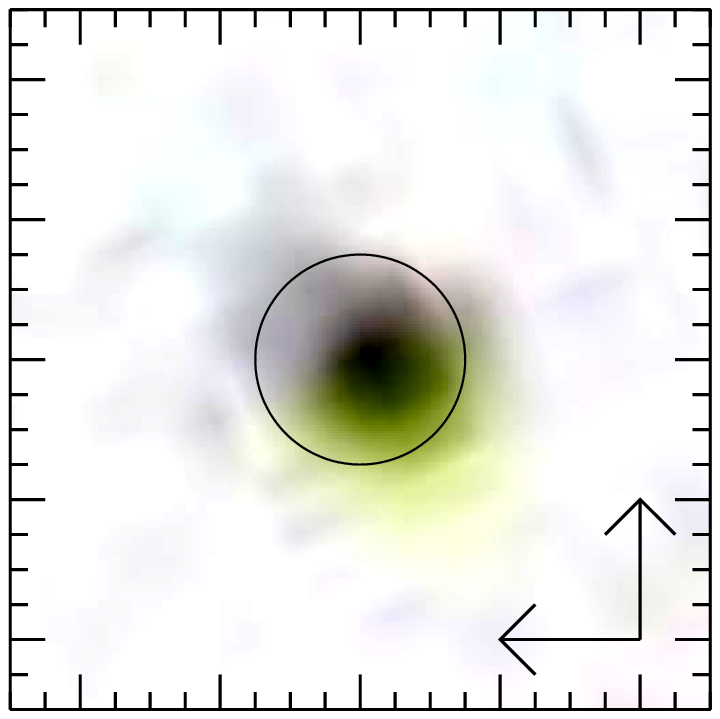} \\

\includegraphics[bb=219 0   425 790,angle=90,clip=,width=0.395\hsize ]{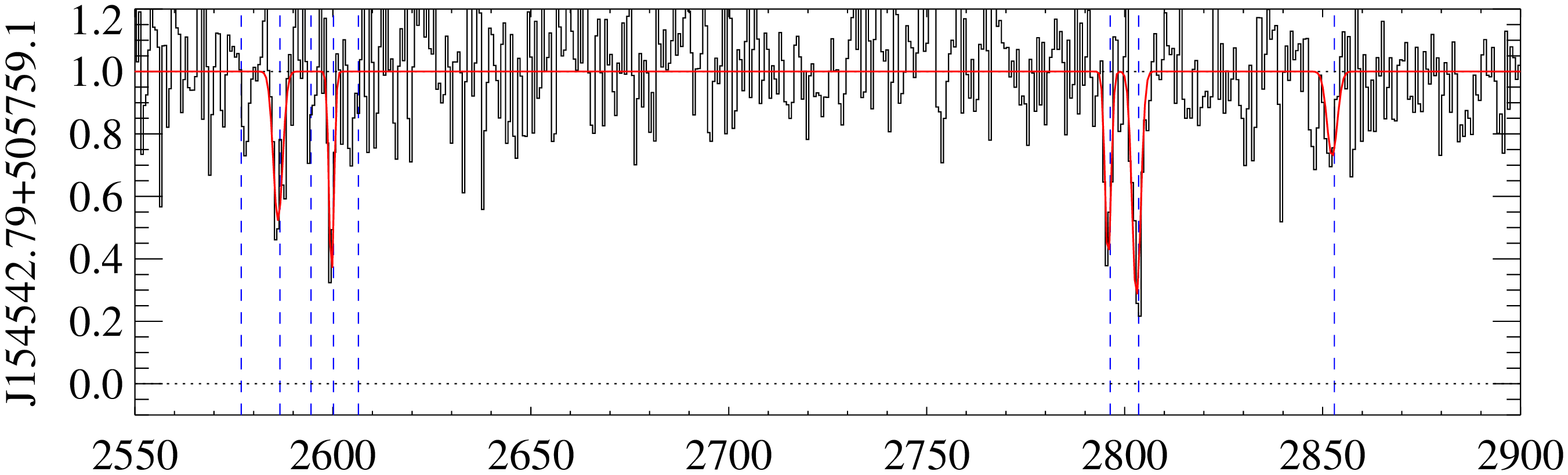} &
\includegraphics[bb=219 0   425 750,angle=90,clip=,width=0.375\hsize ]{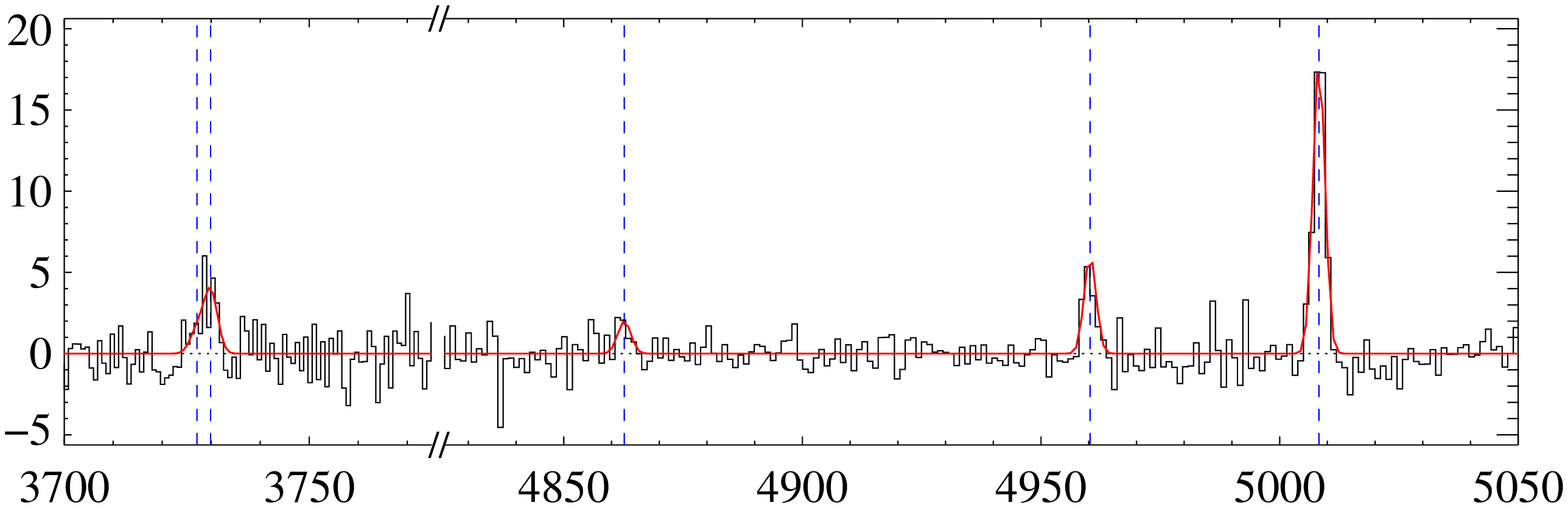} &
\includegraphics[bb=219 519 425 722,angle=90,clip=,width=0.1015\hsize]{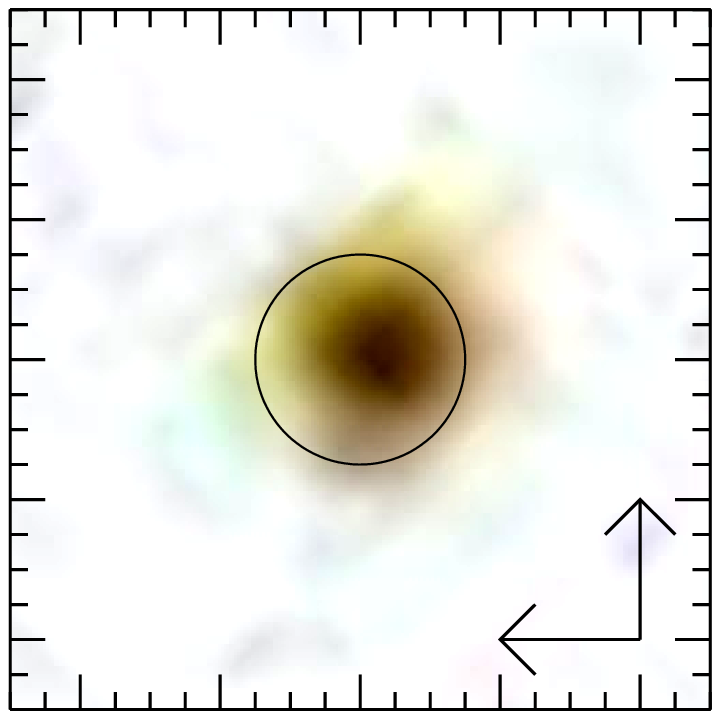} \\

\includegraphics[bb=219 0   425 790,angle=90,clip=,width=0.395\hsize ]{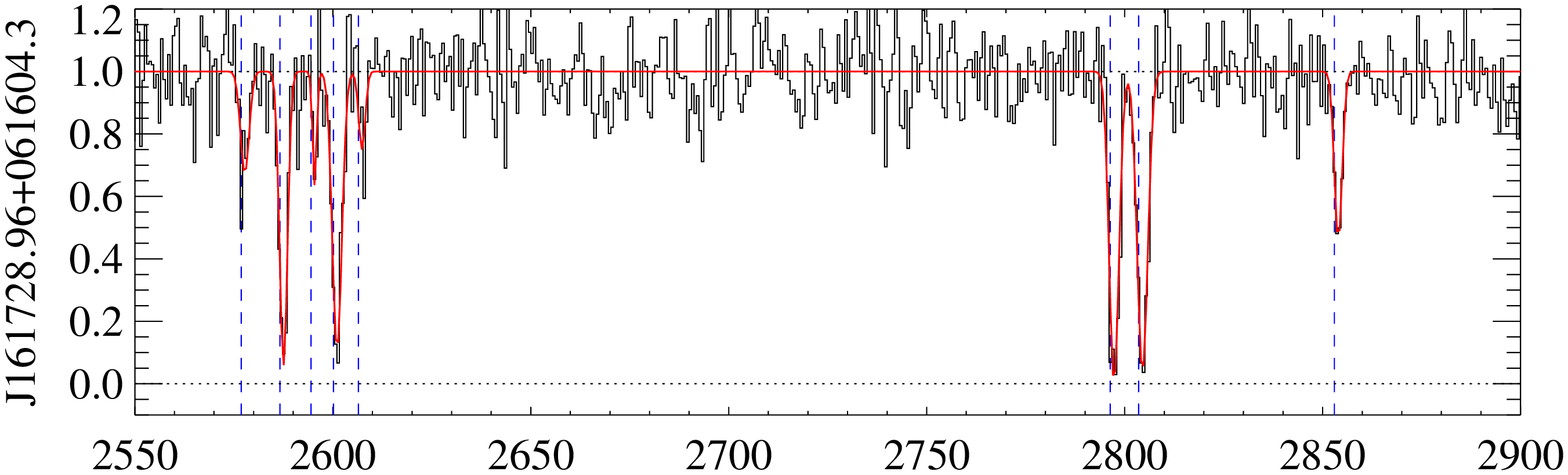} &
\includegraphics[bb=219 0   425 750,angle=90,clip=,width=0.375\hsize ]{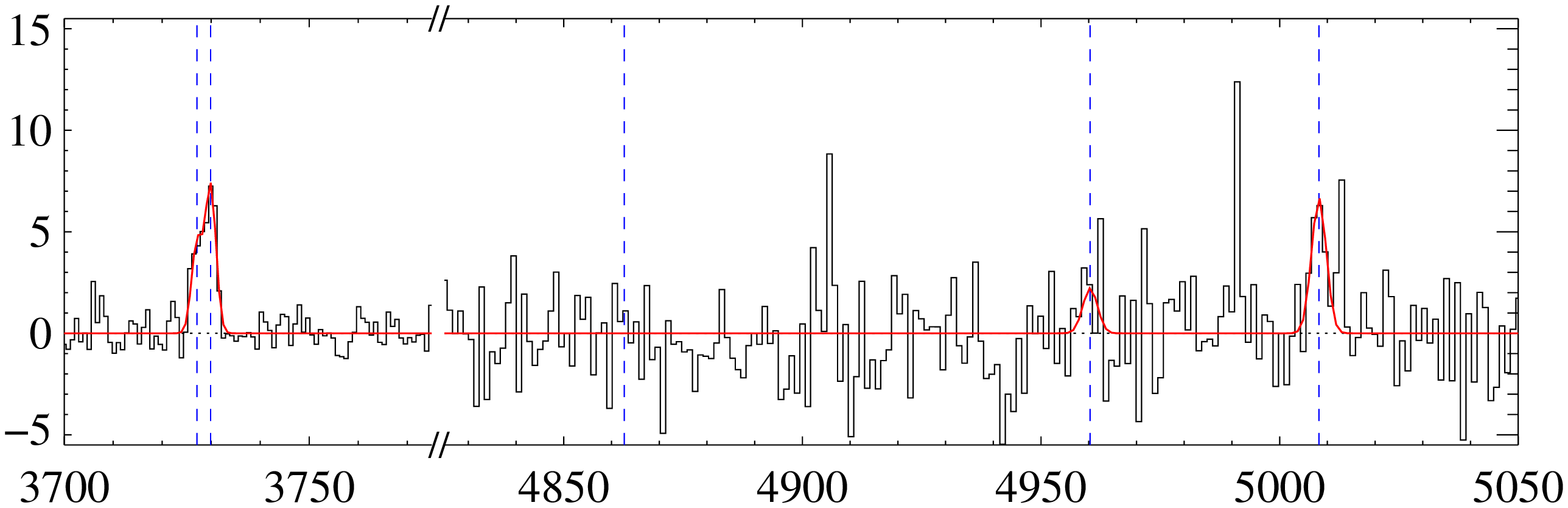} &
\includegraphics[bb=219 519 425 722,angle=90,clip=,width=0.1015\hsize]{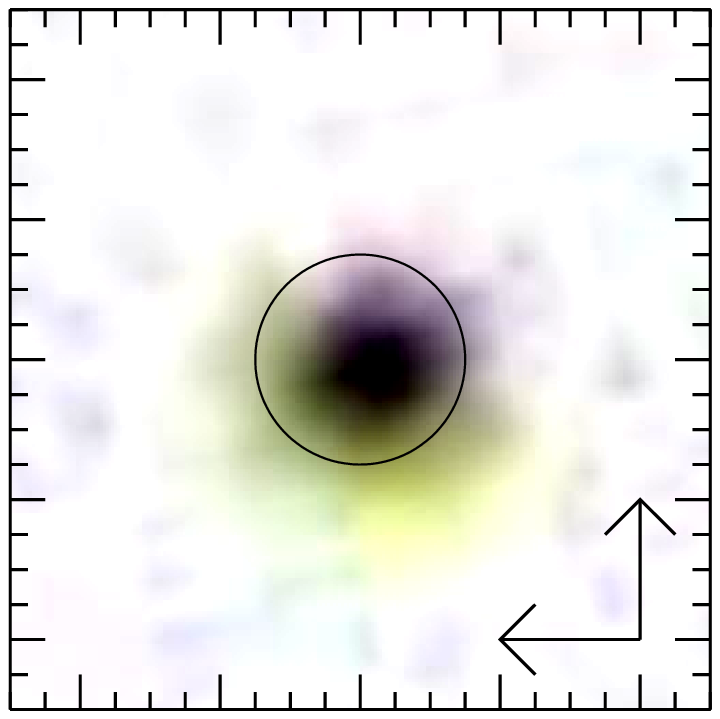} \\

\includegraphics[bb=219 0   425 790,angle=90,clip=,width=0.395\hsize ]{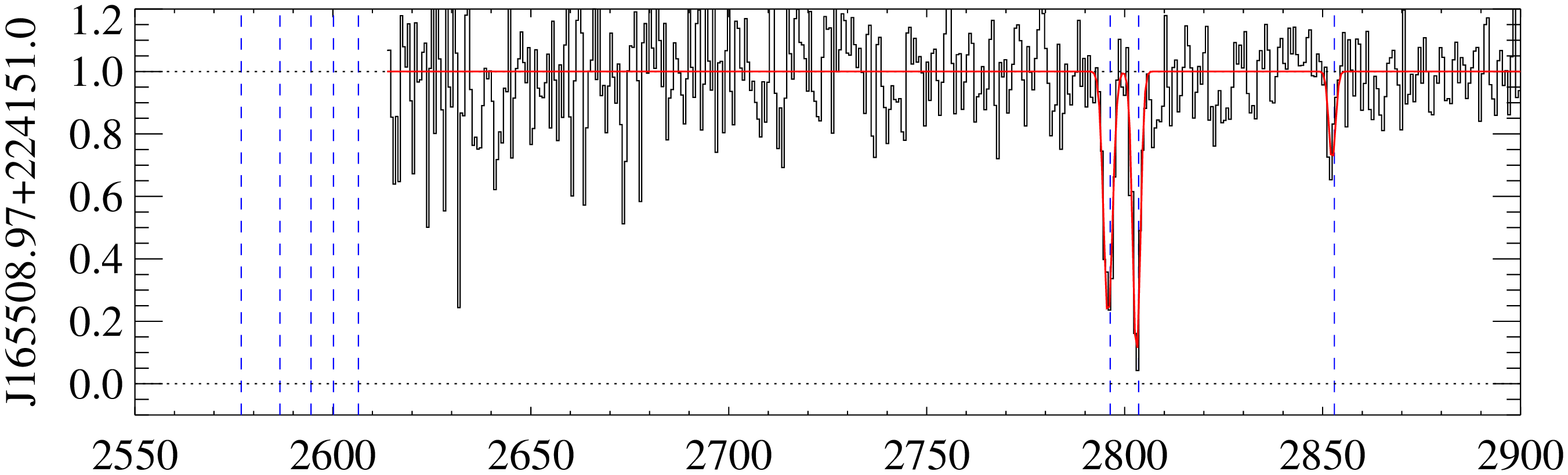} &
\includegraphics[bb=219 0   425 750,angle=90,clip=,width=0.375\hsize ]{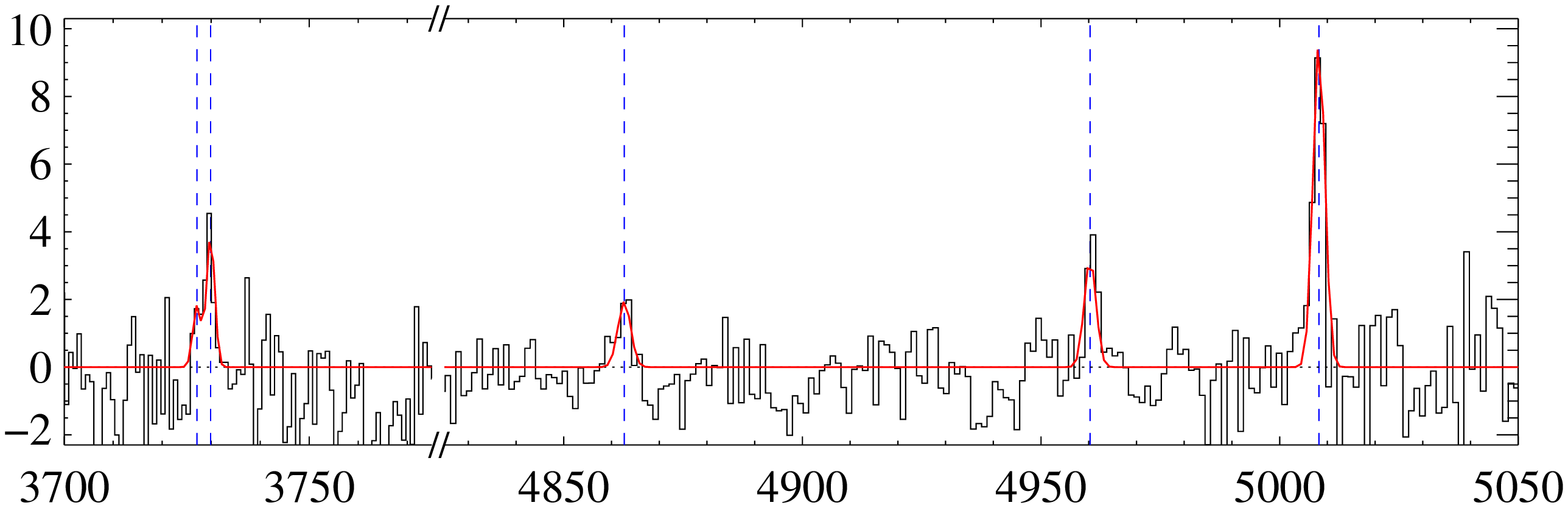} &
\includegraphics[bb=219 519 425 722,angle=90,clip=,width=0.1015\hsize]{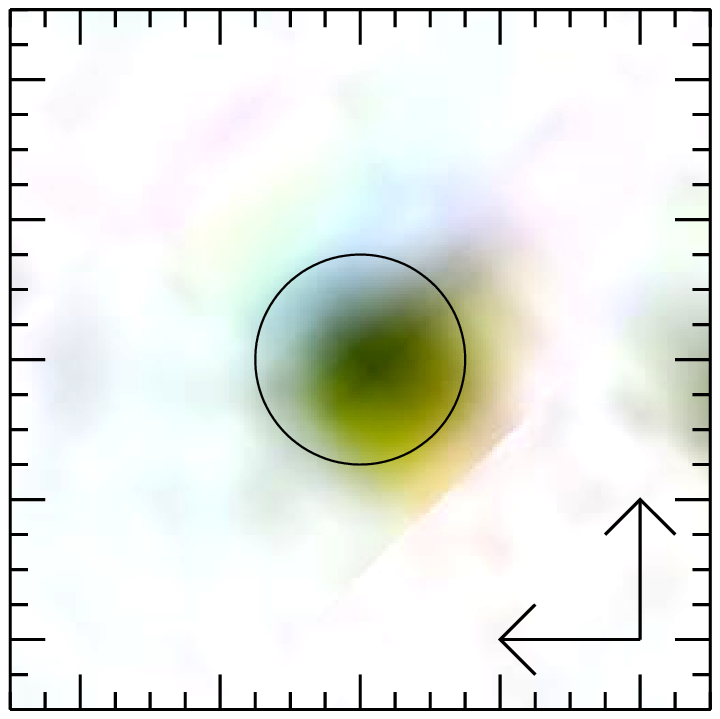} \\

\includegraphics[bb=190  0  425 790,angle=90,clip=,width=0.395\hsize ]{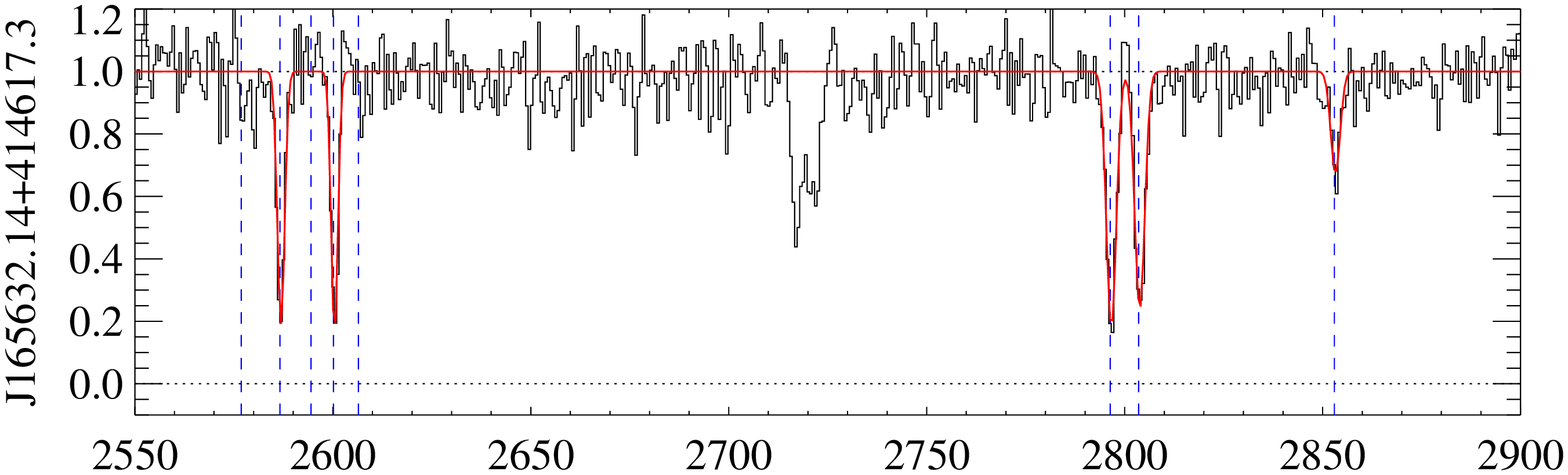} &
\includegraphics[bb=190  0  425 750,angle=90,clip=,width=0.375\hsize ]{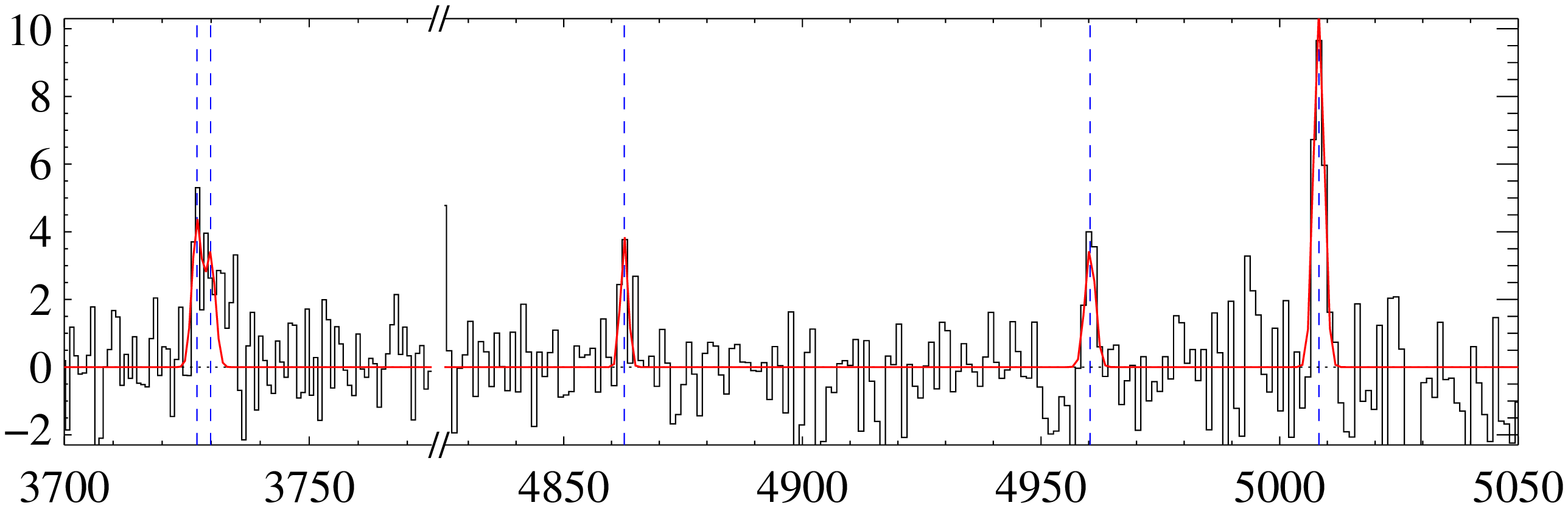} &
\includegraphics[bb=190 519 425 722,angle=90,clip=,width=0.1015\hsize]{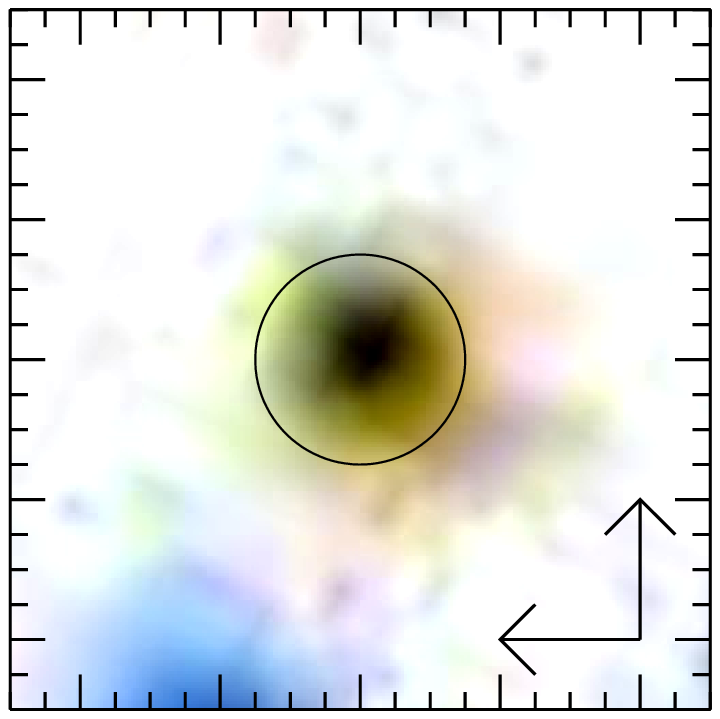} \\

Rest wavelength~({\AA}) & Rest wavelength~({\AA}) &  $\leftarrow$ 10 arcsec $\rightarrow$ \\
\end{tabular}
\caption{continued}

\end{figure*}

\begin{center}
\begin{table*}
\begin{center}
\caption{Star formation rates of the intervening galaxies$^a$ \label{sfr}}
\begin{tabular}{c c c c c c c c c c c}
\hline
\hline
           QSO &$\zgal$   & \multicolumn{3}{c}{$F_{\lambda }$ (10$^{-17}$ erg\,s$^{-1}$\,cm$^{-2}$)}     &\multicolumn{3}{c}{$L$ (10$^{40}$ erg\,s$^{-1}$)$^b$}& \multicolumn{3}{c}{SFR (M$_\odot$ yr$^{-1}$)          }\\
               &          &[O\,{\sc ii}]  &[O\,{\sc iii}]&   H$\beta$   & [O\,{\sc ii}]& [O\,{\sc iii}]&     H$\beta$          & [\OII]$^c$ & \Hb$^c$ & [\OII]+\Hb$^d$ \\
\hline                     
J080808+064108 &  0.433   &   12.1(4.2)  &  25.1(3.0)  &   4.9(3.4)  &   8.3(2.9)  &   17.2(2.8)  &   6.7(4.7)   &   0.4      & 0.8 &    0.9 \\  
J081154+202148 &  0.445   &    $\le 6$   &  22.0(2.6)  &   4.0(1.8)  &   $\le 4.2$ &   16.1(2.5)  &   5.8(2.7)   &   \ldots   & 0.7 &    \ldots \\  
J091417+325955 &  0.444   &    8.3(2.0)  &  12.3(1.9)  &   $\le 5$   &   6.0(1.4)  &    8.9(1.8)  &   $\le 3.6$   &   0.3      & \ldots &    \ldots \\  
J092913+302225 &  0.439   &   16.9(8.2)  &  32.2(2.6)  &   7.4(2.3)  &  12.0(5.8)  &   22.8(2.4)  &  10.4(3.3)   &   0.5      & 1.2 &    1.7 \\   
J094041+341535 &  0.447   &    7.6(3.2)  &  33.7(4.2)  &   5.7(2.4)  &   5.6(2.3)  &   24.8(4.1)  &   8.4(3.6)   &   0.2      & 0.9 &    3.1-4.3\\   
J095228+032616 &  0.419   &   21.1(7.4)  &  29.2(2.5)  &   6.1(3.2)  &  13.4(4.7)  &   18.5(2.1)  &   7.7(4.1)   &   0.6      & 0.9 &    0.5-0.6 \\  
J104223+092708 &  0.592   &   21.4(11.7) &  97.9(5.7)  &  11.5(4.0)  &  31.0(17.0) &  141.9(11.1) &  33.3(11.7   &   1.4      & 3.5 &    7.3-8.0\\    
J113108+202151 &  0.563   &  110.5(9.6)  & 247.3(6.8)  &  47.1(7.5)  & 141.8(12.3) &  317.4(11.7) & 120.8(19.1   &   6.3      & 11.8 &   16.5-17.7 \\   
J120538+604057 &  0.434   &   33.9(7.5)  &  19.1(2.5)  &   6.3(2.7)  &  23.3(5.2)  &   13.1(2.3)  &   8.7(3.7)   &   1.0      & 1.0 &    0.2-0.3 \\   
J120908+022734$^e$ &  0.669   &   42.8(7.2)  &  50.1(6.1)  &   9.6(4.2)  &  83.7(14.0) &   97.8(15.9) &  37.4(16.5)  &   3.7      & 3.9 &    1.4-1.8 \\   
J121510+141802 &  0.421   &   31.2(8.2)  &  34.3(2.8)  &   7.3(2.6)  &  20.0(5.3)  &   22.0(2.4)  &   9.3(3.3)   &   0.9      & 1.0 &    0.4-0.5 \\   
J122752+165522 &  0.565   &   34.5(6.8)  &  43.0(4.1)  &  14.0(6.6)  &  44.6(8.8)  &   55.6(7.0)  &  36.2(17.0)  &   2.0      & 3.8 &    4.7-5.0 \\   
J125339+175832 &  0.401   &   19.4(4.6)  &  53.8(3.3)  &   7.7(2.7)  &  11.1(2.6)  &   30.7(2.6)  &   8.8(3.1)   &   0.5      & 1.0 &    1.2 \\   
J132542+255525 &  0.433   &    6.9(5.5)  &  34.6(2.9)  &  11.7(3.3)  &   4.7(3.8)  &   23.7(2.7)  &  16.0(4.6)   &   0.2      & 1.7 &   18.1-45.2 \\   
J142421+453523 &  0.421   &   16.4(8.0)  &  26.2(3.3)  &  11.9(4.8)  &  10.5(5.2)  &   16.8(2.8)  &  15.3(6.2)   &   0.5      & 1.7 &    5.4-7.2 \\   
J154542+505759 &  0.525   &   18.6(11.8) &  75.8(5.0)  &   6.9(3.2)  &  20.1(12.8) &   82.2(7.2)  &  14.9(7.0)   &   0.9      & 1.6 &    1.7-1.8 \\   
J161728+061604$^e$ &  0.788   &   32.1(4.9)  &  34.3(9.8)  &   $\le 9.5$  &  93.3(14.3) &   99.8(38.1) &  $\le 27.5$   &   4.1      & \ldots &    \ldots \\ 
J165508+224150 &  0.453   &   11.9(4.8)  &  39.7(5.2)  &   6.5(3.5)  &   9.0(3.6)  &   30.2(5.3)  &   9.9(5.4)   &   0.4      & 1.1 &    2.3-2.6 \\  
J165632+414617 &  0.662   &   18.8(5.1)  &  38.5(6.9)  &   7.6(4.7)  &  35.7(9.7)  &   73.2(17.4) &  28.9(17.7)  &   1.6      & 3.0 &    3.8-4.0 \\    
\hline
\end{tabular}
\end{center}
\begin{flushleft}
\footnotesize
$a$ Due to fibre losses, luminosities and star formation rates should be considered as lower limits.
$b$ Luminosities are {\sl not} corrected for dust-extinction. 
$c$ Updated SFR calibrations for [\OII] \citep{Kewley04} and \Ha\ \citep{Kennicutt98} were 
taken from \citet{Argence09}, assuming an intrinsic Balmer decrement of 2.85: \Ha~=~2.85\,\Hb.
$d$ Self-consistent two-lines calibration ([\OII]+\Hb) from \citet{Argence09}.
$e$ (\MgII+[\OII])-selected galaxies.
\end{flushleft}
\normalsize
\end{table*}\end{center}

\subsection{Luminosities}

\begin{figure}
\centering
\includegraphics[bb=70 178 520 580,clip=,width=\hsize]{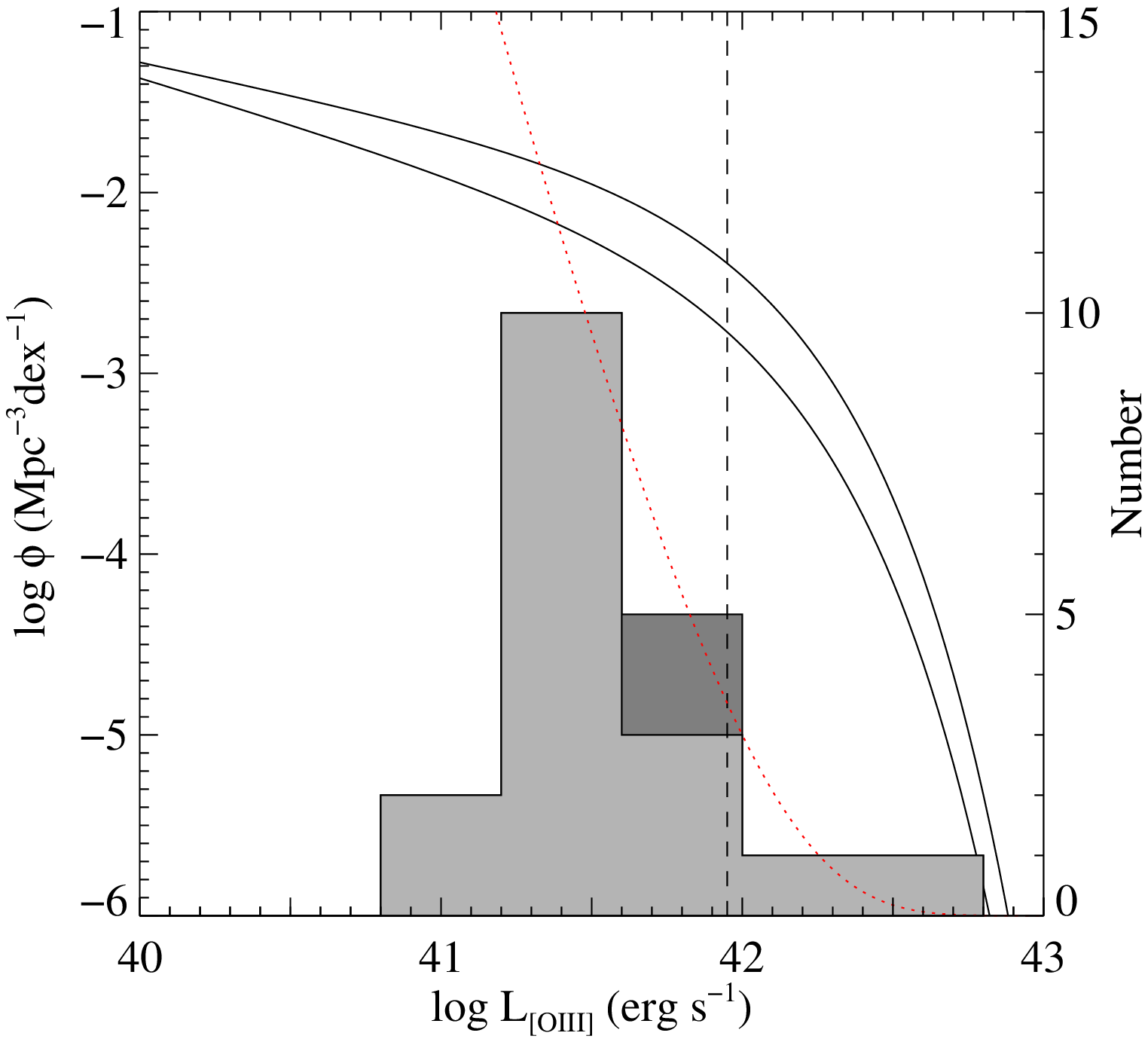}
\caption{The distribution of [\OIII] luminosities for the intervening galaxies  
(light histogram: [\OIII]-selected, dark histogram: (\MgII+[\OII])-selected) is compared 
to the [\OIII] luminosity functions at $z=0.4$ (higher curve) and $z=0.64$ (lower curve),
to be read on the left axis. 
The vertical dashed line marks the position of 
$\lstaro=8.9\times10^{41}$~erg\,s$^{-1}$ \citep{Hippelein03}. {The dotted curve represent 
the [\OIII] luminosity function in linear space and arbitrary scaling to illustrate 
the completeness of the luminosity distribution of the galaxies presented in this paper.}\label{lf}}
\end{figure}

Using the measured redshifts and fluxes we estimate the emission
line luminosities for the cosmological parameters noted above. 
Luminosities for all the
three lines are also summarised in Table~\ref{sfr}. Note that 
we do not apply any correction for the dust reddening or 
the fact that the fibre need not sample the whole galaxy.
This means that the quoted luminosities should be treated as lower  
limits. In Fig.~\ref{lf}, we compare the distribution of the 
measured [\OIII] luminosities of our galaxies and 
the [\OIII] luminosity functions at $z=0.4-0.6$ from \citet{Hippelein03}. 
The vertical dotted line marks the luminosity of a $\lstaro$ galaxy.
As expected we mostly detect galaxies with luminosities in the range 
0.1-3~\lstaro, with a median [\OIII] luminosity $L_{\rm [OIII]} 
\sim$0.2 \lstaro.
The galaxy along the line of sight towards J113108+202151
has $L_{\rm [OIII]} \sim 2 \lstaro$. 
We study this system in detail using our observations with IUCAA Girawali
observatory (IGO) in Section~\ref{IGO}.

The sharp decrease seen in the number of galaxies detected at the low 
luminosity end ($\log L_{\rm [OIII]}<41$) is a consequence of our 
detectability limit, {as can be seen from the departure from 
the dotted curve in Fig.~\ref{lf}}, while the decrease at high luminosities 
($\log L_{\rm [OIII]}>42$) {is a natural consequence of the } decrease 
in the number density of very luminous galaxies.  As there is a factor 30 
spread in the luminosity, this set of galaxies provides a good sample
for various followup studies such as measuring the cross-section
and filling factor of \MgII\ absorbers at low impact parameters (i.e $\le 10$ kpc).

\subsection{Star-formation rates \label{s_sfr}}

In this section, we aim at deriving the star-formation rates of 
intervening galaxies using emission line luminosities. 
It is known that the relationship between the observed luminosity
of a line and star formation rate depends on dust extinction and
metallicity [see for example \citet{Argence09}]. In the redshift
range of galaxies that we focus on in this paper, the \Ha\ line
redshifts into the near-IR wavelengths. Thus we can not use 
the Balmer decrement to get the dust extinction. Also, as the 
continuum of the galaxy is combined with that of the QSO, we 
can not use the SED fitting to get the estimates for reddening. 

Using more than 100\,000 star-forming galaxies from SDSS, \citet{Argence09}
have provided fitting formulae (their Eqs. 23 and 24) that use 
uncorrected [\OII] and [\Hb] luminosities to get the star formation 
rates.  The typical quoted dispersion in the SFR is about 0.23
dex. As noted by \citet{Argence09}, the SFR derived using their Eq.~23 
is weakly sensitive to the variations in dust attenuation.  
However, their Eq.~24 provides a SFR estimate that is weakly sensitive
to the variations in metallicity.  Thus for each galaxy we consider
the SFR estimates based on the above-mentioned two equations to 
provide the realistic range in SFR (last column in Table~\ref{sfr}).
In this table we also give SFR based only on the [\OII] and [\Hb]
calibrators. However, these estimates should be treated as lower limits 
as the SDSS fibres may collect only part of the emission from the galaxies.

It is clear from Table~\ref{sfr} that the star formation rate in the galaxies
in our sample varies between 0.2 and 20~M$_\odot$ yr$^{-1}$. Roughly half of
the galaxies in the sample have a SFR less than 2~M$_\odot$\,yr$^{-1}$. 
Note that in the case of J132542+255525, there is a very large spread 
between the SFRs derived using the different calibrators. We caution that 
the [\OII] and \Hb\ emission lines in this system are close to the detection 
limit and using their ratio may lead to a significant overestimation 
of the SFR and metallicities (see next sub-section).
It is interesting to note that all the 5 \CaII~absorption selected galaxies
studied by \citet{Zych07} have a SFR similar to that of galaxies
in our sample.

\subsection{Emission line metallicities}
We use the R23 ratio, ([\OII]$+$[\OIII])$/$\Hb), as calibrated by 
\citet{Kobulnicky99} to measure the metallicities. This calibration 
provides two solutions for most values of R23, a low and a 
high metallicity estimates. These are generally referred to as 
``lower'' and ``upper'' branches of R23. The use of additional line 
ratios is necessary to break the degeneracy \citep[see][]{Kewley08}. 
Unfortunately, the corresponding lines are not covered by the SDSS spectra.
Therefore, we provide the lower and upper estimates of the metallicity derived using the 
R23 ratio only. 
The oxygen metallicity estimated using uncorrected fluxes are given in Table~\ref{metal}. 
The columns R23$_l$ and R23$_u$ refer to the lower and upper branch of R23. It is known that 
the value of (O/H) estimated with and without dust corrections are consistent with each 
another within 0.1~dex \citep{Moustakas06, Lamareille06}.

\begin{table}
\centering
\caption{Emission-line metallicities}
\label{metal}
\begin{tabular}{c c c c}
\hline
\hline
           QSO &$\zgal$  &   \multicolumn{2}{c}{$\log ($O/H$)+12$} \\
               &         &     R23$_l$&  R23$_u$  \\
\hline                   
J080808+064108 &  0.433  & 8.10 &  8.52 \\
J081154+202148 &  0.445  &\ldots& \ldots\\
J091417+325955 &  0.444  &\ldots& \ldots\\
J092913+302225 &  0.439  & 8.00 &  8.59 \\
J094041+341535 &  0.447  & 7.95 &  8.60 \\
J095228+032616 &  0.419  & 8.22 &  8.45 \\
J104223+092708 &  0.592  & 8.27 &  8.41 \\
J113108+202151 &  0.563  & 8.08 &  8.53 \\
J120538+604057 &  0.434  & 8.36 &  8.37 \\
J120908+022734$^\dagger$ &  0.669  & 8.38 &  8.33 \\
J121510+141802 &  0.421  & 8.33 &  8.37 \\
J122752+165522 &  0.565  & 7.92 &  8.66 \\
J125339+175832 &  0.401  & 8.24 &  8.42 \\
J132542+255525 &  0.433  & 7.40 &  8.86 \\
J142421+453523 &  0.421  & 7.59 &  8.83 \\
J154542+505759 &  0.525  & 8.56 &  8.21 \\
J161728+061604$^\dagger$ &  0.788  &\ldots& \ldots\\
J165508+224150 &  0.453  & 8.06 &  8.54 \\
J165632+414617 &  0.662  & 8.09 &  8.53 \\         
\hline
\end{tabular}

\footnotesize
\begin{flushleft}
For comparison the Solar value is $\log($O/H$)+12=8.69\pm0.05$ \citep{Asplund09}. 
No correction for dust has been applied when measuring R23. 
$\dagger$ (\MgII+[\OII])-selected galaxies.
\end{flushleft}
\normalsize
\end{table}

\citet{Mouhcine06} pointed out that, for the intermediate redshift galaxies 
where the degeneracy is lifted using other line indicators, the (O/H) is found
to be predominantly close to the value obtained for the upper branch.
We find that the (O/H) range in our sample obtained using R23$_u$ 
compares well with that measured in intermediate redshift field
galaxies and cluster galaxies 
\citep[See][]{Kobulnicky03a, Kobulnicky03b, Lilly03, Mouhcine06}.
This confirms that our galaxy selection is not heavily biased 
towards high or low metallicity galaxies. However, we wish to point-out that
the (O/H) determination based on R23$_u$ for the \CaII\ and
DLA-selected galaxies \citep{Zych07} are slightly higher than the 
values we find for the galaxies in our sample.

\section{Analysis of absorption features}

The emission line analysis presented in the previous section 
clearly suggests that the distribution of physical properties of 
the emission line galaxies in our sample are consistent with 
that found for field galaxies at similar redshift range. Thus,
we have an unbiased, albeit small sample of star-forming galaxies
where we will be able to probe the nature of absorption lines
they produce in the spectra of background QSOs that are at 
an impact parameter $\le$ 10 kpc.

In this section, we analyse the absorption lines produced in 
the QSO spectra by the emitting galaxies. For this, each quasar 
spectrum is normalised by {\sl dividing} the observed spectrum by 
the quasar continuum. In principle, the galaxy continuum emission 
should be subtracted prior to normalisation of the spectrum. However, 
continuum emission from the galaxies is expected to be very small 
compared to that of the QSO and should have a negligible effect 
on the measurement of absorption line parameters.
The equivalent widths of metal absorption lines are then obtained by 
simultaneous Gauss-profile fitting, using a single 
absorption redshift (i.e \zabs) to describe all detected absorption 
lines but allowing for it to be different from \zgal\ obtained
from emission lines. 
The normalised spectra and gaussian fits to the absorption lines are also shown 
in Fig.~\ref{sample_fig}. 
The rest equivalent width of 
\FeII\,$\lambda\lambda2586,2600$, \MgII\,$\lambda\lambda2796,2803$ 
and \MgI\,$\lambda2852$ are provided in Table~\ref{abstable}. 
This table also gives the redshift of the galaxy ($\zgal$) and the relative 
velocity shift between the centroids of the absorption and emission lines 
($\Delta {\rm v}/c=(\zabs-\zgal)/(1+\zgal)$).

\begin{table*}
\caption{Absorption line measurements \label{abstable}}
\begin{center}
\begin{tabular}{l c c c c c c c c}
\hline
\hline
\multicolumn{1}{c}{{\Large \strut} QSO}&$\zgal$&$\Delta$v&E(B-V)$^a$&\multicolumn{5}{c}{Rest equivalent widths ({\AA})} \\
               &        &(\kms)    &       &\FeII$\lambda$2586&\FeII$\lambda$2600    &\MgII$\lambda$2796    &\MgII$\lambda$2803    &\MgI$\lambda$2852      \\
\hline 
J080808+064108 &  0.433 &4(16)    &  0.01      &   \ldots   &  \ldots    &  2.5(0.5)  &  1.9(0.4)  &  0.8(0.6)   \\
J081154+202148 &  0.445 &41(15)   &  0.04      &   \ldots   & \ldots     &  1.2(0.4)  &  1.1(0.3)  &  0.3(0.3)   \\
J091417+325955$^b$&0.444& \ldots  &  0.32      &    \ldots  & \ldots     &  \ldots    &  \ldots    &  \ldots     \\
J092913+302225 &  0.439 &-96(13)  &  0.02      &   \ldots   & \ldots     &  2.3(0.8)  &  1.9(0.6)  &  1.3(0.4)   \\
J094041+341535 &  0.447 &-16(16)  & -0.02$^d$  &   \ldots   & \ldots     &  2.1(0.6)  &  2.1(0.5)  &  0.6(0.3)   \\
J095228+032616 &  0.419 &50(18)   &  0.15      &   \ldots   & \ldots     &  2.0(0.7)  &  0.6(0.2)  &  2.5(0.7)   \\
J104223+092708 &  0.592 &-105(16) &  0.02      &  $<$0.6    &  $<$0.6    &  0.6(0.2)  &  0.3(0.2)  & $<$0.25     \\
J113108+202151 &  0.563 &32(7)    & -0.01$^d$  &  2.4(0.4)  &  2.3(0.4)  &  4.1(0.4)  &  3.6(0.3)  & $<$0.3      \\
J120538+604057 &  0.434 &59(21)   &  0.06      &   \ldots   &  \ldots    &  2.5(0.8)  &  1.5(0.5)  &  0.8(0.9)   \\
J120908+022734$^c$& 0.669&11(14)  &  0.01      &  2.1(0.3)  &  2.7(0.3)  &  3.2(0.2)  &  3.2(0.3)  &  1.2(0.2)   \\
J121510+141802 &  0.421 &18(21)   &  0.09      &   \ldots   &   \ldots   &  1.6(0.8)  &  0.6(0.5)  &  0.9(0.7)   \\
J122752+165522 &  0.565 &71(12)   &  0.04      &  2.5(0.6)  &  3.9(0.6)  & $<$12.8    &  $<$5.1    &  \ldots     \\
J125339+175832 &  0.401 &-41(9)   &  0.18      &   \ldots   &  \ldots    &  1.2(0.2)  &  1.4(0.4)  &  0.2(0.3)   \\
J132542+255525 &  0.433 & \ldots  &  0.03      &   \ldots   &  \ldots    &  $<0.6$    &  $<0.6$    &  $<$0.6     \\
J142421+453523 &  0.421 &-97(22)  &  0.04      &   \ldots   &  \ldots    &  3.6(0.7)  &  3.1(0.7)  &  $<$0.7     \\
J154542+505759 &  0.525 &-58(10)  & -0.01$^d$  &  1.4(0.5)  &  1.0(0.5)  &  1.2(0.3)  &  2.0(0.4)  &  0.9(0.5)   \\
J161728+061604$^c$&0.788&104(13)  &  0.07      &  2.4(0.3)  &  2.7(0.3)  &  3.1(0.2)  &  3.2(0.3)  &  1.3(0.3)   \\
J165508+224150 &  0.453 &-57(12)  & -0.02$^d$  &    \ldots  &  \ldots    &  2.2(0.3)  &  2.2(0.3)  &  0.6(0.3)   \\
J165632+414617 &  0.662 &28(1)    &  0.03      &  1.9(0.6)  &  1.8(0.6)  &  2.6(0.7)  &  2.4(0.8)  &  1.0(0.7)   \\
\hline
\end{tabular}
\end{center}
\footnotesize
\begin{flushleft}
$a$ Negative values for E(B-V) are the consequence of intrinsic quasar shape variations. 
$b$ A Lyman-limit system is present at $\zabs\sim4.5$ towards this quasar, preventing absorption 
line measurements at $\lambda<5000$~{\AA}. 
$c$ (\MgII+[\OII])-selected galaxies.
\end{flushleft}
\normalsize
\end{table*}

\subsection{\MgII\ and \FeII\ absorption lines}

We have chosen the redshift range such that the available SDSS spectrum
of each QSO covers the expected wavelength range of \MgII\ 
absorption from the galaxy. In the case of the $\zgal=0.444$ galaxy 
towards J091417+325959 ($\zem = 4.66$), the flux at the expected position of \MgII\ absorption
is completely absorbed by a higher redshift Lyman limit system.
Similarly in the case of the $\zgal=0.565$ galaxy towards J122752+165522
the wavelength range of \MgII\ absorption is blended with the Lyman-$\alpha$
forest absorption from high redshift. In the remaining 15 galaxies 
that are selected mainly through [\OIII] emission we detect \MgII\
absorption with rest equivalent widths greater than 1 {\AA} in 13 
cases (See Fig.~\ref{sample_fig}). This implies a detection rate 
of \MgII\ absorption with $\WMgII \ge 1$~{\AA} to be $\sim$87\%.
Interestingly 9 of these systems have $\WMgII \ge 2$~{\AA} 
(see Table~\ref{abstable}). This corresponds to $\sim$ 60\% detection
rate.

The mean \MgII\ equivalent width of our [\OIII]-selected 
galaxy sample is $\avg{\WMgII}\sim2.1$~{\AA}  
(or $\avg{\WMgII}\sim2.2$~{\AA} when including the two 
[\OII]-selected galaxies).
\begin{figure}
\centering
\includegraphics[bb=70 178 520 580,clip=,width=\hsize]{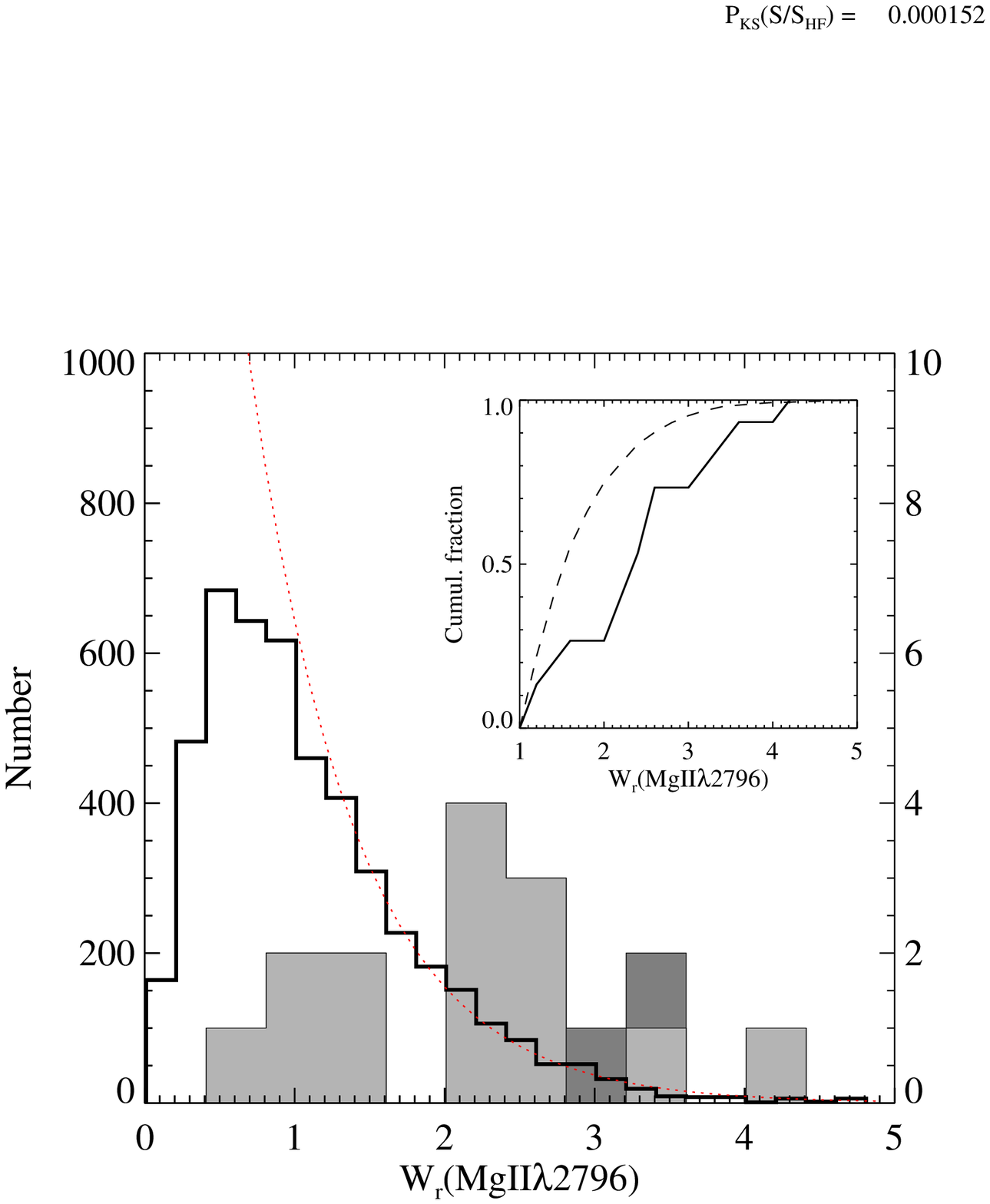}
\caption{\label{histmgii} Distribution of \MgII\,$\lambda2796$ rest 
equivalent widths for the overall SDSS \MgII\ sample (with $\zabs=0.5-0.7$, 
unfilled histogram) and that of the [\OIII]-selected galaxies, (light grey 
histogram). The values for the two (\MgII+[\OII])-selected galaxies are 
represented in dark grey. 
The two distributions are represented with different 
scales for presentation purpose only (SDSS left, galaxy sample right).
{The dotted line represents the parametrisation 
by \citet{Nestor05}, scaled to match the number of systems 
with $\WMgII \ge 1$~{\AA}.}
The cumulative distributions {(starting at \WMgII = 1~{\AA})} are 
shown in the inset figure (\MgII\ from 
SDSS: dashed; \MgII\ from our galaxy sample: solid).}
\end{figure}
In Fig.~\ref{histmgii} we present the distribution of \MgII\,$\lambda2796$ 
equivalent widths from the galaxies in our sample
and that obtained at $0.5<z<0.7$ from our automatic 
search in the SDSS-DR7. Comparing the distribution of \MgII\ 
equivalent widths with the exponential parametrization by \citet{Nestor05}, 
we find that our sample of 
\MgII-selected absorbers is probably complete 
down to $\WMgII = 1$~{\AA}. Since [\OIII]-selected \MgII\ absorbers 
are selected without any a priori information on their absorbing properties, 
we can compare the two distributions for equivalenth widths above this value.
It is clear from the figure that the \MgII\ equivalent widths in our
[\OIII]-selected sample are predominantly distributed towards higher equivalent widths. A double-side 
Kolmogorov-Smirnov test gives a probability $1.5\times10^{-4}$ that the two 
distributions (for $\WMgII \ge 1$~{\AA}) arise from the same parent population.  
Therefore, \MgII\ absorption lines associated to line-emitting galaxies 
are clearly characterised by larger equivalent widths than the \MgII-selected 
absorbers. Below, we address the question of what fraction of strong
\MgII\ absorption systems in the same redshift range are detected in 
our emission line search. 

In our [\OIII]-emission selected sample, we cover the rest wavelength
range of \FeII$\lambda\lambda$2586,2600 for 5 galaxies. We 
detect \FeII\ absorption lines in all the cases when $\WMgII\ge1$~{\AA}.
In all these cases we find $W_{\lambda2600}\ge1$~{\AA} and \MgI\ absorption
is also detected. These systems satisfy the criteria defined by \citet{Rao06} 
on the equivalent widths of \FeII, \MgII\ and \MgI\ to select Damped 
Lyman-$\alpha$ systems \citep[see also][]{Rao00}. We therefore expect 
that more than half of the 
systems in our sample are bona-fide DLAs with $\log N(\HI)\ge20.3$. 
From Fig.~3 of \citet{Steidel95} it is clear that
galaxies associated with DLAs have low impact parameters (i.e $\le$14~kpc) 
\citep[see also][]{Rao03,Chen03}.

\subsection{Dust extinction towards QSOs}

From Fig.~1 of \citet{Argence09}, we can see that the average dust optical 
depth in star forming galaxies in their SDSS sample is $\sim 1.2$.
This corresponds to an A$_V$ of 1.3 and E(B-V) of 0.42 for the 
assumed R$_V$ = 3.1 as in the Galaxy.
Therefore quasar absorbers containing 
large amounts of dust and molecules are likely to be related to star-forming 
regions in the Universe \citep[e.g.][]{Noterdaeme07,Srianand08bump}. Indeed, 
\citet{Wild07} statistically detected the nebular [\OII] emission by stacking 
the spectra of quasars with strong \CaII\ absorbers, which have been proved 
to contain on an average larger amounts of dust than \HI-selected DLAs 
\citep{Wild06, Nestor08}.  
While dusty absorbers are good candidates to search for the host galaxy emission 
lines, it is very interesting to verify whether the reciprocal is also true 
(i.e. whether the absorbers associated to star-forming galaxies within an
impact parameter of 10 kpc are also dusty).

\begin{figure}
\centering
\includegraphics[bb=70 178 500 580,clip=,width=\hsize]{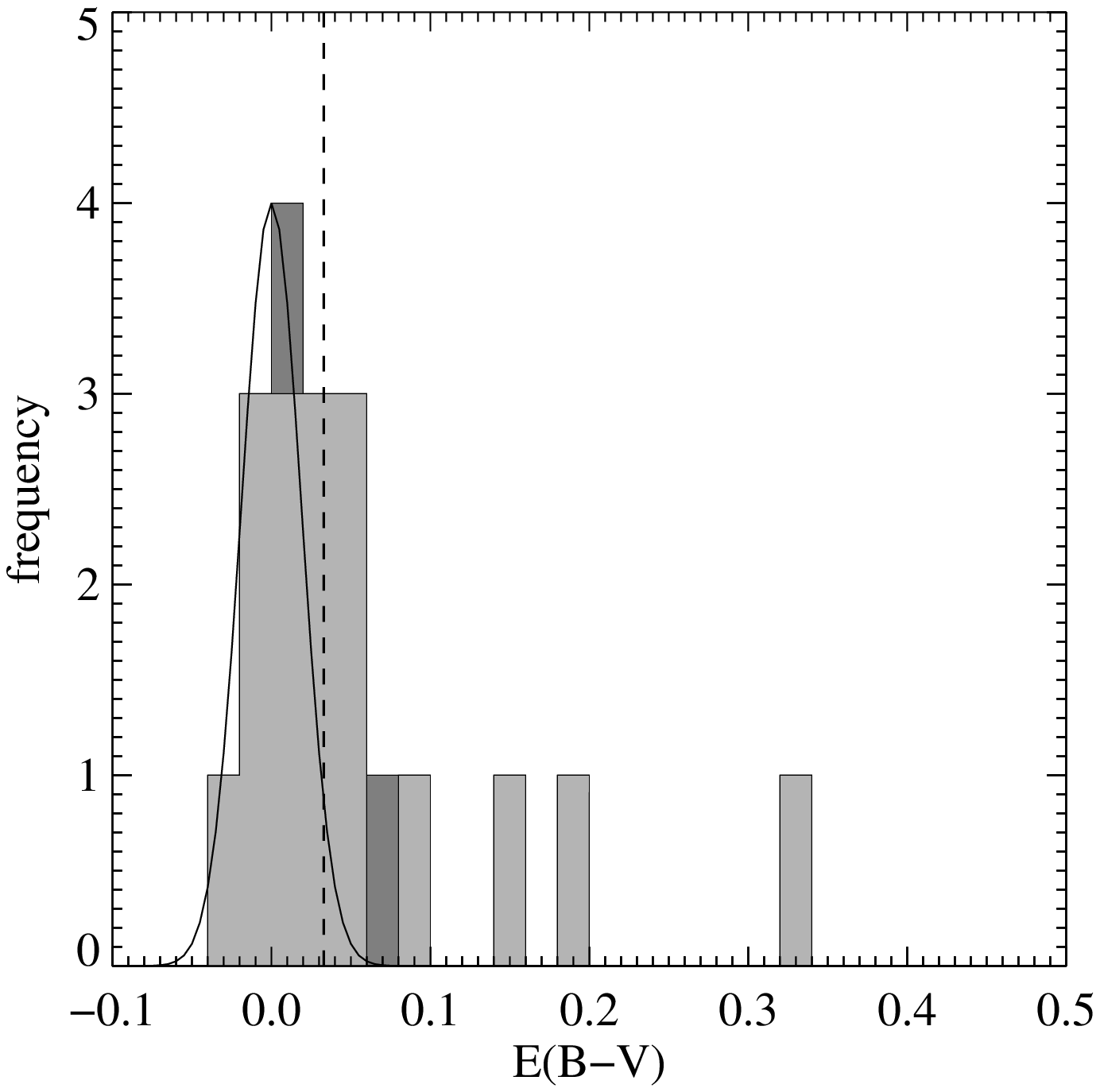}
\caption{Distribution of E(B-V) derived by fitting the SED of the background quasar. The light grey 
histogram represents [\OIII]-selected galaxies while the E(B-V) values for the two (\MgII+[\OII])-selected 
galaxies are represented in dark grey. The dashed line represents the median value for the 19 galaxies. 
{The Gaussian illustrates the $\sim$0.02~mag dispersion towards negative values, presumably due 
to intrinsic shape variations.}
\label{histebv}}
\end{figure}

We aim here at deriving the selective reddening E(B-V) of the background 
QSO produced by the absorbing galaxy. We use the same procedure as described 
in \citet{Srianand08bump} and \citet{Noterdaeme09co}. 
In short, we fit the observed spectrum with a SDSS quasar composite spectrum 
\citep{VandenBerk01}, reddened by an extinction law shifted to the redshift 
of the intervening galaxy. We use the SMC extinction curve given by 
\citet{Gordon03}, which has been shown to reproduce well the average reddening 
due to \MgII\ absorbers \citep[e.g.][see however \citealt{Srianand08bump} for 
individual cases]{Khare05, Menard05, Wild06, York06}.
Other extinction laws (LMC, MW) provide similar results as 
they are very similar in the rest wavelength range of the 
absorbers covered by the SDSS spectra. QSO-to-QSO intrinsic shape variations 
are actually the main source of uncertainties.  The distribution of E(B-V) is 
shown on Fig.~\ref{histebv}. As can be seen from this figure, 
the range in intrinsic QSO UV slopes introduces a scatter of about 0.02~mag in the 
distribution of measured E(B-V).

We found three systems (towards J091417+325955, J095228+032616 and J125339+175832) 
with E(B-V)~$\ge0.15$. For the first 
system, the reddening of the quasar is derived using a limited wavelength-range 
because of the presence of a Lyman-limit system. The quasar also has a 
high redshift ($\zqso=4.66$). Therefore, the quoted value of E(B-V) is unreliable. 
Interestingly,  in the case of J095228+032616 in addition to 
the \MgII\ absorption associated with the [\OIII] emitting 
galaxy at $\zgal = 0.419$  there is a strong \MgII\ system
at $\zabs = 0.977$ with rest equivalent width of the \MgII\
doublets 2.6 and 2.3 \AA\ respectively. This system
also has very strong \FeII\ lines. The QSO SED is reasonably well 
reproduced assuming the reddening is produced in this system. 
However, the best fitted curve under-predicts the flux in the blue
end of the spectrum.
Thus, the E(B-V) value for the galaxy at $\zgal = 0.419$ should be 
considered as an upper limit. In the case of J125339+175832 the QSO 
appears to be very red. We also notice
that high order Balmer lines as well as \CaII\ H and K lines
are seen in absorption at the redshift of the QSO. Thus the
reddening could be mainly due to the QSO host galaxy.

As can be seen  Fig.~\ref{histebv}, the distribution is concentrated around a 
median value E(B-V)~=~0.03.  Even in the top three cases with high E(B-V)
there are indications that the QSO colours are not necessarily reddened 
by to the emission line galaxy alone.
The median value is similar to or slightly higher than what is found for DLAs 
(E(B-V)~$<$~0.02, \citealt{Murphy04}; 
E(B-V)~$<$~0.04, \citealt{Ellison05}) and lower than the E(B-V) found in \CaII\ absorbers \citep{Wild06} and dusty 21-cm and CO absorbers at intermediate
redshifts \citep{Srianand08bump, Noterdaeme09co}. 
{As expected} the measured E(B-V) along the QSO
line of sight is much less than that measured for the SDSS
galaxies using emission line ratios \citep{Argence09}.

\subsection{Relationship between emission and absorption}

\citet{Ledoux06a} have established a correlation between the velocity width 
of low ionisation lines and the metallicity of high redshift DLAs. 
The slope of this relationship is shown to be consistent with the
mass-metallicity relation found in local galaxies \citep{Tremonti04}. 
In this sub-section we explore various possible correlations between star formation
indicators and metallicity indicators from the emission line fluxes and
\MgII\ equivalent width. We make the assumption that the \MgII\ equivalent
width reflects the number of components and the velocity spread
between them and is not due to line saturation. This assumption allow us
to use $\WMgII$ as an indicator of the velocity spread
along the QSO line of sight \citep[see e.g.][]{Nestor03,Ellison06}.

\begin{figure}
\centering
\begin{tabular}{c}
\includegraphics[bb=30 293 550 510,clip=,width=\hsize]{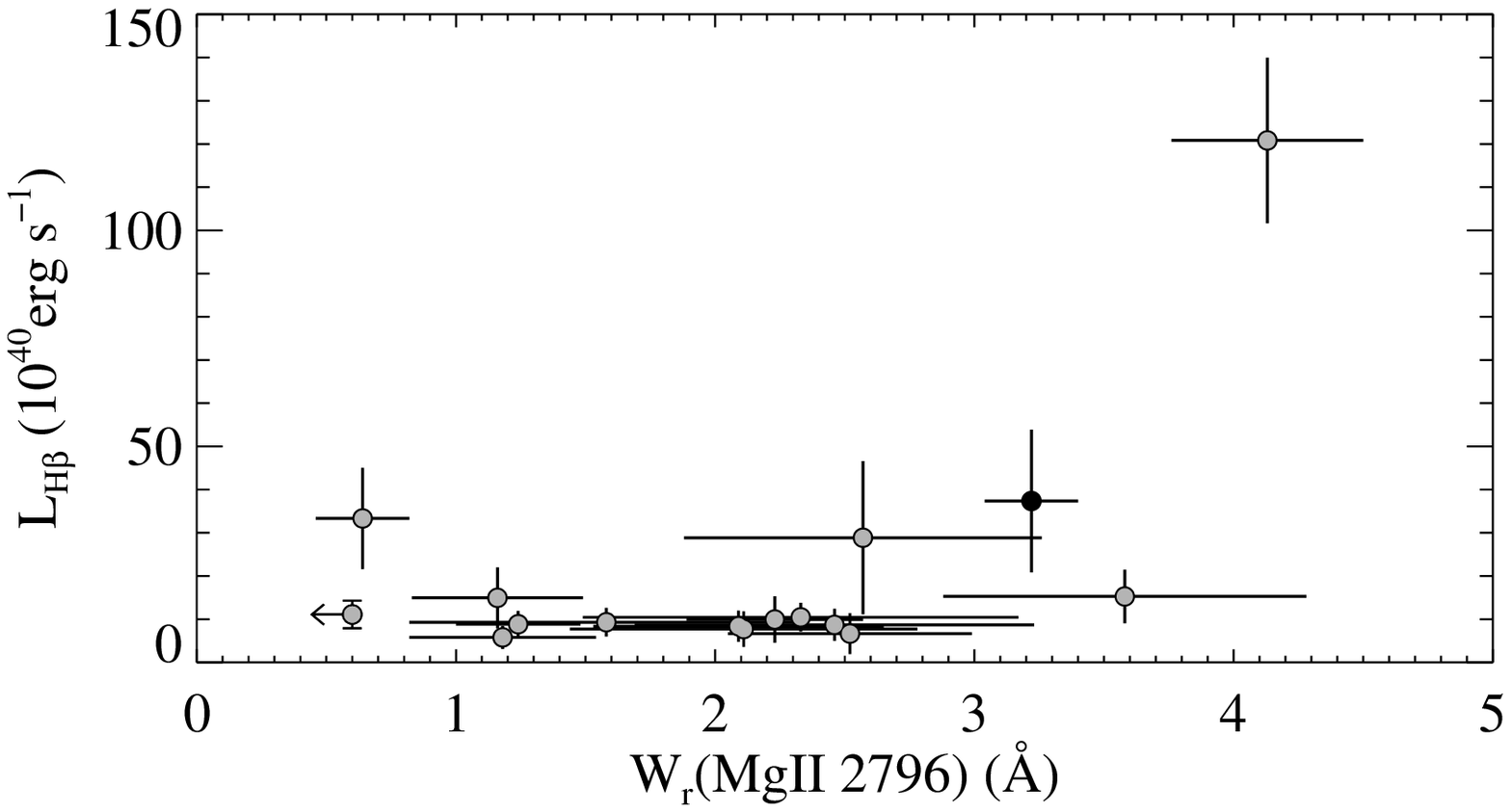}\\
\includegraphics[bb=30 293 550 510,clip=,width=\hsize]{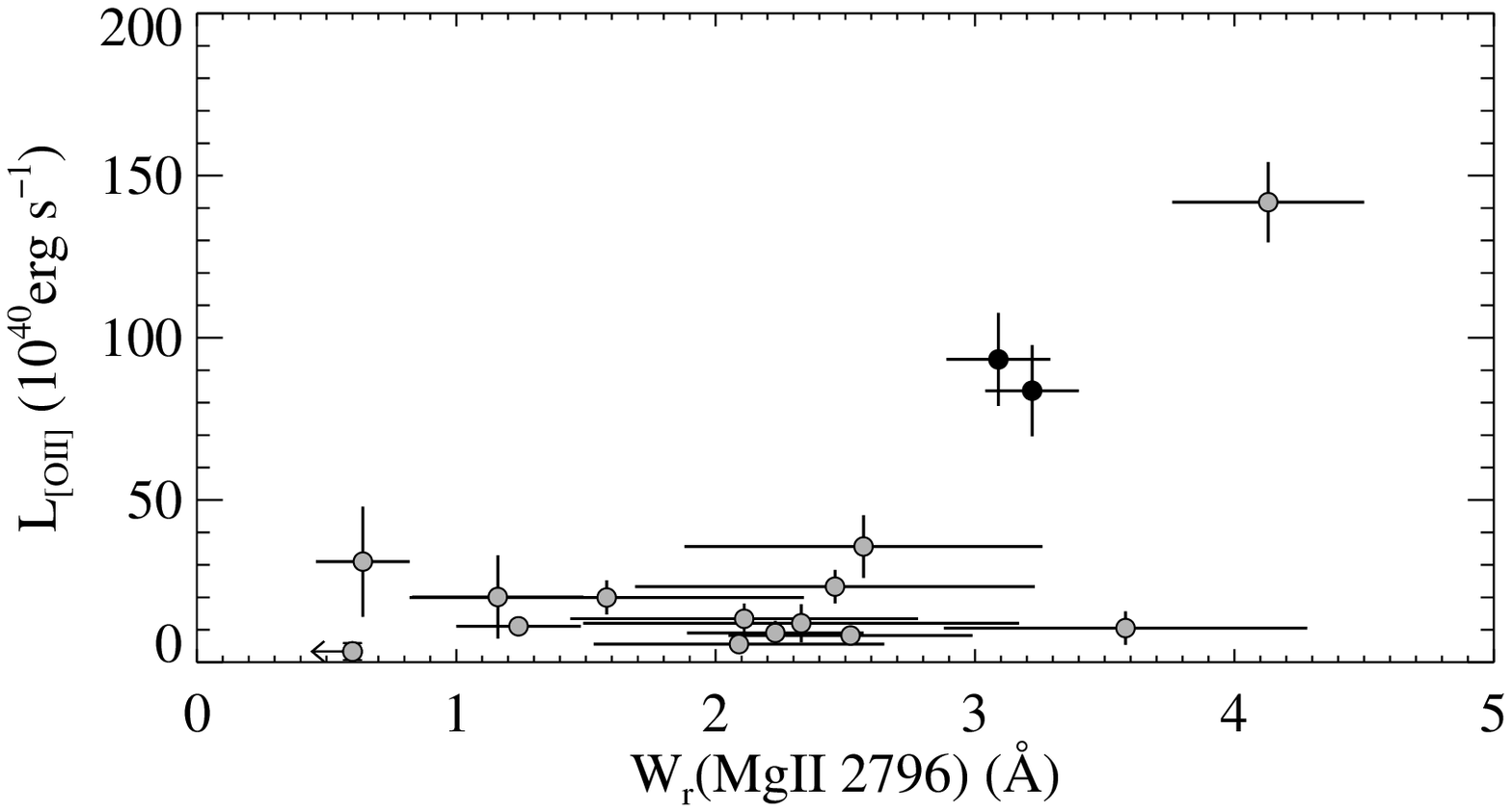}\\
\includegraphics[bb=30 293 550 510,clip=,width=\hsize]{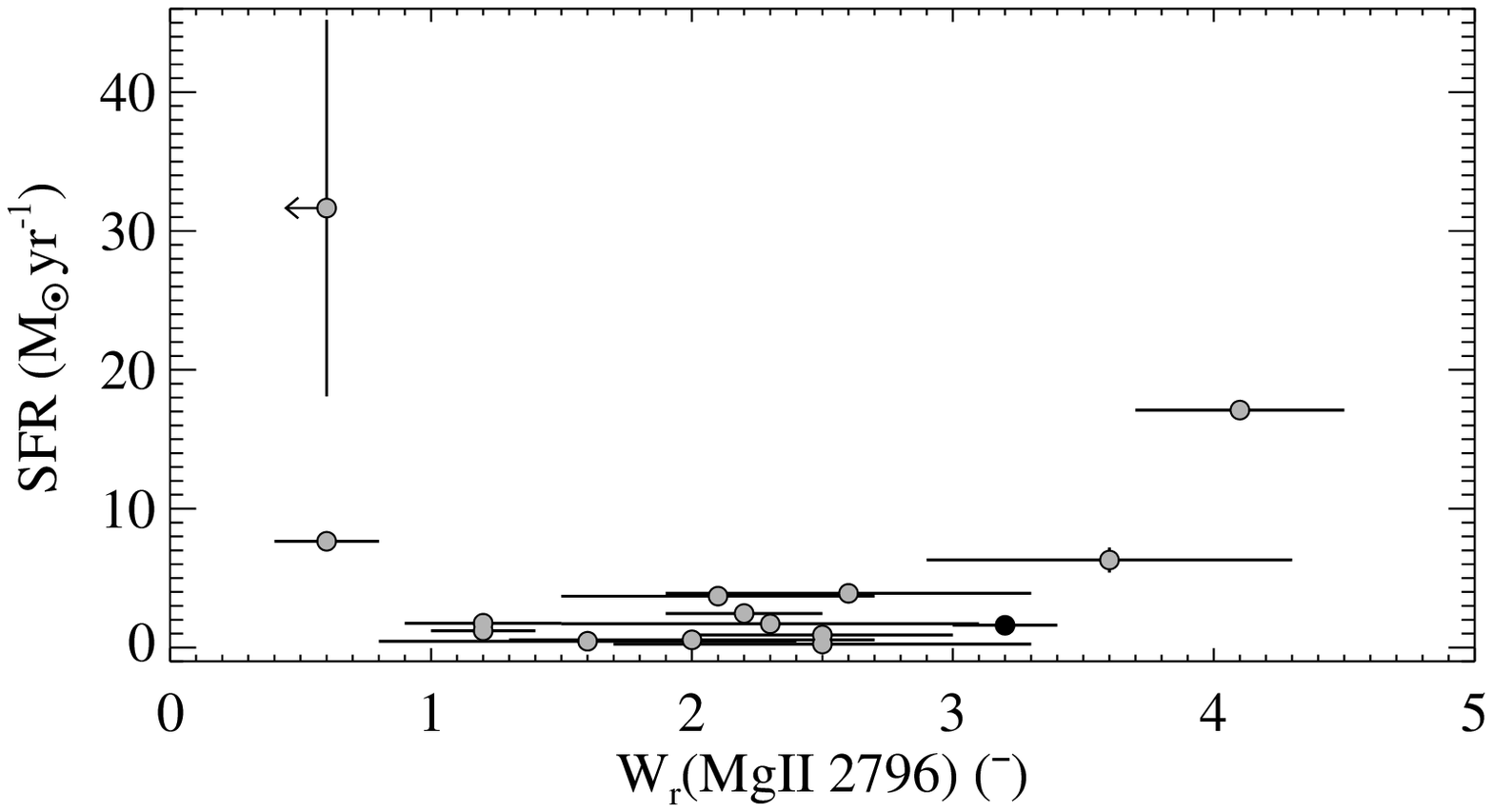}\\
\includegraphics[bb=30 250 550 510,clip=,width=\hsize]{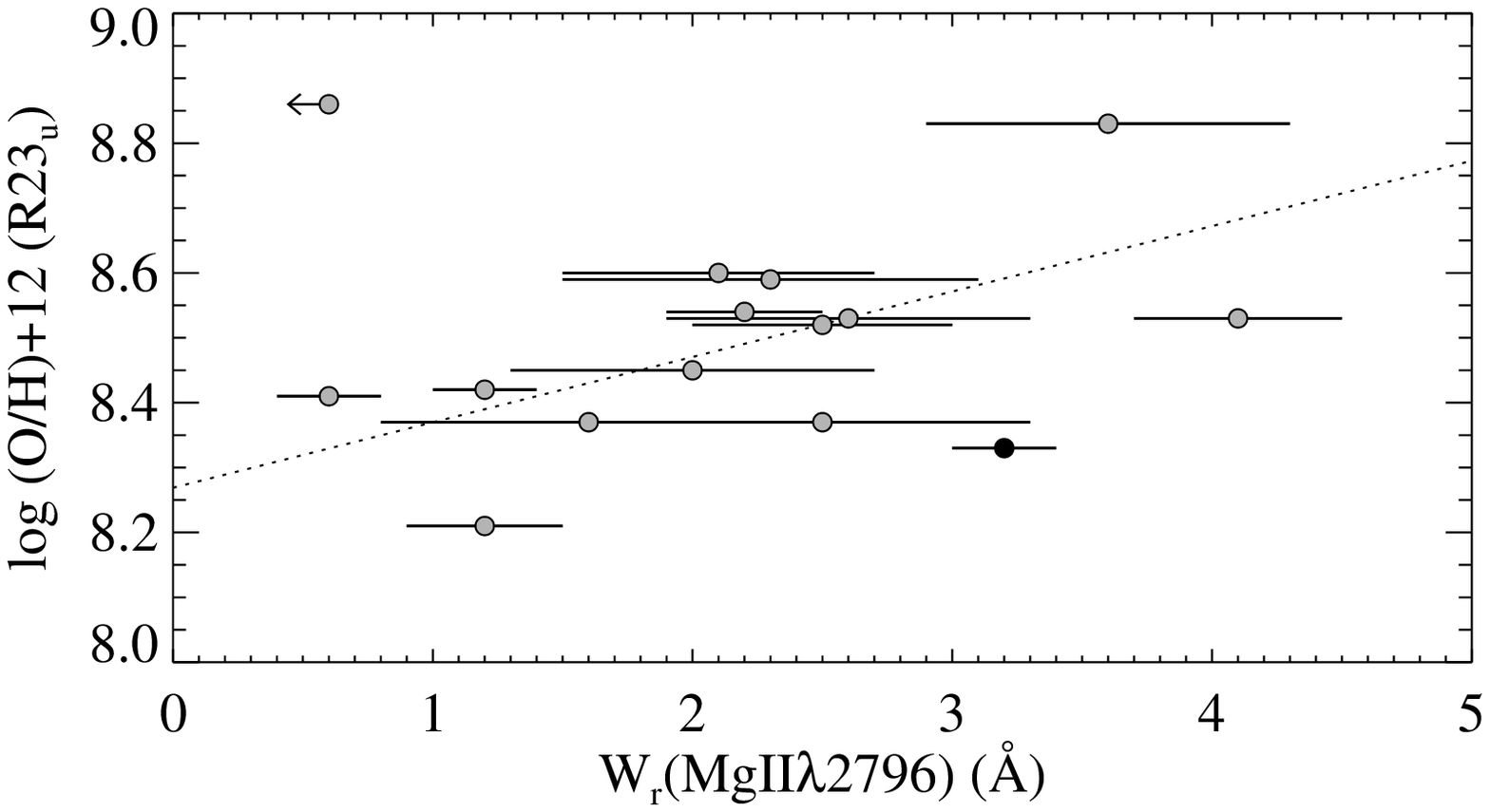}\\
\end{tabular}
\caption{\label{oii_mgii} From top to bottom: [\OII] luminosities, \Hb\ 
luminosities, SFRs and emission-line metallicities vs the \MgII\ equivalent 
width of the associated absorbing gas.
The SFRs are that from the two lines calibration (last column of Table~\ref{sfr}). 
{The dotted line on the bottom panel represents the linear regression 
fit to the metallicity - equivalent width correlation.}
Grey circles represent 
[\OIII]-selected galaxies with black circles the (\MgII+[\OII])-selected 
galaxies.
}
\end{figure}

In the three upper panels of Fig~\ref{oii_mgii} we plot the luminosities of 
[\OII], \Hb\, and the derived star formation rates as a function 
of the rest equivalent width of the \MgII$\lambda$2796 line. No statistically 
significant correlation is seen between [\OII], \Hb\ luminosity and $\WMgII$. 
We also do not find any strong correlation between the SFR 
given in the last column of Table~\ref{sfr} and $\WMgII$. 
However, the largest equivalent width systems are also associated to the largest 
luminosities. Except from the system towards J132542+255525, this is also 
true for the SFR. We remind that since the measured [\OII] and \Hb\ emission line fluxes 
are very low in this galaxy, it is well possible that the SFR and metallicities, 
which depend on the line ratios, are significantly over-estimated (see Sect.~\ref{s_sfr}).

In the bottom panel of Fig.~\ref{oii_mgii}, we plot the upper-branch 
(R23$_u$) estimate of (O/H) as a function of $\WMgII$. 
As pointed-out before, in the absence of additional constraints 
we end up with two degenerate metallicity measurements using R23. 
However, in the five line-emitting galaxies studied by \citet{Zych07} -- with similar 
properties as those presented here -- the upper-branch of R23 is preferred.
A trend for increasing emission-line metallicities with increasing equivalent width 
can be seen. 
The correlation {($\log ({\rm O/H})+12=0.1\WMgII+8.27$)} is significant at the 2\,$\sigma$ level. 
However, to draw a firm conclusion on the velocity metallicity
correlation we need to remove the degeneracy in the (O/H)
estimation and get the velocity spread in the absorbing gas using
high resolution spectroscopy. 
Interestingly, a similar trend is also observed between the 
\MgII\ equivalent width and the gas-phase metallicity measured 
along the quasar line of sight \citep[e.g.][]{Nestor03,Murphy07}.

\begin{figure}
\centering
\includegraphics[bb=60 178 520 580,clip=,width=\hsize]{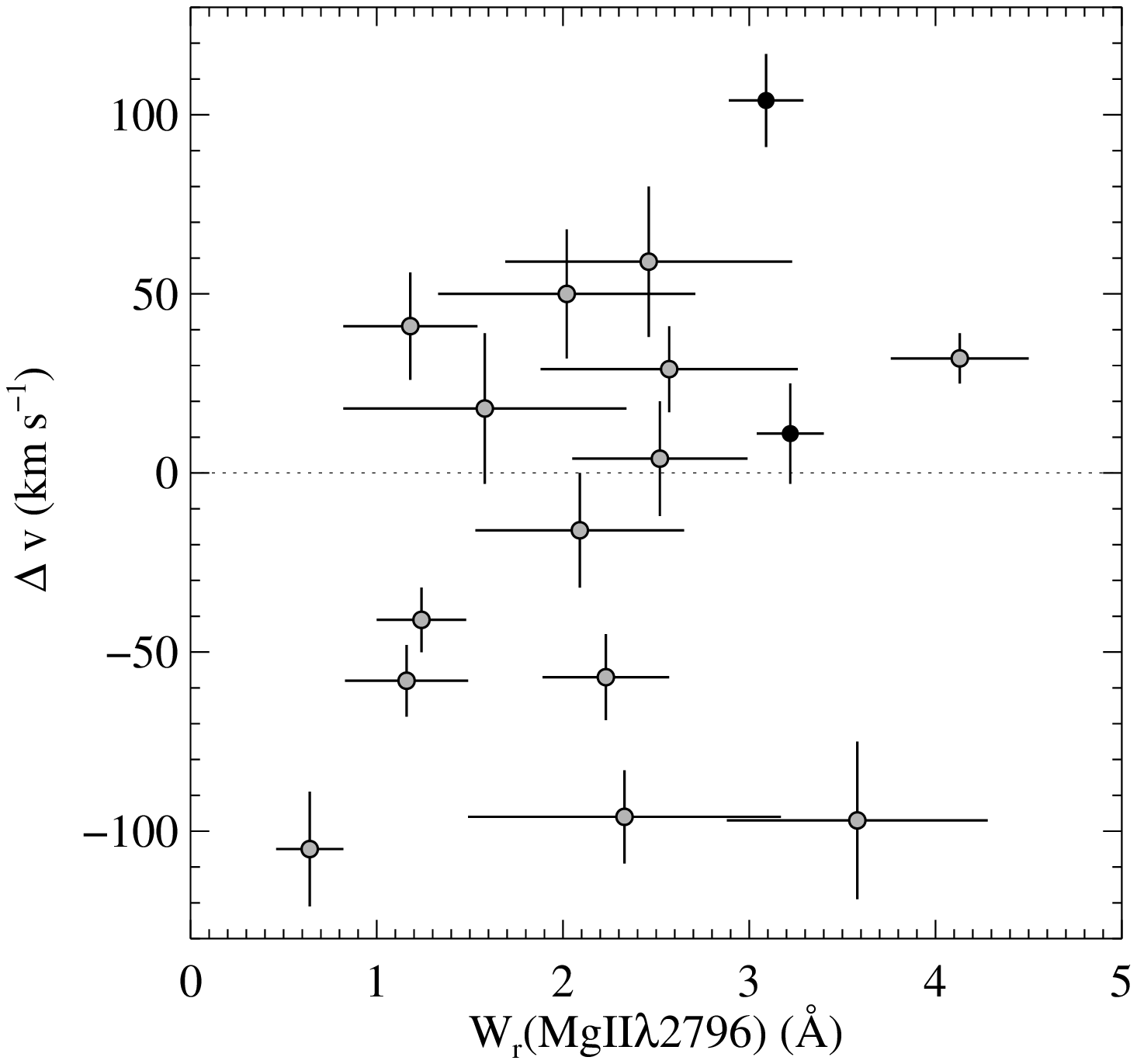}
\caption{ The velocity offset between the absorption and emission line 
redshifts is plotted against the \MgII\ rest equivalent width. Symbols 
are as per Fig.~\ref{oii_mgii}}
\label{kinematics} 
\end{figure}

The velocity shifts ($\Delta v$) between the emission 
and the absorption lines are quite 
small (see Table~\ref{abstable}), at most about 100~\kms. These are consistent
with the expected circular velocities of typical galaxies suggesting the
absorbing gas is bounded to the emission line galaxy.   
We note that the $\Delta v$ we measure here are lower than that measured 
with respect to luminous galaxies at larger impact parameters \citep[100-200~\kms\ 
for impact parameters $\sim$14-75~h$^{-1}$~kpc][]{Steidel02}.
In Fig.~\ref{kinematics} we plot $\Delta v$ against $\WMgII$. We
do not find any trend between the two quantities.
Although the sample is too 
small to conclude, this may be explained by the variety of galaxy morphologies 
found to be associated with low-redshift \MgII\ systems \citep{LeBrun97}.

\section{Average line emission from \MgII\ absorbers}

Based on our automatic search for \MgII\ absorption in SDSS-DR7 we find
2319, 1807, 494 and 118 systems respectively with $\WMgII \le 1$~{\AA}, 
1-2~{\AA}, 2-3~{\AA} and $\ge$ 3~{\AA} in the redshift range
$0.5\le z\le0.7$. It is clear from the previous discussion that while 
most of the emission line galaxies
produce strong \MgII\ absorption not all the strong \MgII\ 
absorption systems are detected in our [\OIII] emission searches. 
Indeed, the number of strong \MgII\ systems (with $\WMgII\ge2$~{\AA}) 
associated with [\OIII]-emitting galaxies detected within the SDSS fibre is 
about a hundred times less than the total number of strong \MgII-absorbers.
This could be due to (i) poor S/N of the spectra {and/or the relative 
brightness between the QSO and the galaxy}, (ii) low star formation
rate in the underlying galaxies or (iii) the impact parameter of the 
emission line regions being larger than $\sim$10~kpc (or angular separations
more than 1.5'').  

\begin{figure}
\centering
\includegraphics[bb=20 180 550 690,width=\hsize]{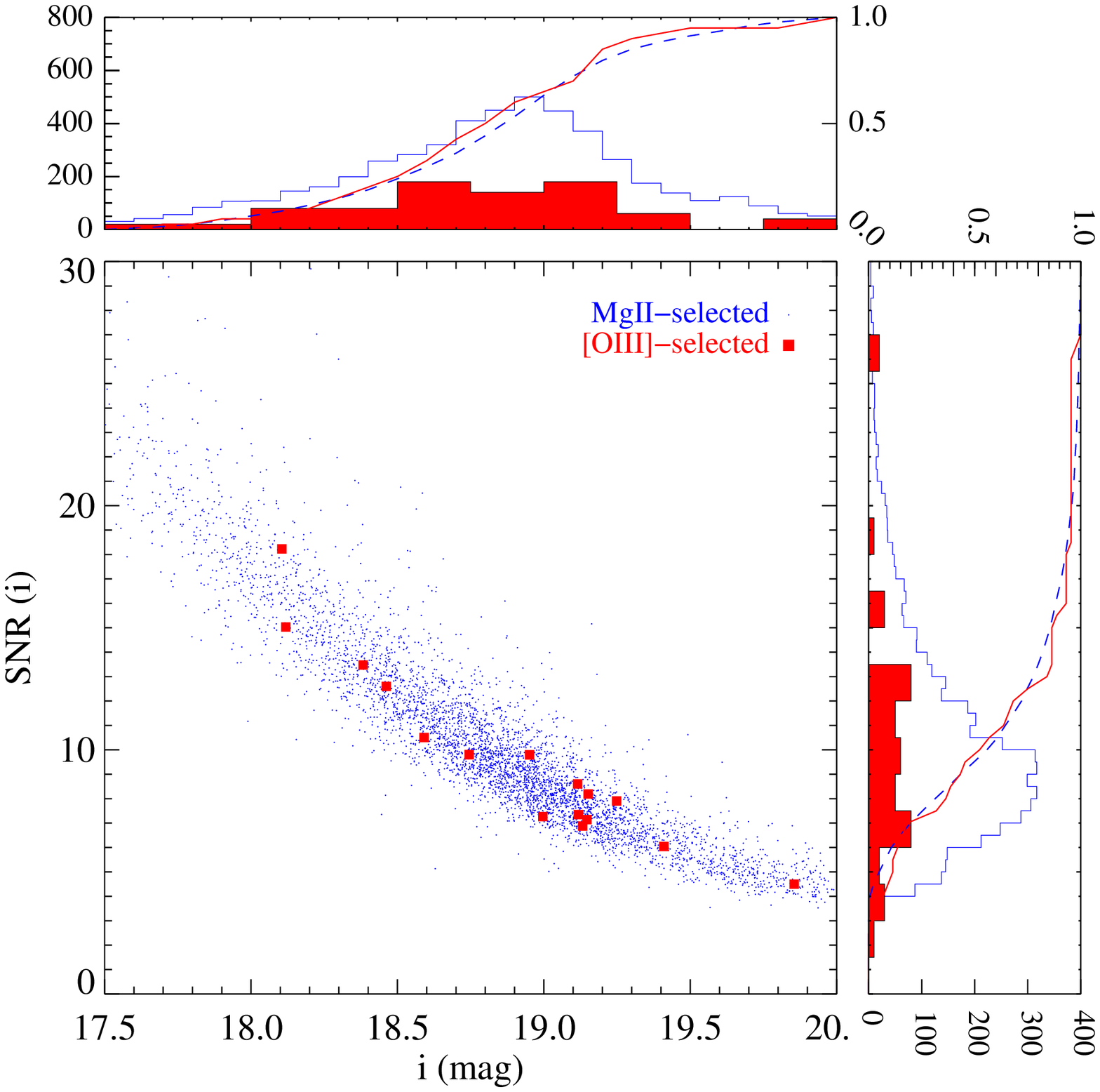}
\caption{$i$-band signal-to-noise ratios and magnitudes for quasars with 
strong ($\WMgII>2$~{\AA}) intervening \MgII-systems (black dots) and 
intervening [\OIII] emission lines (red squares). The top 
and right panels show the distributions and cumulative distributions of 
magnitudes and signal-to-noise, respectively. 
Red filled histograms (solid lines) stand for quasars with intervening 
[\OIII]-emission lines while blue unfilled histograms (dashed line) stand 
for quasars selected for strong intervening \MgII\ absorption lines.
\label{snrmg2}}
\end{figure}

To explore the effect of spectral signal-to-noise ratio and 
galaxy-QSO contrast further, we plot the $i$-band 
magnitude vs the S/N for all QSOs with intervening \MgII\ systems with 
$\WMgII \ge 2$~{\AA} in the redshift range $0.5 \le z \le 0.7$ in 
Fig~\ref{snrmg2}. It is clear from the figure that most of
these QSOs have $i$-band magnitudes in a narrow range 18 to 19.5 mag. The 
spectral signal to noise ratio is constant within a factor 2
in this magnitude range.
To the eye there seems to be a tendency for the [\OIII]-selected galaxies
to prefer slightly fainter $i$-band magnitude (see the upper histogram). 
However, KS tests do not indicate the differences between two populations to be
statistically significant. Thus it seems that there is no clear indication
that we could have missed strong emission from
most of the \MgII\ systems mainly because of the poor signal to noise.
It is also clear from Fig.~\ref{snrmag} that, in the $i$-magnitude
range 18 to 20 mag, the spectral S/N achieved in all the QSO spectra are
good enough to detect emission lines with peak flux in excess of 
10$^{-17}$~erg\,s$^{-1}$\,cm$^{-2}$\,{\AA}$^{-1}$. Thus the lack of direct detection of 
emission lines from the strong \MgII\ systems is consistent with either 
their fluxes being small or the impact parameters being large enough so that 
the emitting regions are not falling inside the fibre.

\begin{figure}
\centering
\includegraphics[bb=85 182 495 400,clip=,width=\hsize]{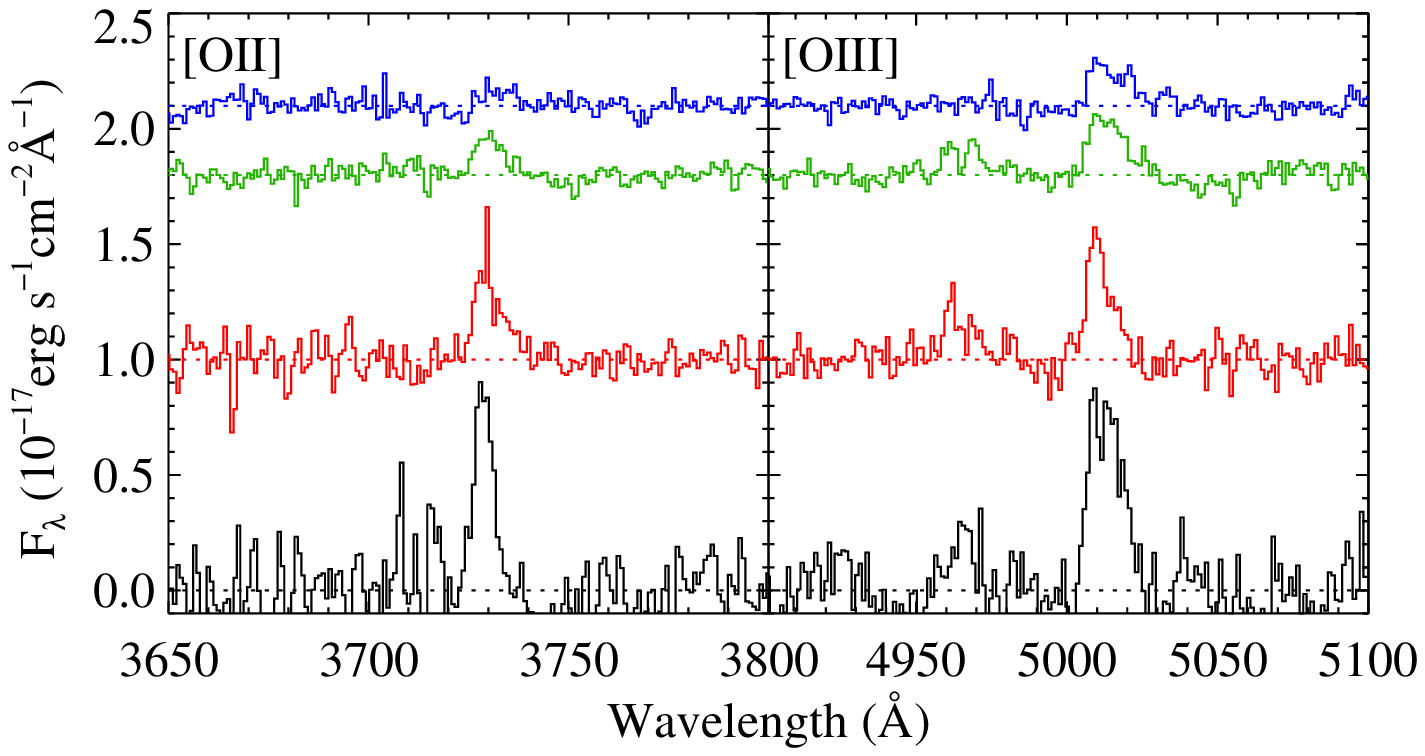}
\caption{\label{stackmgii} [\OII] (left) and [\OIII] (right) emission lines from stacking 
QSO spectra with intervening \MgII\ absorbers at $0.5<\zabs<0.7$ with different $\MgII\,\lambda2796$ 
rest equivalent widths. From top to bottom: $W_{\lambda2796}< 1$; $1\le W_{\lambda2796}< 2$; $2\le W_{\lambda2796}< 3$; 
 $W_{\lambda2796}\ge 3$~{\AA}. The continuum substracted spectra are shifted by a constant for display purpose.}
\end{figure}

In order to investigate the {\sl average} [\OII] and [\OIII] emission 
within 1.5'' of associated intervening \MgII\ absorption systems, 
we build several composite spectra corresponding to different ranges of 
\MgII\ equivalent widths. The quasar spectra featuring 
\MgII\ absorption lines in the range $\zabs=0.5-0.7$ where shifted to the 
absorbers rest frame and combined together using an arithmetic mean. {Note that our 
direct detections are not included in the stacking.}
The continuum flux in the vicinity of the emission lines was removed using a 2nd-order polynomial 
fit. Fig.~\ref{stackmgii} present the resulting composite [\OII] and [\OIII] emission lines from 
intervening \MgII\ absorption systems with rest equivalent width $\WMgII$ in the 
ranges $\le 1$, 1-2, 2-3 and $>3$~{\AA}. 
From this figure, it is clear that the strength of the emission lines increases with the \MgII\ 
equivalent widths. This can also be seen from Table~\ref{tabstack} where we give the average 
luminosity of [\OII] and [\OIII] emission lines for different sub-samples defined using $\WMgII$.
We note that the average [\OII] flux  for $W_{\lambda2796}\sim1$-2~{\AA} 
is similar to that obtained by \citet{Wild07} for \MgII-selected DLAs 
\citep[$1<W_{\lambda2796}/W_{\lambda2600}<2$ and $W_{\lambda2796}>0.6$; see][]{Rao06}.

The average [\OIII] luminosity found by stacking \MgII\ systems
with $\WMgII>2$~{\AA} is roughly $\lstaro$/10 and close to the lowest luminosity 
we directly measure in the [\OIII]-selected galaxies (in $\zgal = 0.444$ 
towards J091417+325955). Note that in this case the detection is 
enabled by the low QSO flux ($i$ = 19.5 mag). As discussed in section~2.1 
this system would not have been detected had the QSO being brighter.
\citet{Zych07} have detected emission lines at the redshift of the
\CaII\ absorber towards J224630.62+131048.5.  The rest equivalent
width of \MgII$\lambda$2796 is 2.2~{\AA}. No clear emission lines
are detected in the SDSS spectrum. However a galaxy is clearly seen
within the area covered by the fibre \citep[See Fig.~5 of][]{Zych07}
and emission lines (with integrated [\OIII] flux of 8$\times10^{-17}$~erg\,$s^{-1}$\,cm$^{-2}$ 
i.e. lower than our detection limit with SDSS) are detected in the VLT/FORS spectra. 
We also note that the [\OIII] luminosity of this galaxy is similar to the 
stacked luminosity obtained using systems with $2<\WMgII<3$~{\AA}.

Direct observations of field galaxies at $z\sim0.6$ show that the sizes of star-forming 
galaxies are typically less than 10~kpc \citep{Dahlen07, DeMello06,Lilly03}. Integral field 
spectroscopic observations of \MgII\ absorbers at $0.8<z<1.2$ by \citet{Bouche07} also 
confirms that \Ha\ emitting regions of \MgII-selected galaxies have similar sizes. This 
means that the galaxies contributing to the stacked emission lines should have impact 
parameters less than about 20~kpc. 
Therefore, it is most unlikely that a bright galaxy at high impact parameter (several tens of kpc) 
will contribute to the average detection of emission lines in the stacked spectrum. 
The stacking method alone does not provide the higher moments 
of the $L_{\rm [OIII]}$-distribution. However, with additional constraints on the galaxy 
sizes and the small number of direct detections, our results 
are consistent with at least part of the strong \MgII\ absorbers arising from 
low $L_{\rm [OIII]}$ luminosity galaxies at low-impact parameters, as seen in the 
case of J224630+131048 \citep{Zych07}.

\begin{center}
\begin{table}
\caption{Luminosities of the emission lines in the stacked spectrum \label{tabstack}}
\begin{center}
\begin{tabular}{lcccc}
\hline
line    & \multicolumn{4}{c}{Average luminosity$^a$ for systems with}\\
        & W$<$1\AA & 1$<$W$<$2 \AA& 2$<$W$<$3\AA & W$>$3 \AA\\
\hline
$[$O\,{\sc ii}$]$  & 0.5(0.2) & 1.4(0.1) & 3.2(0.2) & 5.1(0.6)\\
$[$O\,{\sc iii}$]$ & 2.5(0.3) & 3.5(0.3) & 5.1(0.4) & 10.0(0.9)\\

\hline
\end{tabular}
\end{center}
\begin{flushleft}
$^a$ in units of 10$^{40}$~erg\,s$^{-1}$.\\
\end{flushleft}
\end{table}
\end{center}

It is also interesting to see that the [\OIII] and [\OII] emission lines are
detected in the stacked spectrum even when we consider only 
low equivalent widths. In particular there are roughly 4 times more
systems with $1 \le \WMgII \le 2$~{\AA} compared to that with 
$2 \le \WMgII \le 3$~{\AA}.
However the average [\OIII] and [\OII] luminosities are less only by a factor 2.
Even though our \MgII-selected systems with low rest equivalent widths ($\WMgII<1$~{\AA}) are likely to 
be biased towards QSOs with high S/N spectra, it is interesting to see
that we detect [\OII] emission at 3$\sigma$ level and that [\OIII] is 
detected at $>5 \sigma$. This means that galaxies with low 
impact parameters can also produce low equivalent width absorption lines. This is consistent
with the fact that we do not detect \MgII\ absorption with $\WMgII \ge 1$~{\AA}
in two of our [\OIII]-selected galaxies.

In summary, star-forming galaxies with low [\OIII] luminosities seem to provide 
an important contribution to the population of \MgII\ absorption selected galaxies.

\section{The case of SDSS\,J113108+202151 \label{IGO}}

\begin{figure}
\includegraphics[bb=80 182 487 569,clip=,width=\hsize]{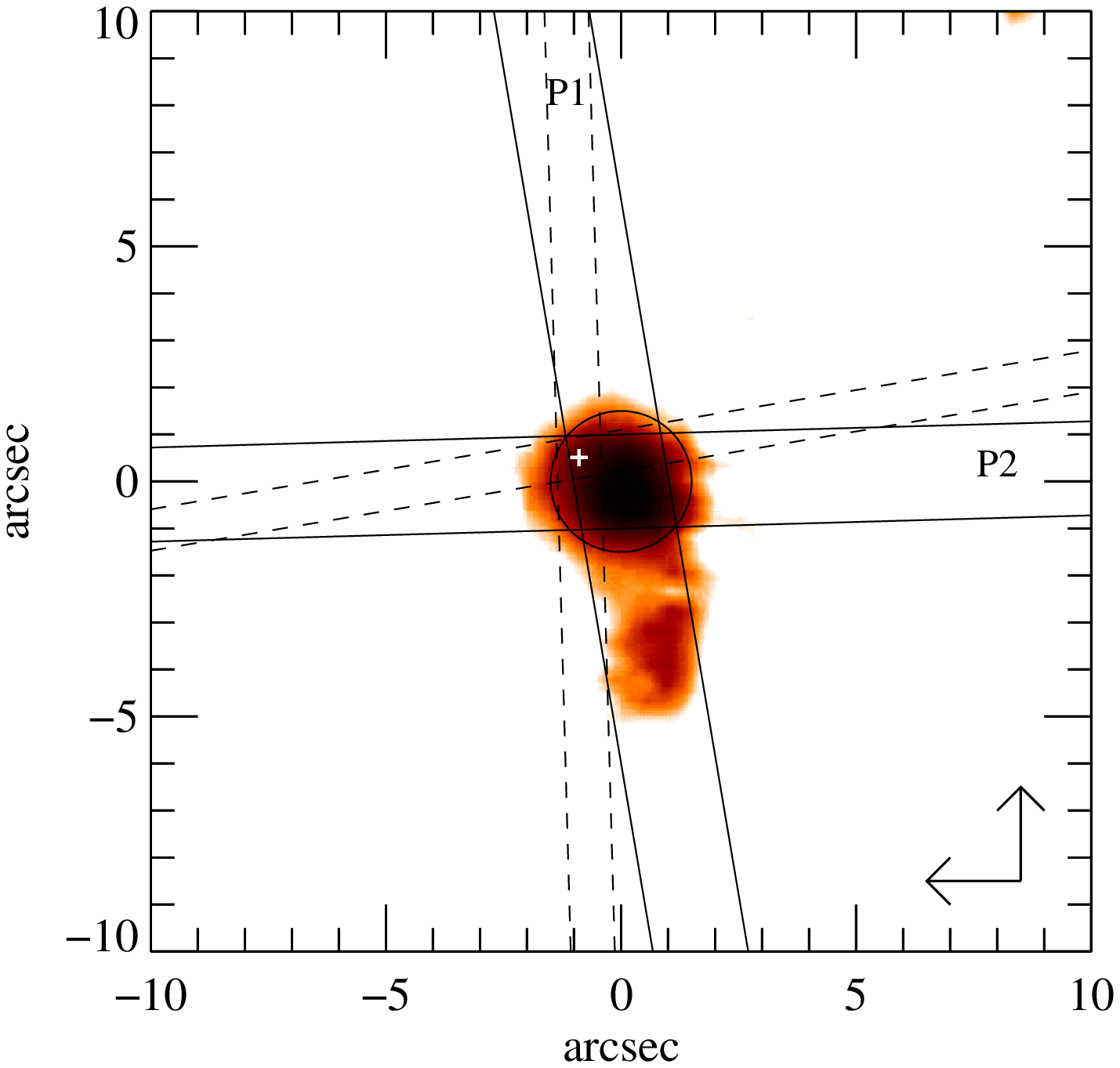}
\caption{Layout of the slits (solid lines) and SDSS fibre (circle) on the quasar SDSS\,J113108+202151. 
The positions (P1 and P2) are indicated on top (as represented in Fig.~\ref{2dspec}) of 
the slits. The dashed lines represent the 1\,$\sigma$ constraints on the centroid of the intervening emission 
line galaxy, as obtained from each slit orientation. The cross marks the centroid of the galaxy. \label{im1131}}
\end{figure}

In this section, we study the galaxy at $\zgal=0.563$ towards J113108+202151 which 
presents the strongest emission lines in our $z>0.4$ sample and also the strongest 
associated \MgII\ absorption lines. An extended galaxy (SDSS\,J113108.31+202147.3)
is clearly visible close to the quasar line of sight in the SDSS image, 
with its centroid located at about 3.5$\arcsec$ from the quasar image 
and extended towards the quasar image (see Fig.~\ref{im1131}). Moreover, the photometric redshift 
provided by SDSS, $z=0.65\pm0.07$, is close to that of the detected emission 
and absorption lines. This is therefore a very good case to study the 
connection between the galaxy and the absorber. 

In order to obtain an accurate measurement of the location of the emission 
region, we performed long-slit spectroscopy with the 2\,m telescope of the 
IUCAA Girawali Observatory. 
Observations were carried out on March 20, 2009 using IUCAA Faint Object 
Spectrograph and Camera (IFOSC). A 2$\arcsec$ slit and GR7 grism covering 
the wavelength range between 3900~{\AA} to 6800~{\AA} were used. The detector 
used is a LN2 cooled thinned 2k$\times$2k CCD camera. Each pixel on the 
CCD covers 0.34$\arcsec$ of sky which corresponds to 1.4~{\AA} using the 
above grism. Three exposures of 2700~s each were taken at slit position 10 
deg from North (P1) and two exposures of 2700\,s each were taken at slit 
position 92 deg from North (P2). The slit positions have been indicated 
in Fig.~\ref{im1131}. For flat fielding Halogen lamp flats were used. 
Helium and Neon lamps were used simultaneously for getting comparison 
spectrum. The IRAF routine {\sc response} was used for 
making a normalised flat. The {\sc doslit} package has been used to extract 
and calibrate the 1D spectrum. Two dimensional analysis was performed
using the method described in \citet{Vivek09}.

\begin{figure*}
\begin{tabular}{ccc}
\multirow{2}{*}{\rotatebox{90}{arcsec}} & \includegraphics[bb=190 90 370 728,clip=,angle=90,width=0.46\hsize]{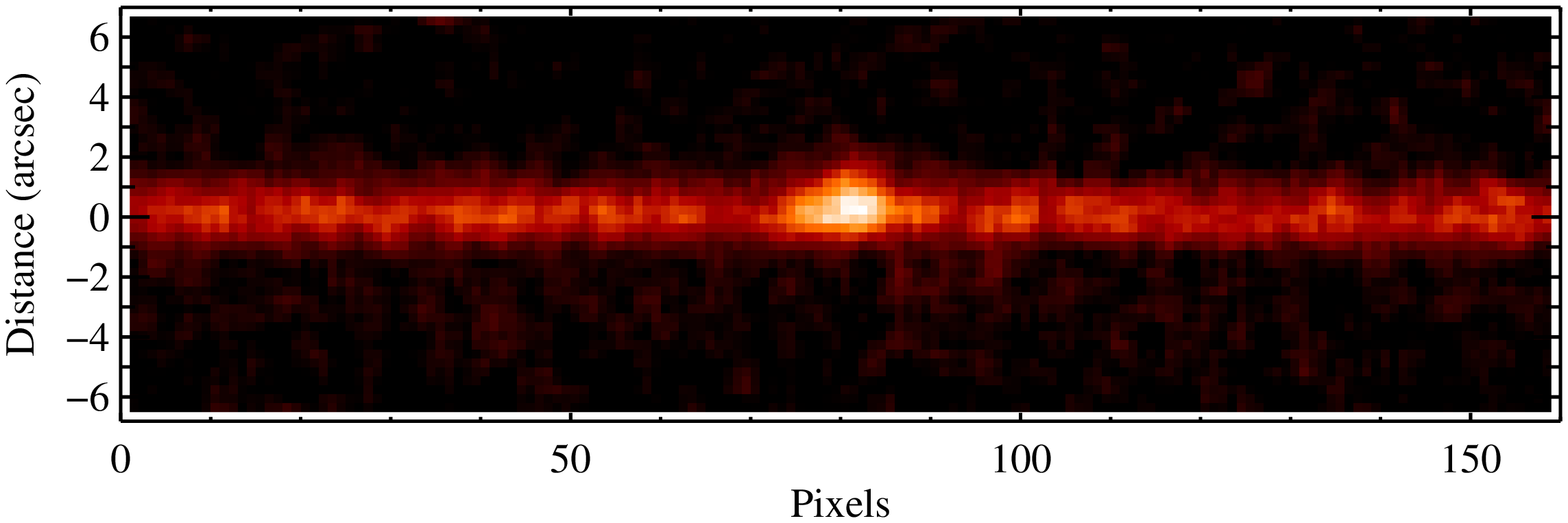}  &
\includegraphics[bb=190 90 370 728,clip=,angle=90,width=0.46\hsize]{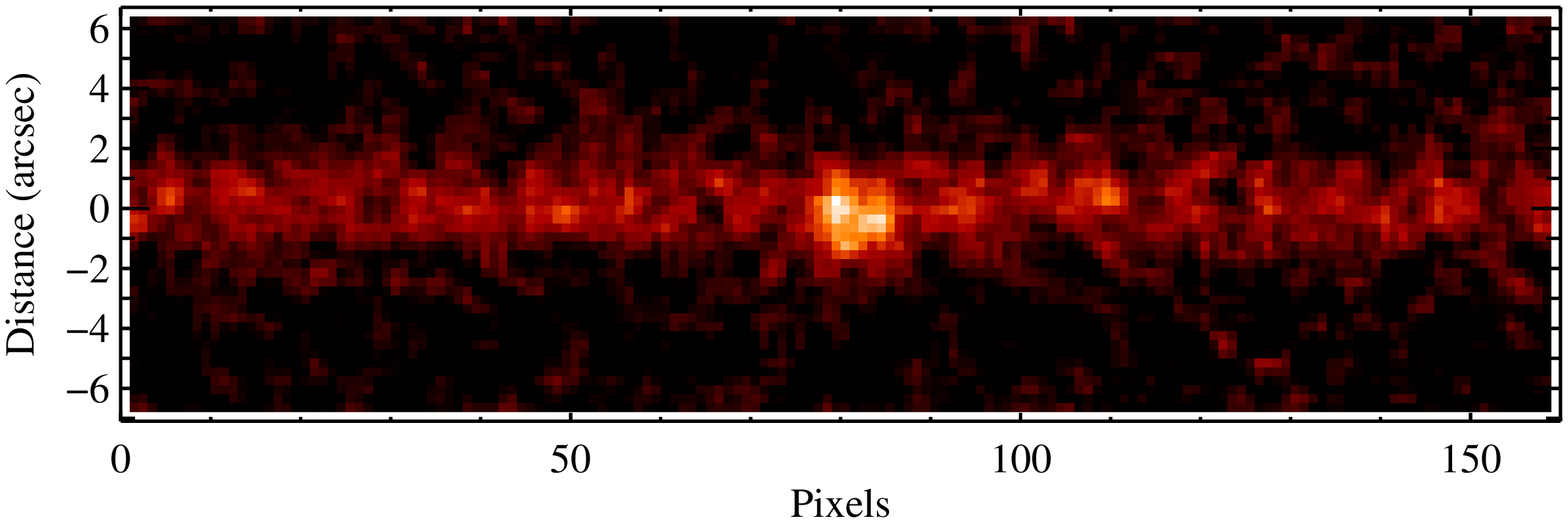}  \\
 & \includegraphics[bb=168 90 370 728,clip=,angle=90,width=0.46\hsize]{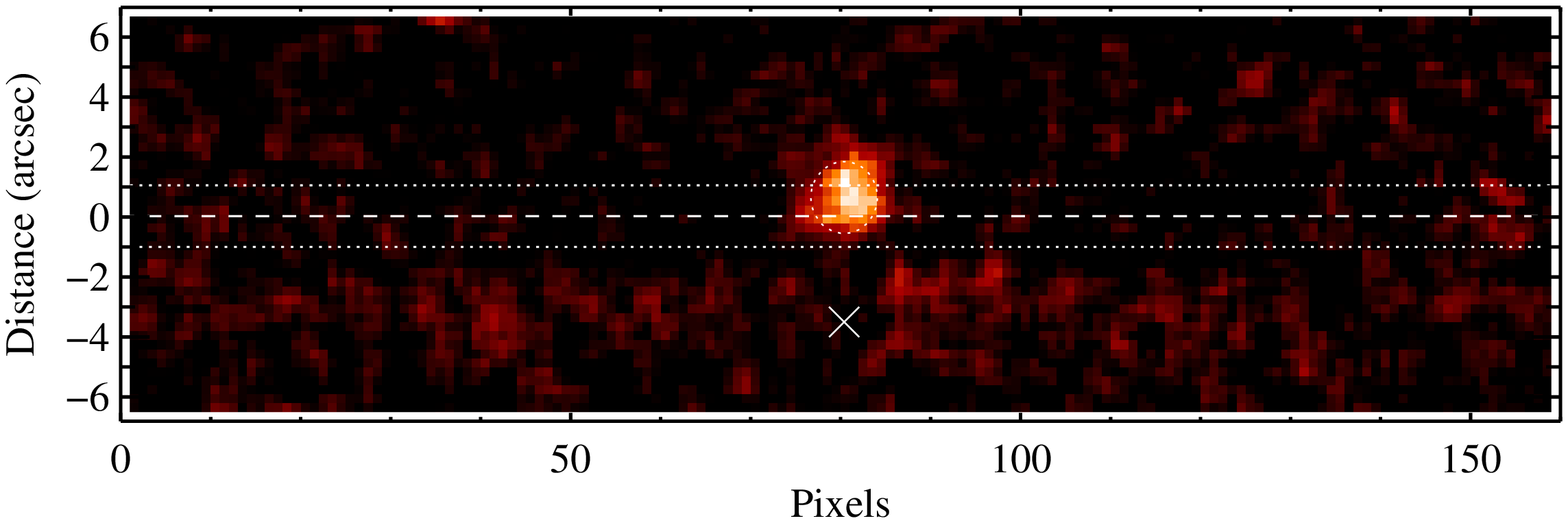} & 
\includegraphics[bb=168 90 370 728,clip=,angle=90,width=0.46\hsize]{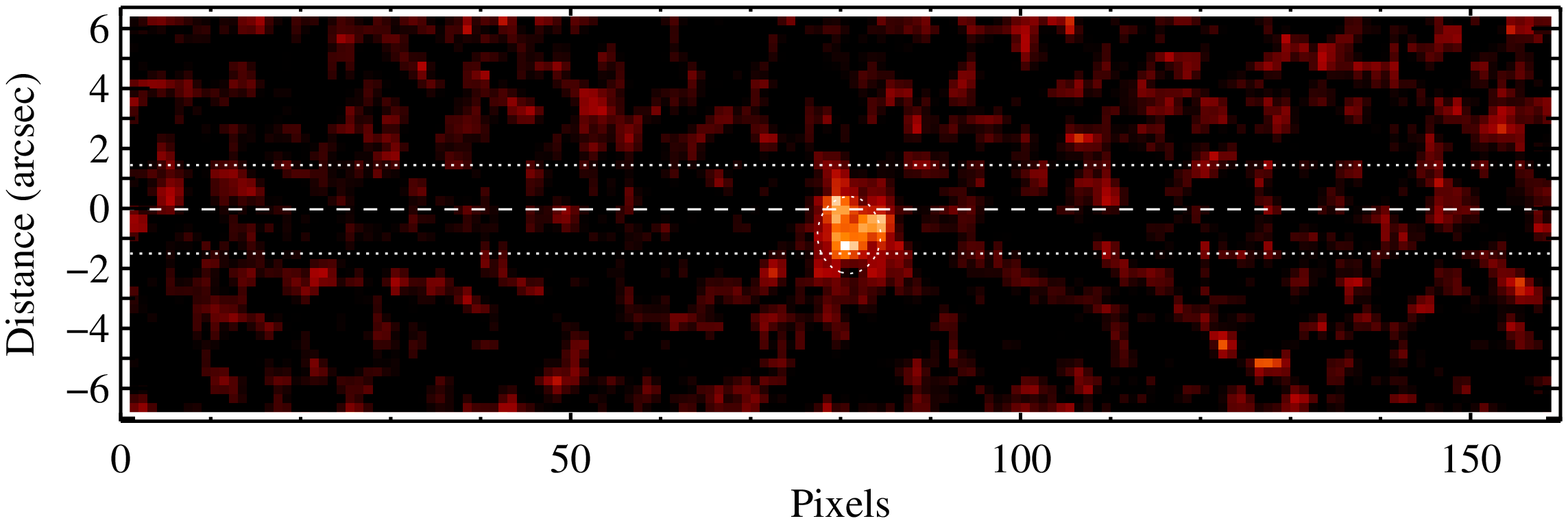}  \\
 & \multicolumn{2}{c}{Dispersion axis (pixels)} \\
\end{tabular}
\caption{The background-subtracted 2D spectra of the quasar SDSS\,J113108+202151 and the galaxy at $\zgal=0.56$. 
Top: Total (quasar+galaxy) spectra obtained with two slit orientations 
(left: P1, 10 degrees from North; right: P2, 92 degrees from North, see Fig.~\ref{im1131}). 
Bottom: Same spectra after removing the quasar trace. The centre of the quasar trace is represented 
by the horizontal dashed lines, and its FWHM by the horizontal dotted lines. The ellipses represent 
the FWHM of a 2D-Gaussian fitted over the galaxy emission line. The 'X' on the bottom left panel 
represent the expected position of a [\OII] emission line located 3.5\,$\arcsec$ south from the quasar, 
i.e., at the centroid of the galaxy resolved by the SDSS. The data has been smoothed $2\times2$ pixels. 
 \label{2dspec}}
\end{figure*}

Fig.~\ref{2dspec} represents the two-dimensional spectra obtained with the two slit orientations as 
illustrated on Fig.~\ref{im1131}. We removed the quasar trace by fitting a Gaussian in the spatial 
direction, whose amplitude is allowed to vary smoothly (2nd-order polynomial) with the wavelength. 
We also left the position of the Gaussian in the spatial direction to vary linearly with the wavelength, to 
take into account the possible misalignment between the quasar trace and the pixels of the CCD. 

The [\OII] emission can be seen as a bright spot in both spectra. Interestingly, 
despite of aligning the slit with the galaxy seen at 3.5\,$\arcsec$ from the quasar (see Fig.~\ref{im1131}), 
there is no emission line at the corresponding position (marked by a 'X') in the 2D spectrum. On the contrary, 
while the second slit angle was chosen to avoid including the galaxy, the [\OII] emission is still detected. 
This demonstrates the galaxy seen on the SDSS image is not responsible for the detected emission lines. 
{Interestingly, this galaxy does not produce any other detectable absorption line system in the 
QSO spectra over the wavelength range covered by the SDSS and IGO spectra.}
We obtained 
the centroid of the [\OII] emission by fitting the line with a two-dimensional Gaussian function. From triangulating 
the positions with the two slits, we are able to put a good constraint on the position of the emitting region.
We measure an impact parameter $D=1.1\arcsec$, which corresponds to 7~kpc at the redshift of the galaxy.
The centroid of the emission region is shown as a white cross on Fig.~\ref{im1131}.
We see the integrated [\OII] flux measured in IGO and SDSS spectra are consistent with one another suggesting 
that the whole line-emitting region is within the SDSS fibre.

\section{Summary and discussion} 

Taking advantage of the available $\sim$100\,000 fibre spectra of quasars in the Sloan Digital Sky Survey-II, DR7, 
we build a unique sample of 46 star-forming galaxies at $z<0.8$ detected through their nebular 
([\OIII] and/or [\OII]) emission lines seen on top of the background quasar spectra. We show the 
the detectability of [\OIII] lines is not biased by the luminosity of the background quasars. 

We study both the emission and absorption properties of a sub-sample of 17 galaxies at $z\ge0.4$ 
for which the expected positions of \MgII\ lines are covered by the SDSS spectra.
The detections show that we are probing a unbiased population of low luminosity [\OIII]-emitting galaxies 
at small impact parameters ($\la 10$~kpc; i.e. the SDSS fibre radius) from the quasar lines of sight. We 
find that typical properties (metallicity, star-formation rates, kinematics) of these galaxies are 
similar to that of normal star-forming galaxies at these redshifts. 
The low E(B-V) we measure along the quasar lines of sight indicates that the absorption lines 
arise from regions relatively free of dust. This implies that quasar absorbers selected 
upon the presence of cold gas and dust features \citep{Srianand08bump, Srianand08, Noterdaeme09co} might 
still be the best way to probe the interstellar medium in the densest regions of normal galaxies in the distant 
Universe.

We find that the equivalent widths of \MgII\ absorption lines arising from the [\OIII]-selected galaxies 
are skewed towards higher equivalent widths than the overall population of \MgII\ absorbers. However, 
the [\OIII]-selected \MgII\ absorbers represent only a small fraction of the overall \MgII\ population. 
From stacking the spectra of quasars featuring strong \MgII\ absorbers, we detect the [\OII] and [\OIII] 
emission lines. The average line fluxes are below our typical detection limit in individual spectrum. 
This suggests that at least part of the strong ($\WMgII>1$~{\AA}) \MgII\ absorption systems arise from 
low luminosity galaxies at small impact parameters.
Also strong \MgII\ systems have been detected at higher rate around clusters \citep{Lopez08}. In such 
cases, the halo sizes of \MgII\ systems are inferred to be less than 10~$h^{-1}$~kpc \citep{Padilla09}.

The absorption properties of the galaxies indicate that at least half of the emission 
line galaxies in our sample, if not all, contain sufficient neutral gas to produce Damped Lyman-$\alpha$ 
absorption \citep{Rao06} as well as 21-cm absorption \citet{Gupta09} along the quasar line of sight. 
Unfortunately, most of the QSOs in our sample do not have sufficient radio fluxes to carry-out 21-cm searches. 

SDSS spectra allowed us to explore the possible connections between various 
parameters of the galaxies (such as metallicity, dust content and kinematics) derived 
from the absorbing gas and that derived from emission lines in a limited redshift 
range. Nevertheless, our representative sample of 46 galaxies presented in Table~\ref{qso} is ideally 
suited for several follow-up observations using space and ground-based telescopes. This 
should allow one to explore various issues such as: the connection between the reddening 
along the QSO line of sight and the dust extinction in the line-emitting region; the comparison between 
the emission and absorption line metallicities; the dependence of the properties of the 
absorbing gas on the galaxy morphology, kinematics and impact parameter, etc.
Finally, we perform long-slit observations of the most luminous galaxy with \MgII\ absorption 
in our sample (SDSS\,J113108+202151). We show that the [\OII] emission 
detected in the SDSS spectrum is not detected in the extended bright galaxy seen on the SDSS image. This 
once again suggests that one should be cautious in associating intervening absorption (or emission) 
to bright galaxies seen in the field with photometric redshift measurements only.

\section*{Acknowledgements}

We gratefully thank the anonymous referee for thorough reading of the paper and 
helpful comments and suggestions. We also thank P. Petitjean and P. Boiss\'e for useful comments 
on the manuscript.
PN acknowledges support 
from the french Ministry of Foreign and European Affairs. We wish to acknowledge the 
IUCAA/IGO staff for their support during our observations. 
We acknowledge the use of the Sloan Digital Sky Survey. 
Funding for the SDSS and SDSS-II has been 
provided by the Alfred P. Sloan Foundation, the Participating Institutions, the 
National Science Foundation, the U.S. Department of Energy, the National Aeronautics 
and Space Administration, the Japanese Monbukagakusho, the Max Planck Society, and 
the Higher Education Funding Council for England. The SDSS Web Site is \url{http://www.sdss.org}.
The SDSS is managed by the Astrophysical Research Consortium for the Participating Institutions. 
The Participating Institutions are the American Museum of Natural History, Astrophysical 
Institute Potsdam, University of Basel, University of Cambridge, Case Western Reserve 
University, University of Chicago, Drexel University, Fermilab, the Institute for Advanced 
Study, the Japan Participation Group, Johns Hopkins University, the Joint Institute for 
Nuclear Astrophysics, the Kavli Institute for Particle Astrophysics and Cosmology, the 
Korean Scientist Group, the Chinese Academy of Sciences (LAMOST), Los Alamos National 
Laboratory, the Max-Planck-Institute for Astronomy (MPIA), the Max-Planck-Institute for 
Astrophysics (MPA), New Mexico State University, Ohio State University, University of 
Pittsburgh, University of Portsmouth, Princeton University, the United States Naval 
Observatory, and the University of Washington.

\def\aj{AJ}%
\def\actaa{Acta Astron.}%
\def\araa{ARA\&A}%
\def\apj{ApJ}%
\def\apjl{ApJ}%
\def\apjs{ApJS}%
\def\ao{Appl.~Opt.}%
\def\apss{Ap\&SS}%
\def\aap{A\&A}%
\def\aapr{A\&A~Rev.}%
\def\aaps{A\&AS}%
\def\azh{AZh}%
\def\baas{BAAS}%
\def\bac{Bull. astr. Inst. Czechosl.}%
\def\caa{Chinese Astron. Astrophys.}%
\def\cjaa{Chinese J. Astron. Astrophys.}%
\def\icarus{Icarus}%
\def\jcap{J. Cosmology Astropart. Phys.}%
\def\jrasc{JRASC}%
\def\mnras{MNRAS}%
\def\memras{MmRAS}%
\def\na{New A}%
\def\nar{New A Rev.}%
\def\pasa{PASA}%
\def\pra{Phys.~Rev.~A}%
\def\prb{Phys.~Rev.~B}%
\def\prc{Phys.~Rev.~C}%
\def\prd{Phys.~Rev.~D}%
\def\pre{Phys.~Rev.~E}%
\def\prl{Phys.~Rev.~Lett.}%
\def\pasp{PASP}%
\def\pasj{PASJ}%
\def\qjras{QJRAS}%
\def\rmxaa{Rev. Mexicana Astron. Astrofis.}%
\def\skytel{S\&T}%
\def\solphys{Sol.~Phys.}%
\def\sovast{Soviet~Ast.}%
\def\ssr{Space~Sci.~Rev.}%
\def\zap{ZAp}%
\def\nat{Nature}%
\def\iaucirc{IAU~Circ.}%
\def\aplett{Astrophys.~Lett.}%
\def\apspr{Astrophys.~Space~Phys.~Res.}%
\def\bain{Bull.~Astron.~Inst.~Netherlands}%
\def\fcp{Fund.~Cosmic~Phys.}%
\def\gca{Geochim.~Cosmochim.~Acta}%
\def\grl{Geophys.~Res.~Lett.}%
\def\jcp{J.~Chem.~Phys.}%
\def\jgr{J.~Geophys.~Res.}%
\def\jqsrt{J.~Quant.~Spec.~Radiat.~Transf.}%
\def\memsai{Mem.~Soc.~Astron.~Italiana}%
\def\nphysa{Nucl.~Phys.~A}%
\def\physrep{Phys.~Rep.}%
\def\physscr{Phys.~Scr}%
\def\planss{Planet.~Space~Sci.}%
\def\procspie{Proc.~SPIE}%
\let\astap=\aap
\let\apjlett=\apjl
\let\apjsupp=\apjs
\let\applopt=\ao
%

\begin{thebibliography}{60}
\expandafter\ifx\csname natexlab\endcsname\relax\def\natexlab#1{#1}\fi

\bibitem[{{Argence} \& {Lamareille}(2009)}]{Argence09}
{Argence}, B. \& {Lamareille}, F., 2009, \aap, 495, 759

\bibitem[{{Asplund} {et~al.}(2009){Asplund}, {Grevesse}, {Sauval}, \&
  {Scott}}]{Asplund09}
{Asplund}, M., {Grevesse}, N., {Sauval}, A.~J., \& {Scott}, P., 2009, \araa,
  47, 481

\bibitem[{{Bechtold} \& {Ellingson}(1992)}]{Bechtold92}
{Bechtold}, J. \& {Ellingson}, E., 1992, \apj, 396, 20

\bibitem[{{Bergeron} \& {Boiss{\'e}}(1991)}]{Bergeron91}
{Bergeron}, J. \& {Boiss{\'e}}, P., 1991, \aap, 243, 344

\bibitem[{{Bouch{\'e}} {et~al.}(2007){Bouch{\'e}}, {Lehnert}, {Aguirre},
  {P{\'e}roux}, \& {Bergeron}}]{Bouche07}
{Bouch{\'e}}, N., {Lehnert}, M.~D., {Aguirre}, A., {P{\'e}roux}, C., \&
  {Bergeron}, J., 2007, \mnras, 378, 525

\bibitem[{{Bowen} {et~al.}(1995){Bowen}, {Blades}, \& {Pettini}}]{Bowen95}
{Bowen}, D.~V., {Blades}, J.~C., \& {Pettini}, M., 1995, \apj, 448, 634

\bibitem[{{Chen} \& {Lanzetta}(2003)}]{Chen03}
{Chen}, H. \& {Lanzetta}, K.~M., 2003, \apj, 597, 706

\bibitem[{{Christensen} {et~al.}(2009){Christensen}, {Noterdaeme}, {Petitjean},
  {Ledoux}, \& {Fynbo}}]{Christensen09}
{Christensen}, L., {Noterdaeme}, P., {Petitjean}, P., {Ledoux}, C., \& {Fynbo},
  J.~P.~U., 2009, \aap, 505, 1007

\bibitem[{{Dahlen} {et~al.}(2007){Dahlen}, {Mobasher}, {Dickinson}, {Ferguson},
  {Giavalisco}, {Kretchmer}, \& {Ravindranath}}]{Dahlen07}
{Dahlen}, T., {Mobasher}, B., {Dickinson}, M., {Ferguson}, H.~C., {Giavalisco},
  M., {Kretchmer}, C., \& {Ravindranath}, S., 2007, \apj, 654, 172

\bibitem[{{de Mello} {et~al.}(2006){de Mello}, {Wadadekar}, {Dahlen},
  {Casertano}, \& {Gardner}}]{DeMello06}
{de Mello}, D.~F., {Wadadekar}, Y., {Dahlen}, T., {Casertano}, S., \&
  {Gardner}, J.~P., 2006, \aj, 131, 216

\bibitem[{{Ellison}(2006)}]{Ellison06}
{Ellison}, S.~L., 2006, \mnras, 368, 335

\bibitem[{{Ellison} {et~al.}(2005){Ellison}, {Hall}, \& {Lira}}]{Ellison05}
{Ellison}, S.~L., {Hall}, P.~B., \& {Lira}, P., 2005, \aj, 130, 1345

\bibitem[{{Gordon} {et~al.}(2003){Gordon}, {Clayton}, {Misselt}, {Landolt}, \&
  {Wolff}}]{Gordon03}
{Gordon}, K.~D., {Clayton}, G.~C., {Misselt}, K.~A., {Landolt}, A.~U., \&
  {Wolff}, M.~J., 2003, \apj, 594, 279

\bibitem[{{Gupta} {et~al.}(2009){Gupta}, {Srianand}, {Petitjean}, {Noterdaeme},
  \& {Saikia}}]{Gupta09}
{Gupta}, N., {Srianand}, R., {Petitjean}, P., {Noterdaeme}, P., \& {Saikia},
  D.~J., 2009, \mnras, 398, 201

\bibitem[{{Hippelein} {et~al.}(2003){Hippelein}, {Maier}, {Meisenheimer},
  {Wolf}, {Fried}, {von Kuhlmann}, {K{\"u}mmel}, {Phleps}, \&
  {R{\"o}ser}}]{Hippelein03}
{Hippelein}, H., {Maier}, C., {Meisenheimer}, K., {et~al.}, 2003, \aap, 402, 65

\bibitem[{{Kacprzak} {et~al.}(2008){Kacprzak}, {Churchill}, {Steidel}, \&
  {Murphy}}]{Kacprzak08}
{Kacprzak}, G.~G., {Churchill}, C.~W., {Steidel}, C.~C., \& {Murphy}, M.~T.,
  2008, \aj, 135, 922

\bibitem[{{Kennicutt}(1998)}]{Kennicutt98}
{Kennicutt}, Jr., R.~C., 1998, \araa, 36, 189

\bibitem[{{Kewley} \& {Ellison}(2008)}]{Kewley08}
{Kewley}, L.~J. \& {Ellison}, S.~L., 2008, \apj, 681, 1183

\bibitem[{{Kewley} {et~al.}(2004){Kewley}, {Geller}, \& {Jansen}}]{Kewley04}
{Kewley}, L.~J., {Geller}, M.~J., \& {Jansen}, R.~A., 2004, \aj, 127, 2002

\bibitem[{{Khare} {et~al.}(2005){Khare}, {York}, {vanden Berk}, {Kulkarni},
  {Crotts}, {Welty}, {Lauroesch}, {Richards}, {Alsayyad}, {Kumar}, {Lundgren},
  {Shanidze}, {Vanlandingham}, {Baugher}, {Hall}, {Jenkins}, {Menard}, {Rao},
  {Turnshek}, \& {Yip}}]{Khare05}
{Khare}, P., {York}, D.~G., {vanden Berk}, D., {et~al.}, 2005, in IAU Colloq.
  199: Probing Galaxies through Quasar Absorption Lines, {P.~Williams,
  C.-G.~Shu, \& B.~Menard}, ed., pp. 427--429

\bibitem[{{Kisielius} {et~al.}(2009){Kisielius}, {Storey}, {Ferland}, \&
  {Keenan}}]{Kisielius09}
{Kisielius}, R., {Storey}, P.~J., {Ferland}, G.~J., \& {Keenan}, F.~P., 2009,
  \mnras, 397, 903

\bibitem[{{Kobulnicky} {et~al.}(1999){Kobulnicky}, {Kennicutt}, \&
  {Pizagno}}]{Kobulnicky99}
{Kobulnicky}, H.~A., {Kennicutt}, Jr., R.~C., \& {Pizagno}, J.~L., 1999, \apj,
  514, 544

\bibitem[{{Kobulnicky} \& {Phillips}(2003)}]{Kobulnicky03a}
{Kobulnicky}, H.~A. \& {Phillips}, A.~C., 2003, \apj, 599, 1031

\bibitem[{{Kobulnicky} {et~al.}(2003){Kobulnicky}, {Willmer}, {Phillips},
  {Koo}, {Faber}, {Weiner}, {Sarajedini}, {Simard}, \& {Vogt}}]{Kobulnicky03b}
{Kobulnicky}, H.~A., {Willmer}, C.~N.~A., {Phillips}, A.~C., {et~al.}, 2003,
  \apj, 599, 1006

\bibitem[{{Lamareille} {et~al.}(2006){Lamareille}, {Contini}, {Brinchmann}, {Le
  Borgne}, {Charlot}, \& {Richard}}]{Lamareille06}
{Lamareille}, F., {Contini}, T., {Brinchmann}, J., {Le Borgne}, J.-F.,
  {Charlot}, S., \& {Richard}, J., 2006, \aap, 448, 907

\bibitem[{{Le Brun} {et~al.}(1997){Le Brun}, {Bergeron}, {Boisse}, \&
  {Deharveng}}]{LeBrun97}
{Le Brun}, V., {Bergeron}, J., {Boisse}, P., \& {Deharveng}, J.~M., 1997, \aap,
  321, 733

\bibitem[{{Ledoux} {et~al.}(2006){Ledoux}, {Petitjean}, {Fynbo}, {M{\o}ller},
  \& {Srianand}}]{Ledoux06a}
{Ledoux}, C., {Petitjean}, P., {Fynbo}, J.~P.~U., {M{\o}ller}, P., \&
  {Srianand}, R., 2006, \aap, 457, 71

\bibitem[{{Lilly} {et~al.}(2003){Lilly}, {Carollo}, \& {Stockton}}]{Lilly03}
{Lilly}, S.~J., {Carollo}, C.~M., \& {Stockton}, A.~N., 2003, \apj, 597, 730

\bibitem[{{Lopez} {et~al.}(2008){Lopez}, {Barrientos}, {Lira}, {Padilla},
  {Gilbank}, {Gladders}, {Maza}, {Tejos}, {Vidal}, \& {Yee}}]{Lopez08}
{Lopez}, S., {Barrientos}, L.~F., {Lira}, P., {et~al.}, 2008, \apj, 679, 1144

\bibitem[{{Ly} {et~al.}(2007){Ly}, {Malkan}, {Kashikawa}, {Shimasaku}, {Doi},
  {Nagao}, {Iye}, {Kodama}, {Morokuma}, \& {Motohara}}]{Ly07}
{Ly}, C., {Malkan}, M.~A., {Kashikawa}, N., {et~al.}, 2007, \apj, 657, 738

\bibitem[{{Markwardt}(2009)}]{Markwardt09}
{Markwardt}, C.~B., 2009, ArXiv e-prints 0902.2850

\bibitem[{{M{\'e}nard} {et~al.}(2005){M{\'e}nard}, {Zibetti}, {Nestor}, \&
  {Turnshek}}]{Menard05}
{M{\'e}nard}, B., {Zibetti}, S., {Nestor}, D., \& {Turnshek}, D., 2005, in IAU
  Colloq. 199: Probing Galaxies through Quasar Absorption Lines, {P.~Williams,
  C.-G.~Shu, \& B.~Menard}, ed., pp. 86--91

\bibitem[{{Mouhcine} {et~al.}(2006){Mouhcine}, {Bamford},
  {Arag{\'o}n-Salamanca}, {Nakamura}, \& {Milvang-Jensen}}]{Mouhcine06}
{Mouhcine}, M., {Bamford}, S.~P., {Arag{\'o}n-Salamanca}, A., {Nakamura}, O.,
  \& {Milvang-Jensen}, B., 2006, \mnras, 369, 891

\bibitem[{{Moustakas} \& {Kennicutt}(2006)}]{Moustakas06}
{Moustakas}, J. \& {Kennicutt}, Jr., R.~C., 2006, \apj, 651, 155

\bibitem[{{Murphy} {et~al.}(2007){Murphy}, {Curran}, {Webb}, {M{\'e}nager}, \&
  {Zych}}]{Murphy07}
{Murphy}, M.~T., {Curran}, S.~J., {Webb}, J.~K., {M{\'e}nager}, H., \& {Zych},
  B.~J., 2007, \mnras, 376, 673

\bibitem[{{Murphy} \& {Liske}(2004)}]{Murphy04}
{Murphy}, M.~T. \& {Liske}, J., 2004, \mnras, 354, L31

\bibitem[{{Nestor} {et~al.}(2008){Nestor}, {Pettini}, {Hewett}, {Rao}, \&
  {Wild}}]{Nestor08}
{Nestor}, D.~B., {Pettini}, M., {Hewett}, P.~C., {Rao}, S., \& {Wild}, V.,
  2008, \mnras, 390, 1670

\bibitem[{{Nestor} {et~al.}(2003){Nestor}, {Rao}, {Turnshek}, \& {Vanden
  Berk}}]{Nestor03}
{Nestor}, D.~B., {Rao}, S.~M., {Turnshek}, D.~A., \& {Vanden Berk}, D., 2003,
  \apjl, 595, L5

\bibitem[{{Nestor} {et~al.}(2005){Nestor}, {Turnshek}, \& {Rao}}]{Nestor05}
{Nestor}, D.~B., {Turnshek}, D.~A., \& {Rao}, S.~M., 2005, \apj, 628, 637

\bibitem[{{Noterdaeme} {et~al.}(2009{\natexlab{a}}){Noterdaeme}, {Ledoux},
  {Srianand}, {Petitjean}, \& {Lopez}}]{Noterdaeme09co}
{Noterdaeme}, P., {Ledoux}, C., {Srianand}, R., {Petitjean}, P., \& {Lopez},
  S., 2009{\natexlab{a}}, \aap, 503, 765

\bibitem[{{Noterdaeme} {et~al.}(2009{\natexlab{b}}){Noterdaeme}, {Petitjean},
  {Ledoux}, \& {Srianand}}]{Noterdaeme09dla}
{Noterdaeme}, P., {Petitjean}, P., {Ledoux}, C., \& {Srianand}, R.,
  2009{\natexlab{b}}, \aap, 505, 1087

\bibitem[{{Noterdaeme} {et~al.}(2007){Noterdaeme}, {Petitjean}, {Srianand},
  {Ledoux}, \& {Le Petit}}]{Noterdaeme07}
{Noterdaeme}, P., {Petitjean}, P., {Srianand}, R., {Ledoux}, C., \& {Le Petit},
  F., 2007, \aap, 469, 425

\bibitem[{{Padilla} {et~al.}(2009){Padilla}, {Lacerna}, {Lopez}, {Barrientos},
  {Lira}, {Andrews}, \& {Tejos}}]{Padilla09}
{Padilla}, N., {Lacerna}, I., {Lopez}, S., {Barrientos}, L.~F., {Lira}, P.,
  {Andrews}, H., \& {Tejos}, N., 2009, \mnras, 395, 1135

\bibitem[{{Rao} {et~al.}(2003){Rao}, {Nestor}, {Turnshek}, {Lane}, {Monier}, \&
  {Bergeron}}]{Rao03}
{Rao}, S.~M., {Nestor}, D.~B., {Turnshek}, D.~A., {Lane}, W.~M., {Monier},
  E.~M., \& {Bergeron}, J., 2003, \apj, 595, 94

\bibitem[{{Rao} \& {Turnshek}(2000)}]{Rao00}
{Rao}, S.~M. \& {Turnshek}, D.~A., 2000, \apjs, 130, 1

\bibitem[{{Rao} {et~al.}(2006){Rao}, {Turnshek}, \& {Nestor}}]{Rao06}
{Rao}, S.~M., {Turnshek}, D.~A., \& {Nestor}, D.~B., 2006, \apj, 636, 610

\bibitem[{{Spergel} {et~al.}(2003){Spergel}, {Verde}, {Peiris}, {Komatsu},
  {Nolta}, {Bennett}, {Halpern}, {Hinshaw}, {Jarosik}, {Kogut}, {Limon},
  {Meyer}, {Page}, {Tucker}, {Weiland}, {Wollack}, \& {Wright}}]{Spergel03}
{Spergel}, D.~N., {Verde}, L., {Peiris}, H.~V., {et~al.}, 2003, \apjs, 148, 175

\bibitem[{{Srianand} {et~al.}(2008{\natexlab{a}}){Srianand}, {Gupta},
  {Petitjean}, {Noterdaeme}, \& {Saikia}}]{Srianand08bump}
{Srianand}, R., {Gupta}, N., {Petitjean}, P., {Noterdaeme}, P., \& {Saikia},
  D.~J., 2008{\natexlab{a}}, \mnras, 391, L69

\bibitem[{{Srianand} {et~al.}(2008{\natexlab{b}}){Srianand}, {Noterdaeme},
  {Ledoux}, \& {Petitjean}}]{Srianand08}
{Srianand}, R., {Noterdaeme}, P., {Ledoux}, C., \& {Petitjean}, P.,
  2008{\natexlab{b}}, \aap, 482, L39

\bibitem[{{Steidel}(1995)}]{Steidel95}
{Steidel}, C.~C., 1995, in QSO Absorption Lines, {Meylan}, G., ed., p. 139

\bibitem[{{Steidel} {et~al.}(2002){Steidel}, {Kollmeier}, {Shapley},
  {Churchill}, {Dickinson}, \& {Pettini}}]{Steidel02}
{Steidel}, C.~C., {Kollmeier}, J.~A., {Shapley}, A.~E., {Churchill}, C.~W.,
  {Dickinson}, M., \& {Pettini}, M., 2002, \apj, 570, 526

\bibitem[{{Storey} \& {Zeippen}(2000)}]{Storey00}
{Storey}, P.~J. \& {Zeippen}, C.~J., 2000, \mnras, 312, 813

\bibitem[{{Tremonti} {et~al.}(2004){Tremonti}, {Heckman}, {Kauffmann},
  {Brinchmann}, {Charlot}, {White}, {Seibert}, {Peng}, {Schlegel}, {Uomoto},
  {Fukugita}, \& {Brinkmann}}]{Tremonti04}
{Tremonti}, C.~A., {Heckman}, T.~M., {Kauffmann}, G., {et~al.}, 2004, \apj,
  613, 898

\bibitem[{{Tripp} \& {Bowen}(2005)}]{Tripp05}
{Tripp}, T.~M. \& {Bowen}, D.~V., 2005, in IAU Colloq. 199: Probing Galaxies
  through Quasar Absorption Lines, {Williams}, P., {Shu}, C.-G., \& {Menard},
  B., eds., pp. 5--23

\bibitem[{{Vanden Berk} {et~al.}(2001){Vanden Berk}, {Richards}, {Bauer},
  {Strauss}, {Schneider}, {Heckman}, {York}, {Hall}, {Fan}, {Knapp},
  {Anderson}, {Annis}, {Bahcall}, {Bernardi}, {Briggs}, {Brinkmann}, {Brunner},
  {Burles}, {Carey}, {Castander}, {Connolly}, {Crocker}, {Csabai}, {Doi},
  {Finkbeiner}, {Friedman}, {Frieman}, {Fukugita}, {Gunn}, {Hennessy},
  {Ivezi{\'c}}, {Kent}, {Kunszt}, {Lamb}, {Leger}, {Long}, {Loveday}, {Lupton},
  {Meiksin}, {Merelli}, {Munn}, {Newberg}, {Newcomb}, {Nichol}, {Owen}, {Pier},
  {Pope}, {Rockosi}, {Schlegel}, {Siegmund}, {Smee}, {Snir}, {Stoughton},
  {Stubbs}, {SubbaRao}, {Szalay}, {Szokoly}, {Tremonti}, {Uomoto}, {Waddell},
  {Yanny}, \& {Zheng}}]{VandenBerk01}
{Vanden Berk}, D.~E., {Richards}, G.~T., {Bauer}, A., {et~al.}, 2001, \aj, 122,
  549

\bibitem[{{Vivek} {et~al.}(2009){Vivek}, {Srianand}, {Noterdaeme}, {Mohan}, \&
  {Kuriakosde}}]{Vivek09}
{Vivek}, M., {Srianand}, R., {Noterdaeme}, P., {Mohan}, V., \& {Kuriakosde},
  V.~C., 2009, \mnras, L321+

\bibitem[{{Wild} {et~al.}(2006){Wild}, {Hewett}, \& {Pettini}}]{Wild06}
{Wild}, V., {Hewett}, P.~C., \& {Pettini}, M., 2006, \mnras, 367, 211

\bibitem[{{Wild} {et~al.}(2007){Wild}, {Hewett}, \& {Pettini}}]{Wild07}
---, 2007, \mnras, 374, 292

\bibitem[{{York} {et~al.}(2006){York}, {Khare}, {Vanden Berk}, {Kulkarni},
  {Crotts}, {Lauroesch}, {Richards}, {Schneider}, {Welty}, {Alsayyad}, {Kumar},
  {Lundgren}, {Shanidze}, {Smith}, {Vanlandingham}, {Baugher}, {Hall},
  {Jenkins}, {Menard}, {Rao}, {Tumlinson}, {Turnshek}, {Yip}, \&
  {Brinkmann}}]{York06}
{York}, D.~G., {Khare}, P., {Vanden Berk}, D., {et~al.}, 2006, \mnras, 367, 945

\bibitem[{{Zych} {et~al.}(2007){Zych}, {Murphy}, {Pettini}, {Hewett},
  {Ryan-Weber}, \& {Ellison}}]{Zych07}
{Zych}, B.~J., {Murphy}, M.~T., {Pettini}, M., {Hewett}, P.~C., {Ryan-Weber},
  E.~V., \& {Ellison}, S.~L., 2007, \mnras, 379, 1409

\end{thebibliography}

\end{document}